\documentclass[]{jfm}

\usepackage{graphicx}
\usepackage{epstopdf,epsfig}
\usepackage{newtxtext}
\usepackage{newtxmath}
\usepackage{natbib}
\usepackage{hyperref}
\usepackage{tikz}
\usepackage{xspace}
\usetikzlibrary{patterns}
\usetikzlibrary{calc}

\tikzset{
    ncbar angle/.initial=90,
    ncbar/.style={
        to path=(\tikztostart)
        -- ($(\tikztostart)!#1!\pgfkeysvalueof{/tikz/ncbar angle}:(\tikztotarget)$)
        -- ($(\tikztotarget)!($(\tikztostart)!#1!\pgfkeysvalueof{/tikz/ncbar angle}:(\tikztotarget)$)!\pgfkeysvalueof{/tikz/ncbar angle}:(\tikztostart)$)
        -- (\tikztotarget)
    },
    ncbar/.default=0.5cm,
}

\tikzset{square left brace/.style={ncbar=0.5cm}}
\tikzset{square right brace/.style={ncbar=-0.5cm}}

\tikzset{round left paren/.style={ncbar=0.5cm,out=120,in=-120}}
\tikzset{round right paren/.style={ncbar=0.5cm,out=60,in=-60}}

\hypersetup{
    colorlinks = true,
    urlcolor   = blue,
    citecolor  = black,
}

\newcommand{\RomanNumeralCaps}[1]
\linenumbers

\newcommand{\vx}{\boldsymbol{x}}
\newcommand{\vy}{\boldsymbol{y}}
\newcommand{\vz}{\boldsymbol{z}}
\newcommand{\vk}{\boldsymbol{k}}
\newcommand{\vvv}{\boldsymbol{\rm v}}
\newcommand{\sv}{{\rm v}}
\newcommand{\spr}{\pi}
\newcommand{\vnabla}{\boldsymbol{\nabla}}
\newcommand{\vell}{\boldsymbol{\ell}}
\newcommand{\vu}{\boldsymbol{u}}
\newcommand{\vU}{\boldsymbol{U}}
\newcommand{\vL}{\boldsymbol{L}}
\newcommand{\vp}{\boldsymbol{p}}
\newcommand{\vq}{\boldsymbol{q}}
\newcommand{\vzero}{\boldsymbol{0}}

\newcommand{\vf}{\boldsymbol{f}}
\newcommand{\vN}{\boldsymbol{N}}
\newcommand{\vJ}{\boldsymbol{J}}

\newcommand{\vrho}{\boldsymbol{\rho}}
\newcommand{\veta}{\boldsymbol{\eta}}
\newcommand{\vvvarrho}{\boldsymbol{\varrho}}
\newcommand{\veps}{\boldsymbol{\varepsilon}}
\newcommand{\vr}{\boldsymbol{r}}
\newcommand{\vxi}{\boldsymbol{\xi}}
\newcommand{\vX}{\boldsymbol{X}}
\newcommand{\tf}{\tilde{f}}

\newcommand{\ie}{{\it i.e.}~}

\newcommand{\sref}[1]{Sec.~\ref{#1}}
\newcommand{\fref}[1]{Fig.~\ref{#1}}

\newcommand{\aref}[1]{Appendix~\ref{#1}}
\newcommand{\Eq}[1]{Eq.~(\ref{#1})}
\newcommand{\eq}[1]{(\ref{#1})}
\newcommand{\Eqs}[2]{Eqs.~(\ref{#1},\ref{#2})}
\newcommand{\crefs}{ }

\usepackage{abbrevs}
\usepackage{etoolbox}
\newabbrev\RG{Renormalisation Group (RG)}[RG]
\newabbrev\FRG{Functional Renormalisation Group (FRG)}[FRG]
\newabbrev\IR{infrared (IR)}[IR]
\newabbrev\UV{ultraviolet (UV)}[UV]
\newabbrev\BMW{Blaizot--Mendez--Wschebor (BMW)}[BMW]
\newabbrev\MSR{Martin--Siggia--Rose--Janssen--de Dominicis (MSRJD)}[MSRJD]
\newabbrev\NS{Navier-Stokes (NS)}[NS]
\newabbrev\DNS{Direct Numerical Simulations (DNS)}[DNS]

\robustify{\RG}
\robustify{\FRG}
\robustify{\UV}
\robustify{\IR}
\robustify{\BMW}
\robustify{\MSR}
\robustify{\NS}
\robustify{\DNS}

\makeatletter
\renewcommand\maybe@space@{%
  \maybe@ictrue 
  \expandafter   \@tfor
    \expandafter \reserved@a
    \expandafter :%
    \expandafter =%
                 \nospacelist
                 \do \t@st@ic
  \ifmaybe@ic 
    \space
  \fi
}
\makeatother

\title{Functional renormalisation group for turbulence}

\author{L\'eonie Canet\aff{1},\aff{2}
  \corresp{\email{leonie.canet@grenoble.cnrs.fr}}}

\affiliation{\aff{1} Universit{\'e} Grenoble Alpes, CNRS, LPMMC, 38000 Grenoble, France.
\aff{2} Institut Universitaire de France, 1 rue Descartes, 75005 Paris, France}

\begin{document}

\maketitle

\graphicspath{{../FIG/}}

\begin{abstract}
Turbulence is a complex nonlinear and multi-scale phenomenon. Although the fundamental underlying equations, the Navier-Stokes equations, have been known for two centuries, it remains extremely challenging to extract from them the statistical properties of turbulence. Therefore, for practical purpose, a sustained effort has been devoted to obtaining some effective description of turbulence, that we may call turbulence modelling, or statistical theory of turbulence. In this respect, the Renormalisation Group (RG) appears as a tool of choice, since it is precisely designed to provide effective theories from fundamental equations by performing in a systematic way the average over fluctuations. However, for Navier-Stokes turbulence, a suitable framework for the RG, allowing in particular for non-perturbative approximations, have been missing,  which has thwarted for long RG applications. This framework is provided by the modern formulation of the RG called functional renormalisation group. The use of the FRG  has rooted important progress in the theoretical understanding of homogeneous and isotropic turbulence. The major one is the rigorous derivation, from the Navier-Stokes equations, of an analytical expression for any Eulerian multi-point multi-time correlation function, which is exact in the limit of large wavenumbers. We propose in this {\it JFM Perspectives} a survey of  the FRG method for turbulence. We provide a basic introduction to the FRG and emphasise  how the field-theoretical framework allows one to systematically and profoundly exploit the symmetries. We stress that the FRG enables one to describe fully developed turbulence forced at large scales, which was not accessible by perturbative means. We show that  it yields  the energy spectrum and second-order structure function with accurate estimates of the related constants, and also the behaviour of the spectrum in the near-dissipative range.  Finally, we expound the derivation of the spatio-temporal behaviour of $n$-point correlation functions, and largely illustrate these results through the analysis of data from experiments and direct numerical simulations.
\end{abstract}


\tableofcontents

\section{Introduction}
\label{sec:intro}

This {\it JFM Perspectives} proposes a guided tour through the recent achievements and  open prospects of the \FRG formalism applied to the problem of turbulence. Why the \RG should be useful at all to study turbulence ? The very essence of the \RG is to eliminate degrees of freedom in a systematic way, by averaging over fluctuations at all scales, starting from the fundamental description of a system. It thereby  provides an effective model  ``from first principles''. This is precisely what is needed, and often lacking, in many turbulence problems. Indeed, the fundamental equations for fluid dynamics are the \NS equations, but the number of degrees of freedom necessary to describe turbulence grows as Re$^{9/4}$. Conceiving an appropriate modelling, \eg devising some reliable effective equations to simplify the problem, is a major goal in many applications. This is of course a very difficult task in general, and the \RG  certainly offers a valuable tool, at least for idealised situations. The program of applying  \RG to turbulence  started in the early eighties, but it was bound to rely on the tools at disposal at that time, which were based on perturbative expansions. However,  there is no small parameter for turbulence, and this program was largely thwarted. The only way out was to introduce in the theoretical description a forcing of the turbulence with a power-law spectrum, \ie acting at all scales, which is  unphysical (\cite{Eyink94}). Results for a physical large-scale forcing are then extrapolated from a certain limit which cannot be justified. This is explained in more details in \sref{sec:RG}.
 
 In the meantime, 
functional and non-perturbative ways of realising the \RG procedure were developed (\cite{Wetterich93,Morris94}). They allow one in particular to incorporate from the onset a physical large-scale forcing,  and thereby to completely circumvent the inextricable issues posed by perturbative expansions. Indeed, the \FRG  offers a very powerful way to implement non-perturbative -- yet controlled -- approximation schemes, \ie approximations not requiring  the existence of a small parameter. The main scheme consists in the use of an ansatz for a central object in field theory which is called effective action (see \sref{sec:fieldtheory}).  The construction of this ansatz relies on the fundamental symmetries of the problem. This approach has yielded many results, allowing one to obtain   very accurate -- although approximate -- estimates for various quantities, including whole functions, not only numbers. We present as examples in \sref{sec:KPZ}  the results for the scaling functions for Burgers equation, and in \sref{sec:fixed-point} the results for the energy spectrum,  the second and third order structure functions  for \NS equations, including an accurate estimate of the Kolmogorov constant. More strikingly, the \FRG can also be used to obtain exact results, which are not based on an ansatz. In turbulence, these results are  explicit expressions for  $n$-point Eulerian spatio-temporal correlation functions. They are obtained as an exact limit at large wavenumbers. Given the very few exact results available for 3D turbulence, this result is particularly remarkable and is the main topic of this {\it  Perspectives}.

 Let us emphasise that all the results obtained so far with \FRG  concern  stationary, homogeneous and isotropic turbulence. This {\it  Perspectives} is thus focused on this idealised situation. However, the scope of application of \FRG is not restricted to this situation. It certainly cannot tackle arbitrarily complicated situations and would be of no use to determine the precise flow configuration around such particular airfoils or in some specific meteorological conditions. However, it could be used, and be efficient, for small-scale modelling, as it is designed to provide an effective description, \eg the form of the coarse-grained effective equations, at any given scale, in particular at the grid scale of large-scale numerical simulations for instance. This direction of research is still in its infancy, and will not be further developed in this contribution, although it definitely maps out a promising program for the future.

 Let us also point out that there can be different level of reading of this {\it Perspectives}. Some parts (\eg \sref{sec:fieldtheory}, \sref{sec:FRG}, \sref{sec:general-flow} and \sref{sec:general-solution}) are rather technical, with the objective of  providing a comprehensive introduction on the basic settings of the \FRG for turbulence, and more importantly on the essential ingredients, in particular symmetries, that enter approximation schemes in this framework. Therefore, the underlying hypotheses are clearly stated to highlight the range of validity of the different results presented.  Other parts (\eg \sref{sec:Burgers1D}, \sref{sec:BurgersD}, \sref{sec:spectrum},  \sref{sec:DNS} and \sref{sec:scalar}) are devoted to illustrating the physical implications of these results, by comparing them with actual data and observations from experiments and \DNS. Therefore one may first wish to focus on the concrete outcomes of \FRG, deferring the  technical aspects. 
 
 In details, this {\it JFM Perspectives}  is organised as follows. We start in \sref{sec:RG} by stressing the interest of \RG  as a mean to study turbulence and its different implementations. We explain in \sref{sec:Langevin} how one can represent stochastic fluctuations as a path integral, to arrive at the field theory for  turbulence in \sref{sec:actions}. A key advantage of this formulation is that it provides a framework to deeply exploit the symmetries. We show  in \sref{sec:Ward} how one can derive exact identities relating different correlation functions from symmetries (and extended symmetries) of the field theory. These identities include well-known relations such as the K\'arm\'an-Howarth or Yaglom relations (\sref{sec:Karman}), but are far more general. The \FRG formalism is introduced in \sref{sec:FRG} with the standard approximation schemes used within this framework. To illustrate the application of this method, we start in \sref{sec:KPZ} with the Burgers equation as a warm-up example. We then  present  in \sref{sec:fixed-point} a first important result for \NS turbulence ensuing from \FRG,  which is the existence of a fixed-point corresponding to the stationary turbulence forced at large scales, and the determination of the associated statistical properties. The main achievement following from \FRG method, addressed in \sref{sec:largep},  is the derivation of the formal general expression of $n$-point spatio-temporal correlation functions at large wavenumbers. The rest of this {\it Perspectives} is dedicated to illustrating these results, for \NS turbulence in \sref{sec:DNS}, and for passive scalar turbulence in \sref{sec:scalar}. Another result, concerning the form of the kinetic energy spectrum in the near-dissipation range of scale is briefly mentioned in \sref{sec:dissipative}. Finally, the conclusion and  perspectives are gathered in \sref{sec:perspectives}.

\section{Scale invariance and the Renormalisation Group} 
\label{sec:RG}  
\subsection{Renormalisation Group and turbulence} 
 
 Let us explain why \RG should be useful to study turbulence  from the point of view of statistical physics. The aim of statistical physics is to determine from a fundamental theory at the microscopic scale the effective behaviour at the macroscopic scale of the system, comprising a large number $\sim N_A$ of particles (in a broad sense).  This requires to  average over stochastic fluctuations (thermal, quantum, {\it etc\dots}). When the fluctuations are Gaussian, and elementary constituents are non-interacting, central limit theorem applies and allows one to perform the averaging (which is how one obtains \eg the equation of states of the perfect gas). However, when the system becomes strongly correlated, this procedure fails since the constituents are no longer statistically independent. This problem appeared particularly thorny for critical phenomena, and have impeded for long progress in their theoretical understanding. Indeed, at a second order phase transition, the correlation length of the system diverges, which means that all the degrees of freedom are correlated and fluctuations develop at all scales. The divergence of the correlation length   leads to the quintessential property of scale invariance, characterised by universal scaling laws, with anomalous exponents, \ie exponents not complying with dimensional analysis.  
 
 The major breakthrough to understand the physical origin of this anomalous behaviour, and primarily to compute it, was achieved with the \RG.
   Although the \RG had already been used under other forms in high-energy physics, it acquired its plain physical meaning from the work of K. Wilson (\cite{Wilson74}). The \RG provides the tool to perform the average over fluctuations, whatever their nature, \ie eliminate degrees of freedom, even in the presence of strong correlations, and thereby to build the effective theory at large scale from the microscopic one. One of the key feature of the \RG  is that all the useful information is encoded in the \RG\; {\it flow}, \ie in the differential equation describing the change of the system under a change of scale. In particular, a critical point, associated with scale invariance, corresponds by essence to a fixed-point of the \RG flow. Let us notice that the notion of large scale is relative to the microscopic one, and it depends on the context. For turbulence, the microscopic scale, denoted $\Lambda^{-1}$, refers to the scale at which the continuum description of a fluid, in terms of \NS equation, is valid, say the order of the mean-free-path in the fluid (much  smaller than the Kolmogorov scale $\eta$). The ``large'' scale of the \RG then refers to typical scales at which the  behaviour of the fluid is observed, \ie   inertial or dissipation ranges, thus including the usual ``small'' scales of turbulence.

 The analogy between critical phenomena and turbulence is obvious, and was early pointed out in (\cite{Nelkin74}), and later refined in \eg (\cite{Eyink94gold}). Indeed, when turbulence is fully developed,  one observes in the inertial range universal behaviours, described by scaling laws with anomalous critical exponents, akin at an equilibrium second-order phase transition. As the \RG had been so successful in the latter case, it early arose as the choice candidate to tackle the former. Concomitantly, the \RG was extended to study not only the equilibrium but also the dynamics of systems (\cite{Martin73,Janssen76,Dominicis76}),  and the first implementations of the ``dynamical \RG'' to study turbulence date back to the early eighties (\cite{Forster77,Dominicis79,Fournier83,Yakhot86}). However, the formulation of the \RG has remained for long intimately linked with perturbative expansions, relying on the existence of a small parameter. This small parameter is generically chosen as the distance $\varepsilon = d_c-d$ to an upper critical dimension $d_c$, which is the dimension where the fixed-point associated with the phase transition under study becomes Gaussian, and the interaction coupling vanishes. 
 In the paradigmatic example of the $\phi^4$ theory which describes the second-order phase transition in the Ising universality class,  the interaction coupling $g$ has a scaling dimension $L^{4-d}$.  Thus it vanishes in the $L\to\infty$ limit for $d\geq d_c=4$, which means that fluctuations become negligible and the mean-field  approximation suffices to provide a reliable description. The Wilson-Fisher fixed-point describing the transition below $d_c$  can then be captured by a perturbative expansion in $\varepsilon = d_c-d$ and the coupling $g$.

 In contrast, in turbulence, the ``interaction'' is the non-linear advection term, whose ``coupling'' is unity, \ie it is not small, and does not vanish in any dimension. Thus, one lacks a small parameter to control perturbative expansion.
  The usual strategy has been to introduce an artificial  parameter $\varepsilon$ through a forcing with power-law correlations behaving in wavenumber space as $p^{4-d-\varepsilon}$, \ie applying a forcing on all scales, which is unphysical (\cite{Eyink94}) \footnote{Note that the letter $p$ is used to denote a wavenumber throughout this {\it Perspectives}. The pressure is denoted with a  letter $\spr$ to avoid any confusion.}. 
 Fully developed turbulence in $d=3$ should then correspond to an \IR dominated spectrum of the stirring force, as it occurs for $\varepsilon \geq 2$, hence for large values  for which the extrapolation of the perturbative expansions is fragile. Moreover, one finds an $\varepsilon$-dependent  fixed-point, with  an energy spectrum $E(p)\propto p^{1-2\varepsilon/3}$. The K41 value is recovered for  $\varepsilon=4$, but this value should somehow freeze for  $\varepsilon$ larger than 4, and such a freezing mechanism can only be invoked within the perturbative analysis (\cite{Fournier83}). In fact, the situation is even worse since it was recently shown that the turbulence generated by a power-law forcing or by a large-scale forcing is simply different (it corresponds to two distinct fixed-points of the \RG, in particular the latter is intermittent whereas the former is not for any value of $\varepsilon$ including $\varepsilon=4$) (\cite{Fontaine2022sabra}). Therefore, \NS turbulence with a large-scale forcing simply cannot be extrapolated from the setting with  power-law forcing.
   Not only recovering the K41 spectrum turns out to be  difficult within this framework, but the first \RG analyses also failed to capture the sweeping effect (\cite{Yakhot89,Chen89,Nelkin90}), and led to the conclusion that one should go to a quasi-Lagrangian framework to obtain a reliable description (\cite{Lvov95a,Lvov97}). However, the difficulties encountered were severe enough to thwart progress in this direction. We refer to (\cite{Smith98,Adzhemyan99,Zhou2010}) for reviews of these developments.

 In the meantime, a novel formulation of the \RG has emerged, which allows for non-perturbative approximation schemes, and thereby bypasses the need of a small parameter. The \FRG is a modern implementation of  Wilson's original conception of the \RG (\cite{Wilson74}). It was formulated in the early nineties (\cite{Wetterich93,Morris94,Ellwanger94}), and widely developed since then (\cite{Berges2002,Kopietz2010,Delamotte2012,Dupuis2021}). One of the noticeable features of this formalism is its versatility, as testified by its wide range of applications, from high-energy physics (QCD and quantum gravity) to condensed matter (fermionic and bosonic quantum systems) and classical statistical physics, including non-equilibrium classical and quantum systems or disordered ones. We refer the interested reader to (\cite{Dupuis2021}) for a recent review. This has led to fertile methodological transfers, borrowing from an area to the other.  The \FRG was moreover promoted to a high-precision method, since it was shown to yield for the archetypical three-dimensional Ising model  results for the critical exponents competing with the best available estimates in the literature (\cite{Balog2019}), and to the most precise ones for the ${\cal O}(N)$ models in general (\cite{Polsi2020,Polsi2021}). 
 
 The \FRG has been applied to study turbulence in several works (\cite{Tomassini97,Monasterio2012,Barbi2013,Canet2016,Canet2017,Tarpin2018,Tarpin2019,Pagani2021}), including a study of
  decaying turbulence within a perturbative implementation of the \FRG (\cite{Fedorenko2013}).  
 This method has turned out to be fruitful in this context, which is the motivation of this {\it Perspectives}.

\subsection{Hydrodynamical equations}

The starting point of field-theoretical methods is a ``microscopic model''. For fluids, this model is the fundamental hydrodynamical description provided by \NS equation
\begin{equation}
  \partial_t \sv_\alpha+ \sv_\beta \partial_\beta \sv_\alpha=-\frac 1\rho 
\partial_\alpha \pi +\nu \nabla^2 \sv_\alpha+f_\alpha
\label{eq:NSeq} 
\end{equation}
where the velocity field $\vvv$, the pressure field $\pi$, and the external force
 $\vf$ depend on the 
space-time coordinates $(t,\vx)$, and with  $\nu$  the kinematic viscosity and 
$\rho$ the density of the fluid.  
 We focus in this review on incompressible flows, satisfying 
\begin{equation}
 \partial_\alpha \sv_\alpha = 0.
\label{eq:incompressibility}
\end{equation}
 The external stirring force is introduced to maintain a stationary turbulent state.
 Since the small-scale (\ie $\ell \ll L$, that is inertial and dissipative) properties are expected to be universal with respect to the large-scale forcing, it can be chosen as a stochastic force, with a Gaussian distribution, of zero average and covariance
\begin{equation}
 \langle f_\alpha(t,\vx)f_\beta(t',\vx')\rangle=2 \delta(t-t')N_{\alpha 
\beta}\left(\frac{|\vx-\vx'|}{L}\right),
\label{eq:forcing}
\end{equation}
where  $\vN$ is concentrated around the integral scale $L$, such that it models the most common physical situation, where the energy is injected at large scales. This is one of the great advantages of the \FRG approach compared to perturbative \RG: it can incorporate any functional form for the forcing, and not necessarily a power-law. Hence, 
one can consider a large-scale forcing, and therefore completely bypass the difficulties encountered in perturbative approaches which arise from trying to access the physical situation as an ill-defined limit from a power-law forcing.

We are also interested in a passive scalar field $\theta(t,\vx)$ advected by a turbulent flow.
 The dynamics of the scalar is governed by the advection-diffusion equation
\begin{equation}
\partial_{t}\theta +\sv_\beta \partial_{\beta}\theta - \kappa_\theta \nabla^{2}\theta =  f_\theta\,,\label{eq:passive-scalar}
\end{equation}
where $\kappa_\theta$ is the molecular diffusivity of the scalar,
 and $f_\theta$ is an external stirring force acting on the scalar, which can also be chosen Gaussian distributed with zero mean and covariance
\begin{equation}
 \langle f_\theta(t,\vx)f_\theta(t',\vx')\rangle=2 \delta(t-t')M\left(\frac{|\vx-\vx'|}{L_\theta}\right)\, ,
\end{equation}
with $L_\theta$ the integral scale of the scalar.

Finally, we consider a simplified model of turbulence, introduced by Burgers (\cite{Burgers48}), which describes the dynamics of a 1D compressible randomly stirred fluid. The Burgers equation reads
\begin{equation}
\p_t \sv + \sv \partial_x \sv = \nu \partial_x^2 \sv + f\, ,
\label{eq:Burgers}
\end{equation}
and can be interpreted as a model for fully compressible hydrodynamics, or pressureless  Navier-Stokes equation (see  \cite{Bec2007} for a review).  $f$ is again a random Gaussian force with covariance
\begin{equation}
 \langle f(t,x)f(t',x')\rangle=2 \delta(t-t')D(x-x')\, .
\end{equation}
It corresponds to model C of (\cite{Forster77}). In fact, there exists an exact mapping between the Burgers equation and the Kardar-Parisi-Zhang equation which describes the kinetic roughening of a stochastically growing interface (\cite{Kardar86}). The KPZ equation gives the time evolution of the  height field  $h(t,\vx)$ of a $(d-1)$-dimensional interface growing in a $d$-dimensional space as
\begin{equation}
\p_t h   = \nu \vnabla^2 h +  \dfrac \lambda 2 \big(\vnabla h\big)^2 +  \eta\, ,
\label{eq:KPZ}
\end{equation}
and has become a fundamental model in statistical physics for non-equilibrium scaling phenomena and phase transitions, akin the Ising model at equilibrium (\cite{Halpin-Healy95,Krug97,Takeuchi18}). For the standard KPZ equation, $\eta$ is interpreted as a microscopic (small-scale) noise which is delta-correlated also in space
\begin{equation}
 \langle \eta(t,\vx)\eta(t',\vx')\rangle=2 D\delta(t-t')\delta^d(\vx-\vx')\, ,
\end{equation}
but it can be  generalised to include a long-range noise $D(|\vx-\vx'|)$.
Defining the velocity field $\vvv = -\lambda \vnabla h$, one obtains from \eq{eq:KPZ} the $d$-dimensional generalisation of the Burgers equation \eq{eq:Burgers} for a potential flow, with forcing $\vf=-\lambda \vnabla \eta$. 

The equations \eq{eq:NSeq}, \eq{eq:passive-scalar} and \eq{eq:Burgers} all yield for some parameters a turbulent regime, where the velocity, pressure, or scalar fields undergo rapid and random variations in space and time. One must account for these fluctuations to build a statistical theory of turbulence. A natural way to achieve this is via a path integral, which includes all possible trajectories weighted by their respective probability. How to write such a path integral is explained in \sref{sec:Langevin}.

\section{Field theoretical formalism for turbulence}
\label{sec:fieldtheory}

\subsection{Path integral representation of a stochastic dynamical equation}
\label{sec:Langevin}

The random forcing in the stochastic partial differential  equations  \eq{eq:NSeq}, \eq{eq:passive-scalar} and \eq{eq:Burgers} acts as a noise source, and thus  these stochastic  equations are formally equivalent to a Langevin equation. The fundamental difference is that, in usual Langevin description, the origin of the noise lies at the microscopic scale, it is introduced to model some microscopic collision processes, and one is usually interested in the   statistical properties of the system at large scales. In the stochastic \NS equation, the randomness is introduced at the integral scale, and one is interested in the statistical properties of the system at small scales (but large with respect to the microscopic scale $\Lambda$). 

Despite this conceptual difference, in both cases, the dynamical fields are fluctuating ones, and there exists a well-known procedure to encompass all the stochastic trajectories within a path integral, which is the \MSR formalism, (\cite{Martin73,Janssen76,Dominicis76}). The idea is simple, and since it is the starting point of all the subsequent analysis, it is useful to describe the procedure in the simplest case of  a scalar field $\varphi(t,\vx)$, following the generic Langevin equation
 \begin{equation}
  \p_t \varphi(t,\vx) = -{\cal F}[\varphi(t,\vx)] + \eta (t,\vx) \, ,
 \label{eq:langevin}
 \end{equation}
 where ${\cal F}$ represents the deterministic forces and $\eta$ the stochastic noise.
  ${\cal F}$ is a functional of the fields and their spatial derivatives.
 The noise   has a Gaussian distribution with zero average and a correlator of the form  
\begin{equation}
 \langle \eta(t,\vx)\eta(t',\vx')\rangle= 2  \delta(t-t') D(|\vx-\vx'|)\,.
  \label{eq:noise}
\end{equation}
The probablity distribution of the noise is thus given by 
\begin{equation}
 {\cal P}[\eta] = {\cal N} \exp\left(-\dfrac{1}{4}\int_{t,\vx,\vx'} \eta(t,\vx)D^{-1}(|\vx-\vx'|)\eta(t,\vx') \right)\, ,
\end{equation}
with ${\cal N}$ a normalisation constant.
Note that the derivation we present can be generalised to include temporal correlations, or a field dependence into the noise correlations \eq{eq:noise}.
The path integral representation of the stochastic equation is obtained in the following way. The probability distribution of the trajectories of the field follows from an 
 average over the noise as
\begin{equation}
 {\cal P}[\varphi] = \int {\cal D} \eta {\cal P}[\eta] \delta(\varphi-\varphi_{\eta})  \, ,
 \label{eq:defProba}
\end{equation}
where $\varphi_\eta$ is a solution of \eq{eq:langevin} for a given realisation of  $\eta$. A change of variable allows one to replace the constraint $\varphi-\varphi_\eta=0$ by the explicit equation of motion
\begin{equation}
  {\cal G}[\varphi(t,\vx)] = 0 = \p_t \varphi(t,\vx) +{\cal F}[\varphi(t,\vx)] - \eta (t,\vx) \, ,
 \end{equation}
 which introduces the functional Jacobian ${\cal J}[\varphi] = \Big|\frac{\delta{\cal G}}{\delta\varphi}\Big|$ \footnote{Two remarks are in order here. First, the existence and uniqueness of  the solution of \eq{eq:langevin} has been implicitly assumed.  Actually, only a solution in a weak sense is required. For the \NS equation, even the existence and uniqueness of weak solutions is a subtle issue from a mathematical viewpoint, and uniqueness may not hold in some cases (\cite{Buckmaster2017}). However, uniqueness is not strictly  required in this derivation, in the sense that for a typical set of initial conditions, there may exist a set of non-unique velocity configurations, provided they are of zero measure. Second, the expression of the Jacobian ${\cal J}[\varphi]$ depends on the discretisation of the Langevin equation \eq{eq:langevin}. In the Ito's scheme, ${\cal J}$ is independent of the fields and can be absorbed in the normalisation of ${\cal P}[\varphi]$, while in the Stratonovich's convention, it depends on the fields, and it can be expressed introducing  two Grassmann anti-commuting fields $\psi$ and $\bar\psi$ as
 ${\cal J}[\varphi]= \left|{\rm det}\frac{\delta{\cal G}(t,\vx)}{\delta\varphi(t',\vx')}\right|= \int {\cal D}\psi {\cal D}\bar\psi e^{\int_{t,t',\vx,\vx'} \bar\psi \frac{\delta {\cal G}}{\delta \varphi} \psi}\, . $
This representation just follows from Gaussian integration of Grassmann variables, which yields the determinant of the operator  in the quadratic form (here $\frac{\delta {\cal G}}{\delta \varphi}$) rather than its inverse, as for standard (non-Grassmann) variables (\cite{Zinn-Justin}).  For an additive noise, which means that the noise part in \Eq{eq:langevin} and its covariance \eq{eq:noise} do not depend on the field $\varphi$, the statistical properties of the system are not sensitive to the choice of the discretisation scheme, and both Ito and Stratonovich conventions yield the same results.  We shall here mostly use the Ito's discretisation for convenience and omit the Jacobian contribution henceforth.}, and leads to
\begin{equation}
 {\cal P}[\varphi] = \int {\cal D} \eta {\cal P}[\eta]{\cal J}[\varphi]  \delta( {\cal G}[\varphi])  \, .\label{eq:probphi}
\end{equation}

 One can then use the Fourier representation of the functional Dirac deltas in \Eq{eq:probphi}, \eg $\delta(\psi) = \int {\cal D}\psi e^{-i \int \bar\varphi \psi}$, where the conjugate Fourier variable is now a field, denoted with an overbar, and called auxiliary field, or response field.  Thus, introducing the auxiliary field 
   $\bar\varphi$ in \Eq{eq:probphi} yields
\begin{align}
 {\cal P}[\varphi] &=  \int {\cal D} \eta {\cal P}[\eta] {\cal D}\bar\varphi 
   e^{-i\int_{t,\vx} \bar \varphi {\cal G}[\varphi]} =  \int {\cal D} \bar\varphi 
   e^{-{\cal S}[\varphi,\bar\varphi]}\,.
 \end{align}
 The second equality stems from  the integration over the Gaussian noise $\eta$, resulting in the action
\begin{equation}
 {\cal S}[\varphi,\bar\varphi] = i \int_{t,\vx} \Big\{ \bar\varphi \big(\p_t\varphi +{\cal F}[\varphi] \big)  \Big\} + \int_{t,\vx,\vx'} \bar \varphi(t,\vx) D(|\vx-\vx'|) \bar\varphi(t,\vx') \, .
\end{equation}
One usually absorbs the complex $i$ into a redefinition of the auxiliary field $\bar\varphi\to -i\bar\varphi$.
The action resulting from the \MSR procedure exhibits a simple structure. The response field appears linearly as Lagrange multiplier for the equation of motion, while the characteristics of the noise,  namely its correlator, 
 are encoded in the quadratic term in $\bar\varphi$.
  
The path integral formulation offers a simple way to compute all the correlation and response functions of the model. Their generating functional is defined by
\begin{align}
{\cal Z}[J,\bar{J}] = \left\langle e^{\int_{t,\vx} \big\{ J\varphi+\bar{J}\bar{\varphi} \big\}}\right\rangle = \int {\cal D}\varphi  {\cal P}[\varphi] e^{\int_{t,\vx} \big\{ J\varphi+\bar{J}\bar{\varphi}\big\}}  =  \int{\cal D}\bar\varphi {\cal D}\varphi e^{-{\cal S}[\varphi,\bar\varphi]+ \int_{t,\vx} \big\{ J\varphi+\bar{J}\bar{\varphi}\big\}}   \, ,
  \label{eq:Zgeneric}
\end{align}
where $J,\bar{J}$ are the sources for the fields $\varphi,\bar\varphi$ respectively. Correlation functions are obtained by taking functional derivatives of ${\cal Z}$ with respect to the corresponding sources, \eg
\begin{equation}
\big\langle  \varphi(t,\vx) \big\rangle= \frac{\delta {\cal Z}}{\delta J(t,\vx)}  \,,
\end{equation}
and functional derivatives with respect to response sources generate response functions (\ie   response to an external drive), hence the name ``response fields''. We shall henceforth use ``correlations'' in a generalised sense including both correlation and response functions.
 In equilibrium statistical mechanics, ${\cal Z}$ embodies the partition function of the system, while in probability theory, it is called the characteristic function, which is the generating function of moments.

 One usually also considers the functional ${\cal W} = \ln {\cal Z}$, which is the analogue of a Helmholtz free energy in the context of equilibrium statistical mechanics. In probability theory, it is the generating function fo cumulants. The generalisation of cumulants for fields rather than variables are called connected correlation functions, and hence ${\cal W}$ is the generating functional of connected correlation functions, for instance,
 \begin{equation}
\big\langle  \varphi(t,\vx)\varphi(t',\vx')\big\rangle_c \equiv  \big\langle  \varphi(t,\vx)\varphi(t',\vx')\big\rangle - \big\langle  \varphi(t,\vx)\big\rangle \big\langle \varphi(t',\vx')\big\rangle  = \frac{\delta {\cal W}}{\delta J(t,\vx)\delta J(t',\vx')}  \,.
\end{equation}
More generally, one can obtain a $n$-point generalised connected correlation function ${\cal W}^{(n)}$ by taking $n$ functional derivatives with respect to either $J$ or $\bar J$ evaluated at $n$ different space-time points $(t_i,\vx_i)$ as
\begin{equation}
\label{eq:defWn}
 {\cal W}^{(n)}(t_1,\vx_1,\cdots,t_n,\vx_n) =  \frac{\delta^n {\cal W}}{\delta {\cal J}_{i_1}(t_1,\vx_1)\cdots \delta {\cal J}_{i_{n}}(t_{n},\vx_n)} \, .
\end{equation}
 where ${\cal J} = (J,\bar J)$ and hence $i_k\in\{1,2\}$. 
 It is also useful to introduce another notation ${\cal W}^{(\ell,m)}$, which specifies that the $\ell$ first derivatives are with respect to $J$ and the $m$ last to $\bar J$, 
 that is
\begin{equation}
\label{eq:defWlm}
 {\cal W}^{(\ell,m)}(t_1,\vx_1,\cdots,t_{\ell+m},\vx_{\ell+m}) =  \frac{\delta^{\ell+m} {\cal W}}{\delta J(t_1,\vx_1)\cdots \delta J(t_\ell,\vx_\ell)\delta \bar J(t_{\ell+1},\vx_{\ell+1})\cdots \delta \bar J(t_{\ell+m},\vx_{\ell+m})}\, .
\end{equation}

A last generating functional which plays a central role in field theory and in the \FRG framework is the Legendre transform of  ${\cal W}$, \ie the analogue of the Gibbs free energy in equilibrium statistical mechanics,
  defined as
 \begin{equation}
 \label{eq:Legendre}
  \Gamma[\langle\varphi\rangle,\langle\bar\varphi\rangle] = {\sup}_{\big\{J,\bar J\big\}} \Bigg[\int_{t,\vx} \Big\{ J \langle \varphi\rangle  +  \bar{J} \langle \bar\varphi\rangle  \Big\} -{\cal W}[J,\bar{J}] \Bigg]\,.
 \end{equation}
$\Gamma$ is called the effective action, it is a functional of the average fields, defined with the usual Legendre conjugate relations
\begin{equation}
\label{eq:Legendre-conjugate}
 \langle \varphi(t,\vx)\rangle = \frac{\delta {\cal W}}{\delta J(t,\vx)}\, ,\quad\quad  J(t,\vx)\ = \frac{\delta \Gamma}{\delta \langle \varphi(t,\vx)\rangle}\, ,
\end{equation}
and similarly for the response fields.
From a field-theoretical viewpoint, $\Gamma$ is the generating functional of one-particle-irreducible correlation  functions, which are obtained by taking functional derivatives of $\Gamma$ with respect to average fields
\begin{equation}
 {\Gamma}^{(n)}(t_1,\vx_1,\cdots,t_n,\vx_n) =  \frac{\delta^n {\Gamma}}{\delta {\Psi}_{i_1}(t_1,\vx_1)\cdots \delta {\Psi}_{i_{n}}(t_{n},\vx_n)} \, .
\end{equation}
 where ${\Psi} = (\langle \varphi\rangle,\langle \bar\varphi\rangle)$ and $i_k\in\{1,2\}$. They are also denoted  ${\Gamma}^{(\ell,m)}$ conforming  to the definition \eqref{eq:defWlm}.
 The one-particle-irreducible correlation functions are also simply called vertices, because in diagrammatic representations {\it \`a la} Feynman they precisely correspond to the vertices of the diagrams, as for instance in \fref{fig:flowgam2}.
 An important point to be highlighted is that both sets of correlation functions $\Gamma^{(n)}$ and ${\cal W}^{(n)}$  contain the exact same information on the statistical properties of the model. Each set can be simply reconstructed from the other {\it via} a sum of tree diagrams. 

To conclude these general definitions, we also consider the Fourier transforms in space and time of all these correlation functions. The  Fourier convention, used throughout this {\it Perspectives}, is
\begin{equation}
 f(\omega,\vq)  = \int_{t,\vx}  f({t,\vx})\; e^{-i \vq \cdot\vx + 
i\omega t} \;\;,\qquad 
 f(t,\vx) =  \int_{\omega,\vq} f (\omega,\vq)\, e^{i \vq \cdot\vx - i\omega 
t},
\end{equation}
where  $\int_{t,\vx}\equiv \int dt d^d \vx$ and  $\int_{\omega,\vq} \equiv\int \frac{d^d 
\vq}{(2\pi)^d}\frac{d\omega}{2\pi}$.
Because of translational invariance in space and time, the Fourier transform of a $n$-point correlation function, \eg $\Gamma^{(n)}$,  takes the form
\begin{equation}
  \Gamma^{(n)}({\omega_1,\vp_1}, \dots,\omega_n, {\vp_{n}}) = (2\pi)^{d+1}\delta\left(\sum_i \omega_i\right)\delta^d\left(\sum_i \vp_i\right) 
  \bar\Gamma^{(n)}({\omega_1,\vp_1}, \dots, {\omega_{n-1},\vp_{n-1}}) \, ,
  \label{eq:defGammabar}
\end{equation}
 that is, the total wavevector and total frequency are conserved, and the last frequency-wavevector arguments $\omega_n,\vp_n$ can be omitted since they are fixed as minus the sum of  the others.

\subsection{Action for the hydrodynamical equations}
\label{sec:actions}

The \MSR procedure can be straightforwardly applied to the ($d$-dimensional) Burgers equation  where the field $\varphi$ is replaced by the velocity field $\vvv$. Introducing the response field $\bar \vvv$, it yields the action
\begin{equation}
\label{eq:actionBurgers}
 {\cal S}_{\rm B} = \int_{t,\vx}\bar \sv_\alpha \Big[\p_t \sv_\alpha + \sv_\beta\p_\beta \sv_\alpha -\nu \nabla^2 \sv_\alpha \Big] -  \int_{t,\vx,\vx'} \bar \sv_\alpha D_{\alpha\beta}(\vx-\vx')\bar\sv_\beta \, .
\end{equation}

For the \NS equation, one has to also include the  incompressibility constraint \eq{eq:incompressibility}. This can be simply achieved by including the factor $\delta(\p_\beta v_\beta)$ in \eq{eq:defProba}. This functional delta can then be exponentiated along with the equation of motion through the introduction of an additional response field $\bar{\spr}$. This yields the following path integral representation		
\begin{equation}
 {\cal Z}[\vJ,\bar{\vJ},K,\bar{K}] =  \int \mathcal{D}\vvv\mathcal{D}\bar{\vvv}\mathcal{D}\spr \mathcal{D}\bar \spr	\,
 \, e^{-{\cal S}_{\rm NS}[\vvv,\bar{\vvv},\spr,\bar \spr] +\int_{t,\vx}\big\{ \vJ\cdot \vvv+\bar{\vJ}\cdot \bar{\vvv}+K \spr+\bar K\bar \spr\big	\}} 
 \, ,\label{eq:ZNS}
\end{equation}
for the fluctuating velocity and pressure fields and their associated response fields, with the \NS action
\begin{align}
 {\cal S}_{\rm NS}[\vvv,\bar{\vvv},\spr,\bar \spr] &= \int_{t,\vx}\Bigg\{\bar \spr \p_\alpha \sv_\alpha + \bar \sv_\alpha\Big[ \p_t 
\sv_\alpha -\nu \nabla^2 \sv_\alpha +\sv_\beta\p_\beta \sv_\alpha+\frac 1\rho \p_\alpha \spr  \Big]\Bigg\}\nonumber\\
 &-\int_{t,\vx,\vx'}\bar 
\sv_\alpha(t,\vx) N_{\alpha\beta}\left(\frac{|\vx-\vx'|}{L}\right)\bar \sv_\beta(t,\vx')\, .
\label{eq:NSaction}
\end{align}
In this formulation, we have kept the pressure field and introduced a response field $\bar{\spr}$ to enforce the incompressibility constraint. Alternatively, 
 the pressure field can be integrated out using the Poisson equation,
such that one obtains a path integral in terms of two fields $\vvv$ and $\bar\vvv$  instead of four, at the price of a resulting action which is non-local (\cite{Barbi2013}).
We here choose to keep the pressure fields since the whole pressure sector turns out to be very simple to handle, as is shown in \aref{app:Ward}.

We now consider a passive scalar field $\theta$ transported by a turbulent \NS velocity flow  according to the advection-diffusion equation \eq{eq:passive-scalar}. 
The associated field theory is simply obtained by adding the two fields
 $\theta, \bar\theta$ and the two corresponding sources $j,\bar j$ to the field multiplet  $\Phi=(\vvv,\bar\vvv,\spr,\bar \spr,\theta,\bar\theta)$  and source multiplet ${\cal J} = (\vJ,\bar\vJ,K,\bar K,j,\bar j)$ respectively. The generating functional then reads
  \begin{equation}
 {\cal Z}[{\cal J}] =  \int \mathcal{D}\Phi\,
 \, e^{-{\cal S}_{\rm NS}[\vvv,\bar{\vvv},\spr,\bar \spr] -{\cal S}_{\theta}[\theta,\bar\theta,\vvv]  + 	\int_{t,\vx} {\cal J}_\ell \Phi_\ell} 
 \, ,\label{eq:Z-NS-scalar}
\end{equation}
with the passive scalar action  given by
 \begin{equation}
{\cal S}_{\theta}[\theta,\bar\theta,\vvv]  =  \int_{t,\vx}\,\bar{\theta}\Big[\partial_{t}\theta+\sv_{\beta}\partial_{\beta}\theta-\kappa_\theta \nabla^{2}\theta\Big]
   -\int_{t,\vx ,\vx'} \bar{\theta}\left(t,\vx\right)M\left(\frac{\vx-\vx'}{L_\theta}\right)\bar{\theta}\left(t,\vx'\right) \, .
   \label{eq:action-NS-scalar}
\end{equation}
 The two actions \eq{eq:NSaction} and \eq{eq:action-NS-scalar} possess fundamental  symmetries, some of which have only been recently identified. They can be used to  derive a set of exact identities relating among each other different correlation functions of the system. These identities, which include as particular cases the K\'arm\'an-Howarth and Yaglom relations, contain fundamental information on the system. We provide in the next \sref{sec:Ward} a detailed analysis of these symmetries, and show how  the set of exact identities can be inferred from Ward identities.
 
\subsection{Symmetries and extended symmetries}
\label{sec:Ward}

We are interested in the stationary state of fully developed homogeneous and isotropic turbulence. We hence assume  translational invariance in space and time, as well as rotational invariance. Beside these  symmetries, the \NS equation possesses other well-known symmetries, such as the Galilean invariance. In  the field-theoretical formulation, these symmetries are obviously carried over, but they are moreover endowed with a systematic framework to be fully exploited. Indeed, the path integral formulation provides  the natural tool to express the consequences of the symmetries on the  correlation functions of the theory, under the form of exact identities which are called Ward identities.

  In fact, the field-theoretical formulation is even more far-reaching, in the sense that the very notion of symmetry can be extended. A symmetry means an invariance of the model (equation of motion or action) under a given transformation of the fields and space-time  coordinates. Identifying the symmetries of a model is very useful because it allows for simplification, and also to uncover invariants, as according to Noether's theorem, continuous symmetries are  associated with conserved quantities. In general, the symmetries considered are global symmetries, \ie their parameters are constants (\eg angle of a global rotation, vector of a global translation).
  However, if one can identify local symmetries, \ie whose parameters are  space and/or time   dependent, this will lead to local conservation laws, which are much richer, much more constraining. In general, the global symmetries of the model cannot be simply promoted to local symmetries by letting their parameters depend on space and/or time.
However, this can be achieved in certain cases at the price of extending the notion of symmetry. More precisely, it is useful to also consider transformations that do not leave the model strictly invariant, but ``almost'', in the sense that they lead to a variation which is at most linear in the fields. We refer to them as extended symmetries. The key is that  one can deduce from them exact local identities, which are thus endowed with a stronger content than the original global identities associated with global symmetries, \ie they provide additional constraints. 

One may wonder why one focuses on linear variations specifically and what is special about them. In fact, in principle, one can consider any transformation of the fields and coordinates and just write down the corresponding variation of the action. Once inserted in the path integral \eq{eq:Zgeneric}, this leads to a relation for the average value of this variation. The reason is that only when it is linear in the fields can this relation be turned into an identity for the generating functionals themselves. In this case, it becomes extremely powerful because it allows one to obtain  an infinite set of exact relations between correlation functions by taking functional derivatives.  The mechanism is very simple, and is exemplified in the simplest case of the symmetries of the pressure sector in \aref{sec:Wardpressure}. 
In the following, we detail two specific extended symmetries, since they play a fundamental role in \sref{sec:largep}, and refer to \aref{app:Ward} for additional ones. Indeed, these two symmetries yield the  identities \eq{eq:wardGalN} and \eq{eq:wardShift21}, which allow for an exact closure at large wavenumbers of the \FRG equations.

\subsubsection{Time-dependent Galilean symmetry}

A fundamental symmetry of the \NS equation is the Galilean invariance, which is the invariance under the global transformation $\vx\to \vx'=\vx+\vvv_0 t$, $\vvv(t,\vx)\to \vvv(t,\vx')-\vvv_0$. In fact, it was  early recognised  in the field-theoretical context that a time-dependent, (also called  time-gauged), version of this transformation leads to an extended symmetry of the \NS action 
 and useful Ward identities (\citep{Adzhemyan94,Adzhemyan99,Antonov96}). Considering an infinitesimal arbitrary time-dependent vector $\veps(t)$, this transformation reads
\begin{align}
 \delta \sv_\alpha(t,\vx)&=-\dot{\varepsilon}_\alpha(t)+\varepsilon_\beta(t) \partial_\beta \sv_\alpha(t,\vx)\nonumber\\
 \delta \Phi_k(t,\vx)&=\varepsilon_\beta(t) \partial_\beta \Phi_k(t,\vx)\label{eq:gaugedGal}
 \end{align}
where  $\dot{\varepsilon}_\alpha =  \frac{d\varepsilon_\alpha}{dt}$, and $\Phi_k$ denotes any other fields $\bar\vvv,\spr,\bar{\spr},\cdots$. The variation $\varepsilon_\beta(t)\partial_\beta (\,\cdot\,)$ just originates from the change of coordinates.
The global Galilean transformation is recovered for a constant velocity $\dot\veps(t) = \vvv_0$. 
 
The overall variation of the \NS action under the transformation \eq{eq:gaugedGal} is
\begin{equation}
 \delta {\cal S}_{\rm NS}= -\int_{t,\vx} {\varepsilon}_\alpha(t) \p_t^2 \bar \sv_\alpha(t,\vx)\,.
\end{equation}
 Since  the field transformation \eq{eq:gaugedGal} is a mere affine change of variable in the functional integral ${\cal Z}$,  it must leave it unaltered.
  Performing this change of variable in \eq{eq:ZNS} and expanding the exponential  to first order in $\varepsilon$, one obtains:
 \begin{equation}
 \Big \langle \delta {\cal S}_{\rm NS}\Big\rangle =  \Big\langle \delta \int_{t,\vx}{\cal J}_\ell \Phi_\ell\Big\rangle = \int_{t,\vx} \Big\{ -\dot\varepsilon_\alpha(t)  J_\alpha(t,\vx) + \varepsilon_\beta(t)  {\cal J}_\ell(t,\vx)  \p_\beta  \Psi_\ell(t,\vx) \Big\} \, .
\end{equation}
Since this identity is valid for arbitrary $\veps(t)$, one deduces
\begin{equation}
\int_{\vx} \Big\{\p_t  J_\alpha(t,\vx) +  {\cal J}_\ell(t,\vx) \p_\alpha  \Psi_\ell(t,\vx) \Big\} =-\int_{\vx}  \p_t^2 \bar u_\alpha(t,\vx) \,.
\label{eq:wardGinit}
\end{equation}
We use throughout this paper  the notation $\vu \equiv\langle \vvv\rangle$ and $\bar\vu \equiv\langle \bar\vvv\rangle$ for the average fields, and generically for the field multiplet $\Psi \equiv \langle \Phi \rangle$. The identity \eq{eq:wardGinit} is integrated over space, but local in time.	In contrast, the identity stemming from the usual Galilean invariance with a constant $\vvv_0$ is integrated over time as well, and the r.h.s. is replaced by zero since the action is invariant under global Galilean transformation.
 
One can express the sources in term of derivatives of $\Gamma$ using \eq{eq:Legendre-conjugate} and rewrite the exact identity \eq{eq:wardGinit} as the following Ward identity for the functional $\Gamma$
 \begin{equation}
 \label{eq:WardGalG}
  \int_{\vx} \Bigg\{ \p_t \frac{\delta \Gamma}{\delta u_\alpha}   
+ \partial_\alpha   \Psi_\ell \frac{\delta \Gamma}{\delta  \Psi_\ell}\Bigg\}\, =-\int_{\vx}  \p_t^2 \bar u_\alpha \,.
\end{equation}
Note that replacing instead the average values of the fields using \eq{eq:Legendre-conjugate}, the identity  \eq{eq:wardGinit} can be equivalently written for the functional ${\cal W}$ as
\begin{equation}
\label{eq:WardGalW}
\int_{\vx} \Big\{\p_t  J_\alpha +  {\cal J}_\ell  \p_\alpha  \frac{\delta {\cal W}}{\delta {\cal J}_\ell} \Big\} =-\int_{\vx}  \p_t^2 \frac{\delta {\cal W}}{\delta \bar J_\alpha} \,.
\end{equation}

The identities  \eq{eq:WardGalG} and \eq{eq:WardGalW}  are functional in the fields.
One can deduce from them, by functional differentiation, an infinite set of exact identities amongst the correlation functions $\Gamma^{(n)}$ or ${\cal W}^{(n)}$. 
Let us express them for the vertices. They are obtained 
 by taking functional derivatives of the identity  \eq{eq:WardGalG} with respect to   $\ell$ velocity and $m$ response velocity fields, and setting the fields to zero, which yields in Fourier space:
\begin{align}
  \Gamma^{(\ell+1,m)}_{\alpha\alpha_1\dots\alpha_{\ell+m}}(\omega,\vp=\vzero, &\,\omega_1,\vp_1, \dots , \omega_{\ell+m},\vp_{\ell+m})\nonumber\\
  &= -\sum_{k=1}^{\ell+m}\frac{p_k^{\alpha}}{\omega}\Gamma^{(\ell,m)}_{\alpha_1\dots\alpha_{\ell+m}}(\omega_1,\vp_1,\cdots,\omega_k+\omega,\vp_k,\cdots, \omega_{\ell+m},\vp_{\ell+m})\nonumber\\ 
 &\equiv {\cal D}_\alpha(\omega)  \Gamma^{(\ell,m)}_{\alpha_1\dots\alpha_{\ell+m}}(\omega_1,\vp_1, \dots, \omega_{\ell+m},\vp_{\ell+m}) \, , \label{eq:wardGalN}
\end{align}
where the $\alpha_i$ are the space indices of the vector fields. We refer to (\cite{Tarpin2018}) for details on the derivation.
The  operator ${\cal D}_\alpha(\omega)$  hence successively shifts by $\omega$ all the frequencies of the function on which it acts.
The identities  \eq{eq:wardGalN} exactly relate an arbitrary  $(\ell+m+1)$-point vertex  function with one vanishing wavevector carried by a velocity field $u_\alpha$  to a lowered-by-one order $(\ell+m)$-point vertex function. It is clear that this type of identity can constitute a key asset  to address the closure problem of turbulence,
 since the latter precisely requires to express higher-order statistical moments in terms of lower-order ones. Of course, this relation only fixes $\Gamma^{(\ell+m+1)}$ in a specific configuration, namely with one vanishing wavevector, so it does not allow one to  completely eliminate this vertex, and it is not obvious {\it a priori} how it can be exploited. In fact, we will show that, within the \FRG  framework,  this type of configurations play a dominant role at small scales, and the exact identities \eq{eq:wardGalN} (together with \eq{eq:wardShift}) in turn yields the closure of the \FRG equations in the corresponding limit.

\subsubsection{Shift of the response fields}
\label{sec:shift-response}

It was also early noticed in the field-theoretical framework that the \NS action is invariant under a constant shift of the velocity response fields (as by integration by parts in \eq{eq:NSaction} this constant can be eliminated). However, it was not identified until recently that this symmetry could also be promoted to a time-dependent one (\cite{Canet2015}). 
 The latter  corresponds to the following infinitesimal coupled transformation of the response fields 
\begin{align}
\label{eq:shiftjauge}
 \delta \bar \sv_\alpha(t,\vx)&=\bar \varepsilon_\alpha(t) \nonumber\\
 \delta \bar p(t,\vx)&= \sv_\beta(t,\vx) \bar \varepsilon_\beta(t) \,.
 \end{align}
 This transformation indeed induces a variation of the \NS action which is only linear in the fields
 \begin{equation}
 \delta {\cal S}_{\rm NS} = \int_{t,\vx}  {\bar\varepsilon}_\beta(t) \partial_t \sv_\beta(t,\vx) +2 \int_{t,\vx,\vx'} {\bar\varepsilon}_\alpha(t) N_{\alpha_\beta} \left(\frac{|\vx-\vx'|}{L}\right) \bar \sv_\beta(t,\vx')\, .
\end{equation}
Hence, interpreted as a change of variable in \eq{eq:ZNS},  this yields 
 the identity $\Big \langle \delta {\cal S}_{\rm NS}\Big\rangle =  \Big\langle \delta \int_{t,\vx}{\cal J}_\ell \Phi_\ell\Big\rangle$, which can be written as 
the following Ward identity for the functional $\Gamma$
 \begin{equation}
 \label{eq:Wardshift}
\int_{\vx} \Bigg\{\frac{\delta \Gamma}{\delta \bar u_\alpha(t,\vx)}
 + u_\alpha(t,\vx) \frac{\delta \Gamma}{\delta \bar p(t,\vx)} \Bigg\}= \int_{\vx}  \p_t u_\alpha(t,\vx) + 2 \int_{\vx,\vx'}N_{\alpha_\beta} \left(\frac{|\vx-\vx'|}{L}\right) \bar u_\beta(t,\vx') \, .
\end{equation}
Note that this identity is again  local in time.
Taking functional derivatives with respect to  velocity and response velocity fields and evaluating at zero fields, one can deduce again exact identities for  vertex functions (\cite{Canet2016}).  They give the expression of any $\Gamma^{(\ell,m)}$ with one vanishing wavevector carried by a response velocity, which simply reads in Fourier space
\begin{equation}
 \Gamma_{\alpha_1\dots\alpha_{\ell+m}}^{(\ell,m)}(\omega_1,\vp_1, 
\dots,\omega_\ell,\vp_\ell, \omega_{\ell+1},\vp_{\ell+1}=\vzero,\dots)= 0 \, ,
  \label{eq:wardShift}
\end{equation}
 for all $(\ell,m)$ except for the two lower-order ones which keep their original form given by ${\cal S}_{{\rm NS},\alpha\beta}^{(1,1)}$ and ${\cal S}_{{\rm NS},\alpha\beta\gamma}^{(2,1)}$ respectively, \ie:
\begin{align}
\Gamma_{\alpha\beta}^{(1,1)}(\omega_1,\vp_1, \omega_2,\vp_2 =\vzero)&=i\omega_1\delta_{\alpha\beta} \;(2\pi)^{d+1}\delta(\omega_1+\omega_2)\delta^d(\vp_1+\vp_2),\nonumber\\
\Gamma_{\alpha\beta\gamma}^{(2,1)}(\omega_1,\vp_1,\omega_2,\vp_2,\omega_3,\vp_3 =\vzero)) &= -i 
 \Big(p_2^\alpha \delta_{\beta\gamma} +i p_1^\beta \delta_{\alpha\gamma} \Big)\nonumber\\
&\times(2\pi)^{d+1}\delta(\omega_1+\omega_2+\omega_3)\delta^d(\vp_1+\vp_2+\vp_3)\label{eq:wardShift21}\, .
\end{align}

Let us emphasise that the analysis of extended symmetries can still be completed. First, for 2D turbulence, additional extended symmetries have recently been unveiled, which are reported in \aref{app:nexttoleading}. One of them is also realised in 3D turbulence but has not been exploited yet in the \FRG formalism. Second, another important symmetry of the Euler or \NS equation is the scaling or dilatation symmetry, which amounts to the transformation $\vvv(t,\vx) \to b^h \vvv(b^z t,b \vx)$ (\cite{Frisch95,Dubrulle2019}). One can also derive from this symmetry, possibly extended, functional Ward identities. This route has not been explored either. Both could lead to future fruitful developments.
 
\subsection{K\'arm\'an-Howarth and Yaglom relations from symmetries}
\label{sec:Karman}		
		
The path integral formulation conveys an interesting viewpoint on 	
 well-known exact identities such as the K\'arm\'an-Howarth relation (\cite{Karman38})
  or the equivalent Yaglom relation for passive scalars (\cite{Yaglom49}). The K\'arm\'an-Howarth relation
  stems from the energy budget equation associated with the \NS equation, upon imposing     stationarity, homogeneity and isotropy. From this relation, one can derive the exact four-fifths Kolmogorov law for the third order structure function $S^{(3)}$
 (see \eg \cite{Frisch95}). It turns out that the K\'arm\'an-Howarth relation also emerges as the Ward identity associated with a {\it spacetime-dependent} shift of the response fields (which is a generalisation of the {\it time-dependent} shift discussed in \sref{sec:shift-response}), and can  thus be obtained in the path integral formulation as a consequence of symmetries (\cite{Canet2015}).
 
 To show this, let us consider the \NS action \eq{eq:NSaction}  with an additional  source $L_{\alpha\beta}$ in \eq{eq:ZNS} coupled to the local quadratic term $\sv_\alpha(t,\vx)\sv_\beta(t,\vx)$ (which is a composite operator in the field-theoretical language), \ie the term $\int_{t,\vx} \sv_\alpha L_{\alpha\beta}\sv_\beta$ is added to the source terms. This implies that a local, \ie at coinciding space-time points, quadratic average of the velocities, can then be simply obtained by taking a functional derivative of ${\cal W}$  with respect to this new source
 \begin{equation}
\big\langle \sv_\alpha(t,\vx) \sv_\beta(t,\vx)\big\rangle = \frac{\delta {\cal W}}{\delta L_{\alpha\beta}(t,\vx)} = - \frac{\delta \Gamma}{\delta L_{\alpha\beta}(t,\vx)}\, ,
\label{eq:composite}
 \end{equation}
 where the last equality stems from the Legendre transform relation \eq{eq:Legendre}. \footnote{Note that to define $\Gamma$,  the Legendre transform is not taken with respect to the source $L_{\alpha\beta}$, \ie both functional ${\cal W}$ and $\Gamma$ depend on $L_{\alpha\beta}$,  hence the relation \eq{eq:composite}.}.

 The introduction of the source for the composite operator $\sv_\alpha \sv_\beta$ allows one to consider a further extended symmetry of the \NS action, namely the time {\it and space} dependent version of the field transformation \eq{eq:shiftjauge}, which amounts to  $\bar \varepsilon_\alpha(t)\to \bar \varepsilon_\alpha(t,\vx)$. The variation of the \NS action  under this  transformation writes
  \begin{equation}
 \langle \delta {\cal S}_{\rm NS} \rangle=\Big\langle \partial_t \sv_\alpha+\frac{1}{\rho}\partial_\alpha \spr -\nu \nabla^2 \sv_\alpha +\partial_\beta(\sv_\alpha \sv_\beta)  -2\int_{\vx'}\Big(N_{\alpha\beta}\left(\frac{|\vx-\vx'|}{L}\right) \bar \sv_\beta(t,\vx')\Big) \Big\rangle,
 \end{equation}
 which is  linear in the fields, but for the local quadratic term. However, this term  can  be expressed as a derivative with respect to $L_{\alpha\beta}$ using  \eq{eq:composite}. The resulting Ward identity can be written equivalently in terms of $\Gamma$ or ${\cal W}$. We here write it  for ${\cal W}$ since it renders more direct the connection with the K\'arm\'an-Howarth relation 
  \begin{equation}
-\partial_t \frac{\delta \cal{W}}{\delta J_\alpha}-\frac{1}{\rho}\partial_\alpha\frac{\delta \cal{W}}{\delta K}
+\nu \nabla^2 \frac{\delta \cal{W}}{\delta J_\alpha}+ \bar J_\alpha + \bar K \frac{\delta \cal{W}}{\delta J_\alpha} 
-\partial_\beta \frac{\delta \cal{W}}{\delta L_{\alpha\beta}}
+  2\int_{\vx'} N_{\alpha\beta}\left(\frac{|\vx-\vx'|}{L}\right)\frac{\delta \cal{W}}{\delta \bar J_\beta(t,\vx')}  =0  .
\label{eq:wardWgaugedshift}
\end{equation}
 We emphasise that, compared to \eq{eq:Wardshift}, this identity is now  {\it fully local}, in space as well as in time, \ie it is no longer integrated over space. It is also an exact  identity for the generating functional $\cal{W}$ itself, which means that it entails an infinite set of exact identities amongst correlation functions. 
 The K\'arm\'an-Howarth relation embodies the lowest order one, which is obtained by  
   differentiating \eq{eq:wardWgaugedshift} with
respect to $J_\gamma(t_y,\vy)$, and evaluating the resulting identity at zero external sources. Note that in the \MSR formalism, the term proportional to $N_{\alpha\beta}$ is  simply equal to a force-velocity correlation $\langle f_\alpha(t,\vx) \sv_\gamma(t_y,\vy) \rangle$ (\cite{Canet2015}). Summing over $\gamma=\alpha$ and specialising to equal time $t_y=t$, one deduces
 \begin{align}
&-\p_t \langle \sv_\alpha(t,\vx) \sv_\alpha(t,\vy)\rangle +  \nu (\Delta_x+\Delta_y) \langle \sv_\alpha(t,\vx) \sv_\alpha(t,\vy)\rangle \nonumber \\
& -\p_\beta^x \langle \sv_\alpha(t,\vx) \sv_\beta(t,\vx) \sv_\alpha(t,\vy)\rangle -\p_\beta^y \langle \sv_\alpha(t,\vy) \sv_\beta(t,\vy) \sv_\alpha(t,\vx)\rangle\nonumber\\
&+ \langle f_\alpha(t,\vx) \sv_\alpha(t,\vy) \rangle+ \langle f_\alpha(t,\vy) \sv_\alpha(t,\vx) \rangle=0,
\label{eq:KH}
\end{align}
which is the K\'arm\'an-Howarth relation.
This fundamental relation is usually expressed in terms of the longitudinal velocity increments $\delta \sv_\parallel(\vell) = (\vvv(\vr+\vell) - \vvv(\ell))\cdot \vell/\ell$ by choosing the two space points as $\vy=\vx+\vell$ and $\vx = \vr$.  Using homogeneity and isotropy, \eq{eq:KH} can be equivalently expressed as
\begin{align*}
 \epsilon(\vell) &\equiv -\dfrac 1 4 \nabla_{\vell}\cdot  \big\langle |\delta \vvv(\vell)|^2 \delta \vvv(\vell) \big\rangle\nonumber\\
  &= -\dfrac 1 2 \p_t \big\langle  \vvv(\vr) \vvv(\vr+\vell) \big\rangle + \big\langle  \vvv(\vr)\cdot \dfrac{\vf(\vr+\vell) +\vf(\vr-\vell)}{2}  \big\rangle +\nu  \nabla_{\ell}^2 \big\langle \vvv(\vr)\cdot \vvv(\vr+\vell) \big\rangle\,.
\end{align*}

Once again, taking other functional derivatives with respect to arbitrary sources yields
 infinitely many exact relations. To give another example, by differentiating twice \Eq{eq:wardWgaugedshift} with respect to $L_{\mu\nu}(t_y,\vy)$ and $J_\gamma(t_z,\vz)$, one obtains the exact relation for a pressure-velocity correlation (\cite{Canet2015})
\begin{align}
&  \nu \langle \sv_\alpha(t,\vx) \Delta_x \sv_\alpha(t,\vx) \vvv^2(t,\vy) \rangle -\frac 1 \rho \p_\alpha^x \langle \vvv^2(t,\vy) \sv_\alpha(t,\vx) \spr(t,\vx)\rangle  \nonumber\\
&+ \langle f_\alpha(t,\vx)  \sv_\alpha(t,\vx) \vvv^2(t,\vy) \rangle  - \frac 1 2  \p^x_\alpha \langle \sv_\alpha(t,\vx)  \vvv^2(t,\vx) v^2(t,\vy) \rangle =0 \, ,
\label{eq:falko}
\end{align}
which was first  derived in (\cite{Falkovich2010}). Taking additional functional derivatives with respect to $\vJ$ or $\vL$ (or of the any other sources) generates new exact relations between higher-order correlation functions, involving in the averages one more $\sv$ or $\vvv^2$ (or any of the other fields) respectively with each derivative compared to \eq{eq:KH}. Although exact, relations for high-order correlation functions may not be of direct practical use since they are increasingly difficult to measure, but they they are important at the theoretical level.

Shortly after  Kolmogorov's derivation of the exact relation for the third-order structure function, Yaglom established the analogous formula  for scalar turbulence (\cite{Yaglom49}). It was shown in (\cite{Pagani2021}) that this relation can also be simply inferred from symmetries of the passive scalar action \eq{eq:action-NS-scalar}. More precisely, it ensues from the  spacetime-dependent shift of the response fields  
\begin{equation}
\bar{\theta}\left(t,\vx\right)\rightarrow\bar{\theta}\left(t,\vx\right)+\bar\varepsilon\left(t,\vx\right),\quad
\bar{\spr}\left(t,\vx\right)\rightarrow \bar{\spr}\left(t,\vx\right)+\bar	\varepsilon\left(t,\vx\right)\theta\left(t,\vx\right)
\end{equation}
in the path integral \eq{eq:Z-NS-scalar}, in the presence of the additional source term
$\int_{t,\vx}L_{\alpha}\sv_{\alpha}\theta$ 
 where $L_\alpha$ is the source coupled to the composite operator $\sv_{\alpha}\theta$.
  Following the same reasoning, one deduces an exact functional Ward identity
 for the passive scalar, which writes in term of ${\cal W}$
 \begin{equation}
\Big(\partial_{t}-{\kappa_\theta}\nabla^{2} \Big)\frac{\delta W}{\delta j\left(t,\vx\right)}+
\partial_{\beta}\frac{\delta W}{\delta L_{\beta}\left(t,\vx\right)} 
-\int_{\vy} M\left(\frac{|\vx-\vy|}{L_\theta}\right)\frac{\delta W}{\delta \bar{j}\left(t,\vy\right)} = 0\,.
\label{eq:Yaglom}
\end{equation}
 The lowest order relation stemming from it is the Yaglom relation
\begin{equation}
-\frac{1}{2}\frac{\partial}{\partial\left(x-y\right)_{\alpha}}\Bigr\langle\left|\theta\left(t,\vx\right)-\theta\left(t,\vy\right)\right|^{2}\left(\sv_{\alpha}\left(t,\vx\right)-\sv_{\alpha}\left(t,\vy\right)\right)\Bigr\rangle  =  2\epsilon_{\theta}\,,
\end{equation}
where $\epsilon_\theta$ is the mean dissipation rate of the scalar. 
As for the K\'arm\'an-Howarth relation, it can be re-expressed using homogeneity and isotropy in the more usual form
\begin{equation}
\Bigr\langle\big|\theta\left(t,\vr\right)-\theta\left(t,\vr+\vell\right)\big|^{2} \delta \sv_\parallel(t,\vell)\Bigr\rangle  =  -\dfrac{4}{3}\epsilon_{\theta}\ell\,,
\end{equation}
 This relation  hence also follows from symmetries in the path integral formulation.
Again, an infinite set of higher-order exact relations between scalar and velocity correlation functions can be obtained by further differentiating \eq{eq:Yaglom} with respect to sources $\vL$ or $j$.

\section{The functional renormalisation group}
\label{sec:FRG}

As mentioned in \sref{sec:RG}, the idea underlying the \FRG is Wilson's original idea of progressive averaging of fluctuations in order to build up the effective description of a system from its microscopic model. This progressive averaging is organised scale by scale, in general in wavenumber space, and thus  leads to a sequence of scale-dependent models, embodied in Wilson's formulation in a scale-dependent Hamiltonian or action (\cite{Wilson74}).
 In the \FRG formalism, one rather considers a  scale-dependent effective action $\Gamma_\kappa$, called effective average action,
   where $\kappa$ denotes the \RG scale. It is a wavenumber scale, which runs from the microscopic \UV scale $\Lambda$ to a macroscopic \IR scale (\eg the inverse integral scale $L^{-1}$). The interest of the \RG procedure is that the information about the properties of the system can be captured by the {\it flow} of these scale-dependent models, \ie their evolution with the \RG scale, without requiring to explicitly carry out the integration of the fluctuations in the path integral. The \RG flow is governed by an exact very general equation, which can take different forms depending on the precise \RG used (Wilson (\cite{Wilson74}) or Polchinski flow equation (\cite{Polchinski84}), Callan-Symanzik flow equation  (\cite{Callan70,Symanzik70}), \dots). The one  at the basis of the \FRG formalism is usually called the Wetterich equation (\cite{Wetterich93}). We briefly introduce it in the next sections  on the example of the generic scalar field theory of \sref{sec:Langevin}, denoting generically $\Phi(t,\vx)$ the field multiplet, \eg $\Phi=(\varphi,\bar\varphi)$, $\Psi=\big\langle \Phi\big\rangle$ the multiplet of average fields, and ${\cal J}$ the multiplet of corresponding sources.
     
\subsection{Progressive integration of fluctuations}

The core of the \RG procedure is to turn the global integration over fluctuations in the path integral \eq{eq:Zgeneric} into a progressive integration, organised by wavenumber shells. To achieve this, one introduces in the path integral a scale-dependent weight $e^{-\Delta{\cal S}_\kappa}$ whose role is to suppress fluctuations below the \RG scale $\kappa$, giving rise to a new, scale-dependent generating functional
\begin{equation}
{\cal Z}_\kappa[{\cal J}]  =  \int{\cal D}\Phi  e^{-{\cal S}[\Phi] - \Delta{\cal S}_\kappa[\Phi]+ \int_{t,\vx}  {\cal J}_\ell \Phi_\ell }   \, .
\label{eq:Zgenerick}
\end{equation}
 The new term $\Delta{\cal S}_\kappa$ is chosen quadratic in the fields
 \begin{equation}
 \Delta{\cal S}_\kappa[\Phi] =\frac 1 2 \int_{t,\vx,\vx'} \Phi_m(t,\vx) {\cal R}_{\kappa,m m'}(|\vx-\vx'|)\Phi_{m'}(t,\vx')\, ,
 \end{equation}
where ${\cal R}_\kappa$ is called the regulator, or cut-off, matrix. Note that it has been chosen here proportional to $\delta(t-t')$, or equivalently independent of frequencies. This means that the selection of fluctuations is operated in space, and not in time, as in equilibrium. \footnote{The selection can in principle be operated also in time, although it poses some technical difficulties, not to violate causality (\cite{Canet2011heq}) and symmetries involving time -- typically the Galilean invariance. A spacetime cutoff was implemented only in (\cite{Duclut2017}) for Model A, which is a simple, purely dissipative, dynamical extension of the Ising model.
In the following, we restrict ourselves to frequency-independent regulators.}

The precise form of the  elements $[{\cal R}]_{\kappa,ij}$  of the cutoff matrix is not important, provided they satisfy the following requirements:
\begin{align}
 &R_{\kappa}(\vp)\;\sim \; \kappa^2\quad\quad\quad  \hbox{for\;}|\vp|\lesssim \kappa\nonumber \\
&R_{\kappa}(\vp) \longrightarrow 0 \quad\quad\quad \hbox{for\;}|\vp|\gtrsim\kappa	\, ,
\label{eq:cutoff}
\end{align}
where $R_k$ denotes a generic non-vanishing  element of the matrix ${\cal R}\kappa$.
The first constraint endows the low wavenumber modes with a large ``mass'' $\kappa^2$, such that these modes are damped, or filtered out,  for their contribution in the functional integral to be suppressed. The second one ensures that
 the cutoff vanishes for large wavenumber modes, which are thus unaffected. Hence,  
 only these modes are integrated over, thus achieving the progressive averaging.
Moreover, $R_{\kappa=\Lambda}$ is required to be very large such that all fluctuations are frozen at the microscopic (\UV) scale, and $R_{\kappa=0}$
 to vanish such that all fluctuations are averaged over in this (\IR) limit.
 
It follows that the free energy functional ${\cal W}_\kappa = \ln{\cal Z}_\kappa$  also becomes scale-dependent. One defines the scale-dependent effective average  action through the modified Legendre transform
\begin{equation}
  \Gamma_\kappa[\Psi] + \Delta{\cal S}_\kappa[\Psi] = {\sup}_{{\cal J}} \Bigg[\int_{t,\vx} {\cal J}_\ell \Psi_\ell -{\cal W}_\kappa[{\cal J}] \Bigg]\,.
  \label{eq:Legendrek}
 \end{equation}
The regulator term is added in the relation in order to enforce that, at the scale $\kappa=\Lambda$, the effective average action coincides with the microscopic action
 $\Gamma_{\kappa=\Lambda}= {\cal S}$. For fluid dynamics, the scale $\Lambda^{-1}$ represents a very small scale, typically a few mean-free-paths, where the description  in term of a continuous equation becomes valid, and the ``microscopic'' action is the Navier-Stokes action ${\cal S}_{\rm NS}$. This scale is thus much smaller than the Kolmogorov scale $\eta$, and the effect of fluctuations (the renormalisation) is already important at scales $\ell\simeq \eta$, \ie the statistical properties of the turbulence in the dissipative range are non-trivial.
  In the opposite limit $\kappa\to 0$ (or equivalently $\kappa\ll L^{-1}$) the standard effective action $\Gamma$, encompassing all the fluctuations, is recovered since the regulator is removed in this limit  $\Delta{\cal S}_{\kappa=0}=0$. Thus, the sequence of $\Gamma_\kappa$ provides an interpolation between the microscopic action [the NS action] and the full effective action [the statistical properties of the fluid].
 
 \subsection{Exact flow equation for the effective average action}
 
 The evolution of the generating functionals ${\cal W}_\kappa$ and $\Gamma_\kappa$ with the \RG scale  $\kappa$ obeys an exact differential equation, which can be simply inferred from \eq{eq:Zgenerick} and \eq{eq:Legendrek} since
  the dependence on $\kappa$ only comes from the regulator term $\Delta{\cal S}_\kappa$.
  The derivation is very general and can be found in standard references, \eg Sec.~2.3.3. of (\cite{Delamotte2012}). One finds the following exact flow equation for ${\cal W}_\kappa$ 
  \begin{equation}
\partial_{s}{\cal W}_{\kappa}\left[{\cal J}\right]  = -\frac{1}{2}\mbox{Tr}\left[\partial_{s}{\cal R}_{\kappa,mn}\left(\frac{\delta^{2}{\cal W}_{\kappa}\left[{\cal J}\right]}{\delta {\cal J}_{m}\delta {\cal J}_{n}}+\frac{\delta {\cal W}_{\kappa}\left[{\cal J}\right]}{\delta {\cal J}_{m}}\frac{\delta {\cal W}_{\kappa}\left[{\cal J}\right]}{\delta {\cal J}_{n}}\right)\right]\,,
\label{eq:flowW}
\end{equation}
where $s\equiv\log (\kappa/\Lambda)$. This equation is very similar to Polchinski equation (\cite{Polchinski84}).  
 Some simple algebra then leads to the Wetterich equation for $\Gamma_\kappa$ 
 \begin{equation}
\partial_{s}\Gamma_{\kappa}\left[\Psi\right]  = \frac{1}{2}\mbox{Tr}\left[\partial_{s}{\cal R}_{\kappa,mn} G_{\kappa,mn}  \right]\equiv \frac{1}{2}\mbox{Tr}\left[\partial_{s}{\cal R}_{\kappa,mn}\left(\Gamma_{\kappa}^{\left(2\right)}\left[\Psi\right]+{\cal R}_{\kappa}\right)_{mn}^{-1} \right]\,,
\label{eq:flowGamma}
\end{equation}
where  $G_\kappa\equiv {\cal W}^{(2)}_\kappa$ is the propagator, and is the inverse of $\big(\Gamma_\kappa^{(2)}+{\cal R}_\kappa\big)$ from the Legendre relation.
To alleviate notations, the indices $m,n$ refer to  the field indices within the multiplet, as well as other possible  indices (\eg vector component) and space-time coordinates. Accordingly, the trace includes the summation over all internal indices as well as the integration over all spacial and temporal coordinates (conforming to deWitt notation, integrals are implicit).

While \Eq{eq:flowGamma} or \eq{eq:flowW}  are exact, they are functional partial differential equations which cannot be solved exaclty in general. Their functional nature implies that they encompass an infinite set of flow equations for the associated correlation functions or vertices. For instance, taking one functional derivative of \eq{eq:flowGamma} with respect to  a given field and evaluating the resulting expression at a fixed background field configuration (say $\Psi(t,\vx)=0$) yields the flow equation for the one-point vertex $\Gamma_\kappa^{(1)}$. This equation  depends on the three-point 
 vertex $\Gamma_\kappa^{(3)}$. More generally, the flow equation for the $n$-point vertex  $\Gamma_\kappa^{(n)}$ involves the vertices
   $\Gamma_\kappa^{(n+1)}$ and  $\Gamma_\kappa^{(n+2)}$, such that one has to consider an infinite hierarchy of flow equations.
  This pertains to the very common closure problem of non-linear systems. It means that one has to devise some approximations. 

 \subsection{Non-perturbative approximation schemes}
  \label{sec:approx} 
   
   This part may appear very technical for readers not familiar with \RG methods. Its objective is to explain the rationale underlying the approximation schemes used in the study of turbulence, which are detailed in the rest of the {\it Perspectives}.
   
  In the \FRG context, several approximation schemes have been developed and are commonly used (\cite{Dupuis2021}).  Of course one can implement a perturbative expansion, in any available small parameter, such as a small coupling  or an infinitesimal distance to a critical dimension $\varepsilon=d - d_c$. One then retrieves results obtained from standard perturbative \RG techniques. However, the key  advantage of the \FRG formalism is that it is suited to the implementation of non-perturbative approximation schemes. The most commonly used is the derivative expansion, which consists in expanding the effective average action $\Gamma_\kappa$ in powers of gradients and time derivatives (thus yielding an ansatz for $\Gamma_\kappa$). This is equivalent to an expansion around zero external wavenumbers $|\vp_i|=0$ and frequencies $\omega_i =0$ and is thus adapted to describe the long-distance long-time properties of a system. One can in particular obtain universal properties of a system at criticality (\eg critical exponents), but also non-universal properties, such as phase diagrams. Even though it is non-perturbative in the sense that it does not rely on an explicit small parameter, it is nonetheless controlled. It can be systematically improved, order by order (adding higher-order derivatives), and an error can be estimated at each order, using properties of the cutoff (\cite{Polsi2022}). The convergence and accuracy of the derivative expansion have been studied in depth for archetypal models, namely the Ising model and ${\cal O}(N)$ models (\cite{Dupuis2021}). The outcome is that the convergence is fast, and most importantly that  the results obtained for instance for the critical exponents are very precise. 
  For the 3D Ising model,  the derivative expansion has been pushed up to the sixth order ${\cal O}(\partial^6)$  and the results for the critical exponents   compete in accuracy with the best available estimates in the literature (stemming from conformal bootstrap methods) (\cite{Balog2019}). For the ${\cal O}(N)$ models, the \FRG  results   are the most accurate ones available  (\cite{Polsi2020,Polsi2021}). 
     
  The derivative expansion is by construction restricted to describe the zero wavenumber and frequency sector, it does not allow one to access the full space-time dependence of generic correlation functions. To overcome this limitation,  one can resort to another 
   approximation scheme, which consists in a vertex expansion.
 The most useful form of this approximation is called the \BMW scheme (\cite{Blaizot2006,Blaizot2007,Benitez2008}). It lies at the basis of all the \FRG studies dedicated to turbulence. 
  The \BMW scheme essentially exploits an intrinsic property of the Wetterich flow equation conveyed by the presence of the regulator. The Wetterich equation has a one-loop structure. To avoid any confusion, let us emphasise that this does not mean that it is equivalent to a one-loop perturbative expansion: the propagator entering \Eq{eq:flowGamma} is the full (functional) renormalised propagator of the theory,  not the bare one. 
  This simply means that the flow equation involves only one internal, or loop, (\ie integrated over) wavevector $\vq$ and frequency $\omega$. This holds true for the flow equation of any vertex  $\Gamma_\kappa^{(n)}$,  which depends on the $n$ external wavevectors $\vp_i$ and frequencies $\varpi_i$, but always involves only one internal (loop) wavector and frequency $(\omega,\vq)$. 
  
  The key feature of the Wetterich equation is that the internal wavevector $\vq$ is controlled by the scale-derivative of the regulator. Because of the requirement \eq{eq:cutoff}, 
  $\partial_s {\cal R}_\kappa(\vq)$ is exponentially vanishing for $q\equiv |\vq| \gtrsim \kappa$, which implies that the loop integral is effectively cut to $q \simeq \kappa$, with $\kappa$ the \RG scale. This points to a specific limit where this property can be exploited efficiently, namely the large wavenumber limit. Indeed, if one considers large external wavenumbers $p_i\equiv |\vp_i|\gg \kappa$, then $|\vq|\ll |\vp_i|$ is automatically satisfied, or otherwise stated $|\vq|/|\vp_i|\to 0$. Hence the limit of $|\vp_i|\to\infty$ is formally equivalent to the limit $|\vq|\to 0$ (because the $\vp_i$ and $\vq$ always come as sum or product in the propagator and vertices). 
  Moreover, the presence of the regulator ensures that all the vertices are analytical functions of their arguments at any finite $\kappa$, and can thus be Taylor expanded (\cite{Berges2002}). 
  The \BMW approximation precisely consists in expanding the vertices  entering the flow equation around $\vq\simeq 0$, which can be interpreted as an expansion at large $p_i$. This expansion becomes formally exact in the limit where all $p_i\to \infty$. Besides, one has that a generic $\Gamma_\kappa^{(n)}$ vertex with one zero wavevector can be expressed as a derivative with respect to a constant field of the corresponding vertex $\Gamma_\kappa^{(n-1)}$ stripped of the field with zero wavevector.
   This allows one in the original \BMW approximation scheme to close the flow equation at a given order (say for the two-point function), at the price of keeping a dependence in a background constant field $\Psi_0$, \ie $\bar \Gamma_\kappa^{(2)}(\varpi,\vp;\Psi_0)$, which can be cumbersome. This approximation can also in principle be improved order by order, by achieving the $\vq$ expansion in the flow equation for the next order $(n+1)$ vertex instead of the $n$th one, although it becomes increasingly difficult. 
  
  In the context of  non-equilibrium statistical physics, the \BMW approximation scheme 
  was implemented successfully to study the  Burgers or equivalently Kardar-Parisi-Zhang equation \eq{eq:KPZ}, which is detailed in \sref{sec:KPZ}. For the \NS equation, the \BMW approximation scheme turns out to be remarkably efficient in two respects which will become clear in the following: technically, the symmetries and extended symmetries allow one to get rid of the background field dependence, and thus to achieve the closure exactly. Moreover, physically, the large wavenumber limit is not trivial for turbulence, which is unusual. In critical phenomena exhibiting standard scale invariance, the large wavenumbers simply decouple from the \IR properties and this limit carries no independent information (see \sref{sec:KPZ}).
   
\section{A warm-up example: Burgers-KPZ equation}   
\label{sec:KPZ}  
   
  The Burgers equation is often considered as a toy model for classical hydrodynamics. In the inviscid limit, its solution develops shocks after a finite time even for smooth initial conditions (\cite{Bec2007}). Most studies consider potential flows, for which case the (irrotational $d$-dimensional) Burgers equation exactly maps to the KPZ equation \eq{eq:KPZ}. The forcing is  assumed to be a power-law $D(\vp)\sim p^\beta$, which mainly injects energy at small scales (\UV modes) for $\beta>0$, while it acts on large scales (\IR modes) for $\beta<0$. 
   
   One is generally interested in the space-time correlations of the velocity 
   \begin{equation}
    C^{(n)}(\tau,\ell) = \Big\langle \big[\vvv(t+\tau,\vx+\vell)-\vvv(t,\vx) \big]^n \Big\rangle\,.
   \end{equation}
 The structure functions, which correspond to the equal-time correlations, behave as power-law in the inertial range
 $S_n(\ell) \sim \ell^{\zeta_n}$. Besides,
 the two-point correlation function is expected to endow a scaling form
 \begin{equation}
  C^{(2)}(\tau,\ell) \equiv C(\tau,\ell) = \ell^{2(\chi-1)} F(\tau/\ell^z)\,,
  \label{eq:defF}
 \end{equation}
where $\chi$ and $z$ are universal critical exponents called the roughness and dynamical exponent in the context of interface growth, and $F$ is a universal scaling function.
 The roughness exponent is simply related to $\zeta_2$ as $\zeta_2 = 2(\chi-1)$. The dynamical and roughness exponents are related in all dimensions by the exact identity $z+\chi=2$ stemming from Galilean invariance. 

 In one dimension $d=1$,  the critical exponents are known exactly. For $\beta>0$,
 one has (\cite{Medina89,Janssen99,Kloss2014a})
 \begin{equation}
  \chi = \max\left(\dfrac 1 2, -\dfrac \beta 3 +1\right) \, .
  \label{eq:chiBurgers}
 \end{equation}
 Moreover, shocks are overwhelmed by the forcing and there is no intermittency, \ie $\zeta_n = n \zeta_2/2 = n(\chi-1)$ (\cite{Hayot96}).
  For $\beta<-3$, the stationary state contains a finite density of shocks and the scaling of the velocity increments in the regime where $|\ell|$ is much smaller than the average distance between shocks but larger than the shock size can be estimated by a simple argument, yielding (\cite{Bec2007}) 
 \begin{equation}
  \zeta_n = \min\left(1, n\right) \, .
 \end{equation}
While the intermediate regime $-3<\beta<0$ is not well understood, a wealth of exact results are available for $\beta=2$, which have been obtained in the context of the KPZ equation (\cite{Corwin12}). In particular, the analytical form of the scaling function $F$ in \eq{eq:defF} has been determined exactly (\cite{Praehofer00}).

 In higher dimensions $d>1$, very few is known.  The most studied case corresponds to the KPZ equation $\beta=2$.  In contrast with $d=1$ where the interface always roughens, in dimensions $d>2$, the KPZ equation exhibits a (non-equilibrium) continuous phase transition between a smooth phase and a rough phase, depending on the  amplitude  $\lambda$  of the non-linearity in \eq{eq:KPZ}. For $\lambda <\lambda_c$, the interface remains smooth, it is then described by the Edwards-Wilkinson equation which is the linear, non-interacting, $\lambda=0$, version of  KPZ, and which has a simple diffusive behaviour with $z=2$ and $\chi=(2-d)/2$. For $\lambda >\lambda_c$, the non-linearity is dominant and the interface becomes rough, with $\chi>0$. This is the KPZ phase. Contrarily to $d=1$, no exact result has been obtained for $d\neq 1$, and the critical exponents are known only numerically (\cite{Pagnani15}).
 
 Since the early days of the KPZ equation, perturbative \RG techniques (usually referred to as dynamical \RG) have been applied to determine the critical exponents (\cite{Kardar86,Medina89,Janssen99}). However,  the KPZ equation constitutes a striking example where the perturbative \RG flow equations are known  to all orders in perturbation theory, but nonetheless fails in $d\geq 2$ to find the strong-coupling fixed-point expected to govern the KPZ rough phase (\cite{Wiese98}). In contrast,  the \FRG framework allows one to access this fixed point in all dimensions even at the lowest order of the derivative expansion (\cite{Canet2005b,Canet2010}). This fixed-point turns out to be genuinely non-perturbative, \ie not connected to the Edwards-Wilkinson fixed-point (which is the point around which perturbative expansions are performed) in any dimension, hence explaining the failure of perturbation theory, to all orders. We now briefly review these results.
 
 \subsection{FRG for the Burgers-KPZ equation}
 
 The correlation function \eq{eq:defF} can be computed from the two-point functions $\Gamma^{(2)}$.
 The general \FRG flow equation for these functions is obtained by differentiating twice the exact flow equation \eq{eq:flowGamma} for the effective average action $\Gamma_\kappa$. Assuming translational invariance in space and time,  it can be evaluated  at zero fields and then reads
\begin{eqnarray}
\partial_\kappa \bar\Gamma^{(2)}_{\kappa,\ell m}(\varpi,\vp)\! &=& \! {\rm Tr}\! \int_{\omega,\vq} 
\partial_\kappa {\cal R}_\kappa(\vq) \cdot \bar G_\kappa(\omega,\vq) \cdot
\!\bigg(\!\!-\!\frac{1}{2}\, \bar\Gamma^{(4)}_{\kappa,\ell m}(\varpi,\vp,-\varpi,-\vp,\omega,\vq) \nonumber\\
&& \hspace{-2.5cm} +\;\bar\Gamma^{(3)}_{\kappa,\ell}(\varpi,\vp,\omega,\vq) \cdot \bar G_\kappa(\varpi+\omega,\vp+\vq) 
\cdot
\bar\Gamma^{(3)}_{\kappa,m}(-\varpi,-\vp,\varpi+\omega,\vp+\vq) \bigg) \cdot G_\kappa(\omega,\vq)
\label{eq:dkgam2}
\end{eqnarray}
where the last arguments of the $\bar\Gamma_\kappa^{(n)}$ are implicit since they are determined by frequency and wavevector conservation according to \eq{eq:defGammabar}. We used a matrix notation, where only the external field index $\ell$ and $m$ are specified, \ie $\bar\Gamma^{(3)}_{\kappa,\ell}$ is the $2\times2$ matrix of all three-point vertices with one leg fixed at $(\ell,\varpi,\vp)$ and the other two spanning all fields, and similarly for $\bar\Gamma^{(4)}_{\kappa,\ell m}$. It can be represented diagrammatically as in \fref{fig:flowgam2}.
\begin{figure}
\begin{tikzpicture}
 \node[left, scale=1] at (-1.4,0) {$\partial_{s}$};
 \node[draw,circle,minimum size=0.1cm,pattern=north east lines] (Gam) at (0,0){};
  \node at (0,-0.5) {$\bar\Gamma_\kappa^{(2)}$};
    \draw (-180:0.8) -- (Gam);
    \draw (0:0.8) -- (Gam);
    \node[left, scale=1] at (3,0) {$=\;-\displaystyle\frac{1}{2}$};
 \node[draw,circle,minimum size=0.1cm,pattern=north east lines] (Gam1) at (4.3,0){};
  \node at (4.3,-0.5) {$\bar\Gamma_\kappa^{(4)}$};
    \draw (Gam1) -- ++ (-180:0.8);
    \draw (Gam1) -- ++ (0:0.8);
\draw [thick] (4.42,0.) arc (-80:260:0.65);
\draw (4.3,1.28) node {$\boldsymbol{\times}$};
\node[left, scale=1] at (6.3,0) {+};
 \node[draw,circle,minimum size=0.1cm,pattern=north east lines] (Gam2) at (8.,0) {};
 \node at (7.75,-0.4) {$\bar\Gamma_\kappa^{(3)}$};
 \node[draw,circle,minimum size=0.1cm,pattern=north east lines] (Gam3) at (9.4,0) {};
 \node at (9.7,-0.4) {$\bar\Gamma_\kappa^{(3)}$}; 
      \draw (Gam2) -- ++ (-180:0.8);
    \draw (Gam3) -- ++ (0:0.8);
\draw  [thick] (9.4,0.11) arc (05:175:0.7);
\draw  [thick] (9.4,-0.11) arc (-05:-175:0.7);
\node[left, scale=1] at (8.9,0.75) {$\boldsymbol{\times}$};
\end{tikzpicture}
  \caption{Diagrammatic representation of the flow of $\bar\Gamma_{\kappa}^{(2)}$ given by \Eq{eq:dkgam2}. The bold lines represent the  propagator $\bar G_\kappa$, the crosses the derivative of the regulator $\p_\kappa {\cal R}_\kappa$, and the hatched dots the vertices $\bar \Gamma^{(n)}_\kappa$ with $n$ the number of legs. The loop indicates that all the internal indices are summed over along the loop, and the internal momentum $\vq$ and frequency $\omega$ circulating in the loop  are integrated over.}
   \label{fig:flowgam2}
 \end{figure}
In order to make explicit calculations, one has to make some approximation. Here,
 we  resort to an ansatz for the effective average action  $\Gamma_\kappa$, which then enables one to  compute all the $n$-point functions entering the flow equation. A crucial point for the choice of this ansatz is to preserve the symmetries of the model. The fundamental symmetry of the Burgers equation is the Galilean invariance. As a consequence, the $n$-point vertices $\Gamma^{n}$ are related by similar identities as \eq{eq:wardGalN} (simply removing the pressure fields). Thus, a satisfactory ansatz for the Burgers effective average action should automatically satisfy all these identities. To devise such an ansatz is not a simple task in general. However, the construction can be understood in a rather intuitive way. 
 
 Let us define a function $f(t,\vx)$ as a scalar density under the Galilean transformation \eq{eq:gaugedGal} if its infinitesimal variation under this transformation is $\delta f(t,\vx) = \varepsilon_\alpha(t)\p_\alpha f(t,\vx)$ -- \ie it varies only due to the change of coordinates. This then  implies that $\int_{\vx} f(t,\vx)$ is invariant under a Galilean transformation. According to \eq{eq:gaugedGal},   $\bar \vvv$ is  a scalar density, while $\vvv$ is not. However, one can build two scalar densities from it, which are $\partial_\alpha \vvv$ and the Lagrangian time derivative 
 $D_t \vvv \equiv \partial_t \vvv+ \sv_\alpha \p_\alpha \vvv$.
  One can then easily show that combining scalars  together through sums and products  preserve the scalar density property, as well as applying a gradient. While  applying a time derivative $\partial_t$ spoils the scalar density property, the  Lagrangian time derivative preserves it, which identifies $D_t$ as the covariant time derivative for Galilean transformation. These rules allow one to construct an ansatz which is manifestly invariant under the Galilean transformation. 
 
 The more advanced approximation which has been implemented so far  for the Burgers-KPZ equation consists in truncating the effective average action $\Gamma_\kappa$ at second order (SO) in the response field, and thus neglecting higher-order terms in this field which could in principle be generated by the \RG flow. This means that the noise probability distribution is kept to be Gaussian as in the original Langevin description, which  sounds  reasonable. The most general ansatz at this SO order compatible with the symmetries of the Burgers equation, \ie using the previous construction, reads   
\begin{align}
\Gamma_\kappa[\vu,\bar \vu] &= \int_{t,\vx}\Bigg\{\bar 
u_\alpha f_{\kappa}^\rho(D_t,\nabla) D_t u_\alpha   -\bar u_{\alpha} f^{\nu}_{\kappa}(D_t,\nabla)  \nabla^2 u_\alpha  -\bar u_\alpha f^{D}_{\kappa}(D_t,\nabla) \bar  u_\alpha\Bigg\}\, .
 \label{eq:anzSO}
\end{align}
It is parametrised by three scale-dependent renormalisation functions  $f^\nu_{\kappa}$, $f^D_{\kappa}$ and $f^\rho_{\kappa}$, which are arbitrary functions of the covariant operators $\nabla$ and $D_t$, beside depending on the \RG scale $\kappa$ \footnote{Note that $f^\nu_{\kappa}$, $f^D_{\kappa}$ and $f^\rho_{\kappa}$ are scalar functions because they correspond to the longitudinal parts of  $f^\nu_{\kappa,\alpha\beta}$, $f^D_{\kappa,\alpha\beta}$ and $f^\rho_{\kappa,\alpha\beta}$. The transverse parts vanish since the flow is potential.}.  Their initial condition at $\kappa=\Lambda$ is  $f^\nu_\Lambda=\nu$, $f^D_\Lambda=(D_{\alpha\beta})^\parallel$ (longitudinal part of $D_{\alpha\beta}$) and $f^\rho_\Lambda=1$ for which the initial Burgers-KPZ action \eq{eq:actionBurgers} is recovered.
The ansatz \eq{eq:anzSO} encompasses the most general dependence in wavenumbers and frequencies, and also in velocity fields, of the two-point functions compatible with Galilean invariance. Indeed, arbitrary powers of the velocity field $\vu$ are included through the operator $D_t$. Thus $\Gamma_\kappa$ is truncated only in the response field, not in the field itself.  

The choice of an ansatz allows one to compute explicitly all the vertices $\Gamma_\kappa^{(n)}$, their expression  can be found in \cite{Canet2011kpz}. For example, it yields for the two-point functions in Fourier space 
\begin{eqnarray}
\bar{\Gamma}_\kappa^{(1,1)}(\omega,\vp) &=& i\omega f_\kappa^\rho\left(\omega, \vp\right) + \vp\,^2 f_\kappa^\nu(\omega,\vp)  \nonumber\\
\bar{\Gamma}_\kappa^{(0,2)}(\omega,\vp) &=& -2 f_\kappa^D(\omega,\vp)\, . 
\end{eqnarray}
The flow equations for the three  functions  $f^\nu_\kappa$, $f^D_\kappa$ and $f^\rho_\kappa$ can then be computed from the flow equations \eq{eq:dkgam2}. More details on the calculation, in particular on the choice of the regulator, are provided in \sref{sec:fixed-point} for \NS, and can be found for Burgers-KPZ in \eg (\cite{Canet2011kpz}). 

\subsection{Scaling dimensions}
\label{sec:scalingKPZ}

With the aim of analyzing the fixed point structure,  one defines dimensionless and renormalised quantities. 
First, one introduces scale-dependent coefficients $\nu_\kappa\equiv f_\kappa^\nu(0,0)$ and $D_\kappa\equiv f_\kappa^D(0,0)$\footnote {equivalently $D_\kappa$ is defined   from $f_\kappa^D$ evaluated at a finite $\vp$ if $f_\kappa^D$ is non-analytic at $p=0$.}, which  identify at the microscopic scale $\kappa=\Lambda$ with the parameters $\nu$ and $D$ of the action \eq{eq:actionBurgers} \footnote{The scaling dimension of $f_\kappa^\rho$ is fixed to 1 by the symmetries so there is no need to introduce a coefficient $\rho_\kappa$ associated with this function.}. 
 They are associated with scale-dependent anomalous dimensions  as
\begin{equation}
 \eta_\kappa^D = -\partial_s \ln D_\kappa \quad,\quad\quad \eta^\nu_\kappa = -\partial_s \ln \nu_\kappa \, ,
\end{equation}
where $s=\ln(\kappa/\Lambda)$ is the RG ``time'' and $\p_s = \kappa\p_\kappa$.
Indeed, one expects that if a fixed point is reached, these coefficients behave as
power laws $D_\kappa\sim \kappa^{-\eta_*^D}$ and
$\nu_\kappa\sim \kappa^{-\eta_*^\nu}$ where the $*$ denotes fixed-point quantities. The physical critical exponents can then be deduced from  $\eta^D_*$ and $\eta^\nu_*$. 
For this, let us determine the scaling dimensions of the fields from \Eq{eq:anzSO}.  The three functions $f^\nu_\kappa$, $f^D_\kappa$ and $f^\rho_\kappa$ have  respective scaling dimensions $\nu_\kappa$, $D_\kappa$ and $\rho_\kappa\equiv1$ by definition of these coefficients. Since $\Gamma_\kappa$ has no dimension, one deduces from the  term $\propto \p_t$ in $\Gamma_\kappa$ that $[u_\alpha \bar u_\alpha] = \kappa^d$ and from the viscosity term that $[u_\alpha \bar u_\alpha]=\kappa^d \omega \kappa^{-2}\nu_\kappa^{-1}
$. Equating the two terms, this implies that the scaling dimension of the frequency is
$[\omega] = \nu_\kappa \kappa^2  \equiv \kappa^z$, which   defines the dynamical critical  exponent as $z=2-\eta^\nu_*$. 
The scaling dimensions of the fields can then be deduced from the forcing term of $\Gamma_\kappa$ as
 \begin{equation}
  [u_\alpha] = \big(\kappa^{d-2}D_\kappa \nu_\kappa^{-1}\big)^{1/2}  \quad,\quad\quad [\bar u_\alpha] = \big(\kappa^{d+2}\nu_\kappa D_\kappa^{-1}\big)^{1/2} \, .
 \end{equation}
 This implies in particular that the roughness critical exponent $\chi$ is related to $\eta_*^\nu$ and $\eta_*^D$ as $\chi = (2-d-\eta_*^\nu+\eta_*^D)/2$.
 
 To introduce dimensionless quantities, denoted with a hat symbol, wavevectors are measured in units of $\kappa$, \eg $\vp =\kappa \hat{\vp}$
 and frequencies in units of $\nu_\kappa \kappa^2$, \eg $\omega = \nu_\kappa \kappa^2 \hat{\omega}$. Dimensionless fields are defined as $\hat \vu =\big(\kappa^{d-2}D_\kappa \nu_\kappa^{-1}\big)^{-1/2} \vu$, $\hat{\bar\vu} = \big(\kappa^{d+2}\nu_\kappa D_\kappa^{-1}\big)^{-1/2}\bar\vu$, and dimensionless functions  as
\begin{equation}
 \hat{f}_\kappa^\nu(\hat\omega,\hat\vp)\equiv\frac{1}{\nu_\kappa} f_\kappa^\nu\left(\frac{\omega}{\nu_\kappa \kappa^2},\frac{\vp}{\kappa}\right)\,,\quad \hat{f}_\kappa^D(\hat\omega,\hat\vp) \equiv\frac{1}{D_\kappa}  f_\kappa^D\left(\frac{\omega}{\nu_\kappa \kappa^2},\frac{\vp}{\kappa}\right)\,,\quad  \hat{f}_\kappa^\rho(\hat\omega,\hat\vp) \equiv f_\kappa^\rho\left(\frac{\omega}{\nu_\kappa \kappa^2},\frac{\vp}{\kappa}\right)\,.
 \label{eq:deffkappa}
 \end{equation}
 The non-dimensionalisation of the fields introduces a scaling dimension in front  of the advection term.   
 We  define the corresponding dimensionless coupling as $\hat{\lambda}_\kappa = \big(\kappa^{d-4}D_\kappa\nu_\kappa^{-3}\big)^{1/2}\lambda$ (where $\lambda$ is the KPZ parameter, $\lambda=1$ for the Burgers equation). The Ward identity for Galilean symmetry yields that
 $\lambda$ is not renormalised, \ie it stays constant throughout the \RG flow, or otherwise stated  its flow is zero $\p_s \lambda =0$. One thus obtains for the flow of $\hat{\lambda}_\kappa$ 
 \begin{equation}
 \label{eq:dslambda}
  \partial_s \hat{\lambda}_\kappa = \frac{\hat\lambda_\kappa}{2} \big(d -4+ 3 \eta_\kappa^\nu -\eta_\kappa^D\big)\, .
 \end{equation}
 One can deduce from this equation that if a non-trivial fixed-point exists ($\hat\lambda_*\neq 0$), then the fixed point values of the anomalous dimensions satisfy the exact relation
\begin{equation}
d -4+ 3 \eta_*^\nu -\eta_*^D=0\, .
\label{eq:expoGal}
\end{equation}
Using the definitions of the critical exponents, this relation writes $\chi+z=2$ in all dimensions, which is the exact identity mentioned in the introduction of this section.

\subsection{Results in $d=1$}
\label{sec:Burgers1D}

Let us focus on the case $\beta=2$, \ie $D_{\alpha\beta}(\vp)\equiv D p_\alpha p_\beta$ in \eq{eq:actionBurgers}, which corresponds to the KPZ equation, because  exact results are available for this case in one dimension.
In $d=1$, there exists an additional discrete time-reversal symmetry which greatly simplifies the problem. This is one of the reasons why exact results could be obtained in this dimension only. In particular, one can show that it implies a particular form of fluctuation-dissipation theorem which writes for the two-point functions (\cite{Frey94,Canet2011kpz})
 \begin{equation}
2\mathrm{Re}\bar\Gamma_\kappa^{(1,1)}(\omega,\vp) =-\frac{\nu}{D}p^2\bar\Gamma_\kappa^{(0,2)}(\omega,\vp)\,,
\end{equation}
and in turn entails that $D_\kappa=\nu_\kappa$, $f_\kappa^D = f_\kappa^\nu\equiv f_\kappa$, and $f_\kappa^\rho\equiv 1$ \footnote{Note that the viscosity term in \eq{eq:anzSO} has to be symmetrised to exactly preserve the Ward identities ensuing from the time-reversal symmetry for all the $n$-point vertices, see (\cite{Canet2011kpz}).}.
 Thus there remains only one renormalisation function $f_\kappa$ and one anomalous dimension $\eta_\kappa^\nu=\eta_\kappa^D\equiv \eta_\kappa$ to compute in $d=1$. This implies that the fixed-point value of the latter is fixed exactly by \eqref{eq:expoGal} to $\eta_*=1/2$, which yields the well-known values $\chi=1/2$ and $z=3/2$.
The flow equation for the associated dimensionless function $\hat{f}_\kappa$ reads:
\begin{equation}
\partial_s \hat{f}_\kappa(\hat{\omega},\hat\vp) = \Big[\eta_\kappa +(2-\eta_\kappa) \hat{\omega} \;\p_{\hat{\omega}} + \hat\vp \;\p_{\vp} \Big]\hat{f}_\kappa(\hat{\omega},\hat\vp) +\frac{1}{D_\kappa}  \partial_s f_\kappa(\omega,\vp)\,,
\label{eq:flowf}
\end{equation}
where the first term in square bracket comes from the non-dimensionalisation \eq{eq:deffkappa} and the last term corresponds to the nonlinear loop term calculated from the flow equations \eqref{eq:dkgam2}.

This flow equation can be integrated numerically, from the initial condition $\hat{f}_{\Lambda}(\hat\omega,\hat\vp) =1$ at scale $\kappa=\Lambda$. This has been performed in (\cite{Canet2011kpz}) to which we refer for details. One obtains that the flow reaches a fixed-point, where all quantities stop evolving. The function $\hat{f}_\kappa$ converges to a fixed form $\hat{f}_*(\hat\omega,\hat\vp)$, which 
 behaves as a power-law at large $\vp$ and large $\omega$ as 
 \begin{equation}
  \hat{f}_{\kappa}(\hat\omega,\hat\vp) \stackrel{|\hat\vp|\gg 1}{\sim} |\vp|^{-\eta_*}\,,\quad \hat{f}_{\kappa}(\hat\omega,\hat\vp) \stackrel{\omega\gg 1}{\sim} \omega^{-\eta_*/(2-\eta_*)}\,.
 \end{equation}
In fact, this is expected from a very important property of the flow equation called decoupling. Decoupling means that the loop contribution $\frac{\partial_s  f_{\kappa}}{D_\kappa}$  in the flow equation \eq{eq:flowf}  becomes negligible in the limit of large wavenumbers and/or frequencies compared to the linear terms. Intuitively, a large wavenumber is equivalent to a large mass, and degrees of freedom with a large mass are damped and do not contribute in the dynamics at large  scales. This means that the \IR (effective) properties are not affected by the \UV (microscopic) details, and this pertains to the mechanism for universality.
Moreover, one can prove that the decoupling property entails scale invariance.
Indeed, if the loop contribution $\partial_s \hat f_{\kappa}$ is negligible compared to the linear terms, one can readily prove that the general solution of \eqref{eq:flowf} takes the scaling form
\begin{equation}
\hat{f}^*(\hat{\omega},\hat\vp) = \frac{1}{|\vp|^{\eta_*}} \hat{\zeta}\left(\frac{\hat{\omega}}{|\hat{\vp}|^{z}}\right)
\label{eq:solf}
\end{equation} 
where $\hat{\zeta}$ is a universal scaling function, which can be determined by explicitly integrating the full flow equation \eqref{eq:flowf}. The physical correlation function \eq{eq:defF} can be deduced from the ansatz as
\begin{equation}
C(\omega,\vp) = - \frac{\Gamma^{(0,2)}(\omega,\vp)}{|\Gamma^{(1,1)}(\omega,\vp)|^2} = \frac{2 f(\omega,\vp)}{\omega^2 + p^4 f(\omega,\vp)^2}\, .
\end{equation}
Replacing the function $f$ by its fixed-point form \eq{eq:solf} and using the value $\eta_*=1/2$, one obtains
\begin{equation}
\hat C(\hat\omega,\hat\vp) = \frac{2}{\hat p^{7/2}}  \frac{ \hat{\zeta} \left(\frac{\hat{\omega}}{\hat{p}^{3/2}}\right)}{{\hat{\omega}^2}/{\hat{p}^3}+\hat{\zeta}^2 \left(\frac{\hat{\omega}}{\hat{p}^{3/2}}\right)} 
\equiv  \frac{2}{\hat p^{7/2}}  \mathring{F}\left(\frac{\hat\omega}{\hat p^{3/2}}\right) \label{eq:defFrond}\, .
\end{equation}
The physical (dimensional) correlation function $C(\omega,\vp)$ takes the exact same form up to normalisation constants which are fixed in terms of the KPZ parameters $\nu$, $D$ and $\lambda$ and the fixed point value of the dimensionless coupling $\hat\lambda_*$.  

\begin{figure}
\includegraphics[width=0.48\linewidth]{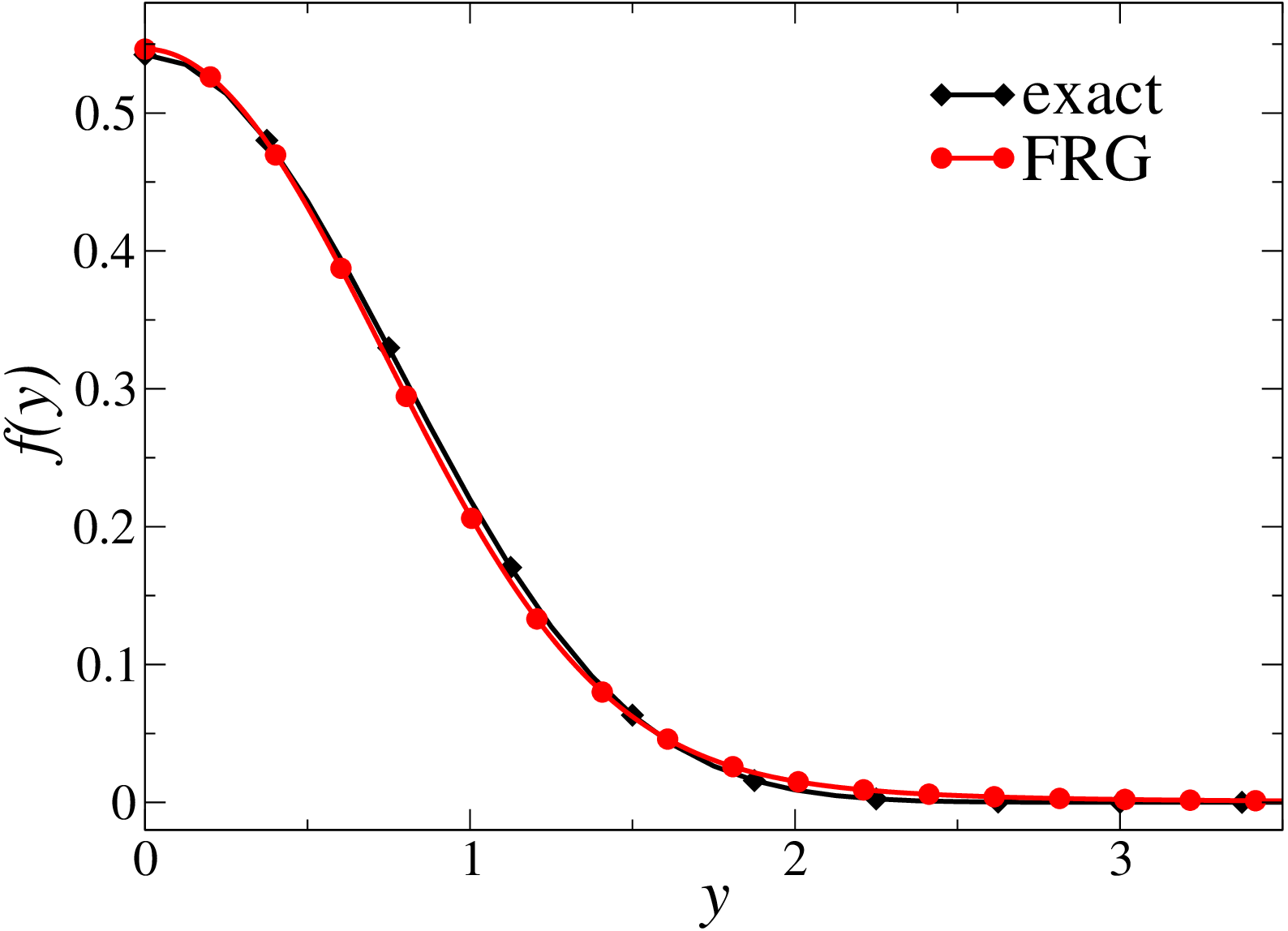}\hspace{0.5cm}
\includegraphics[width=0.49\linewidth]{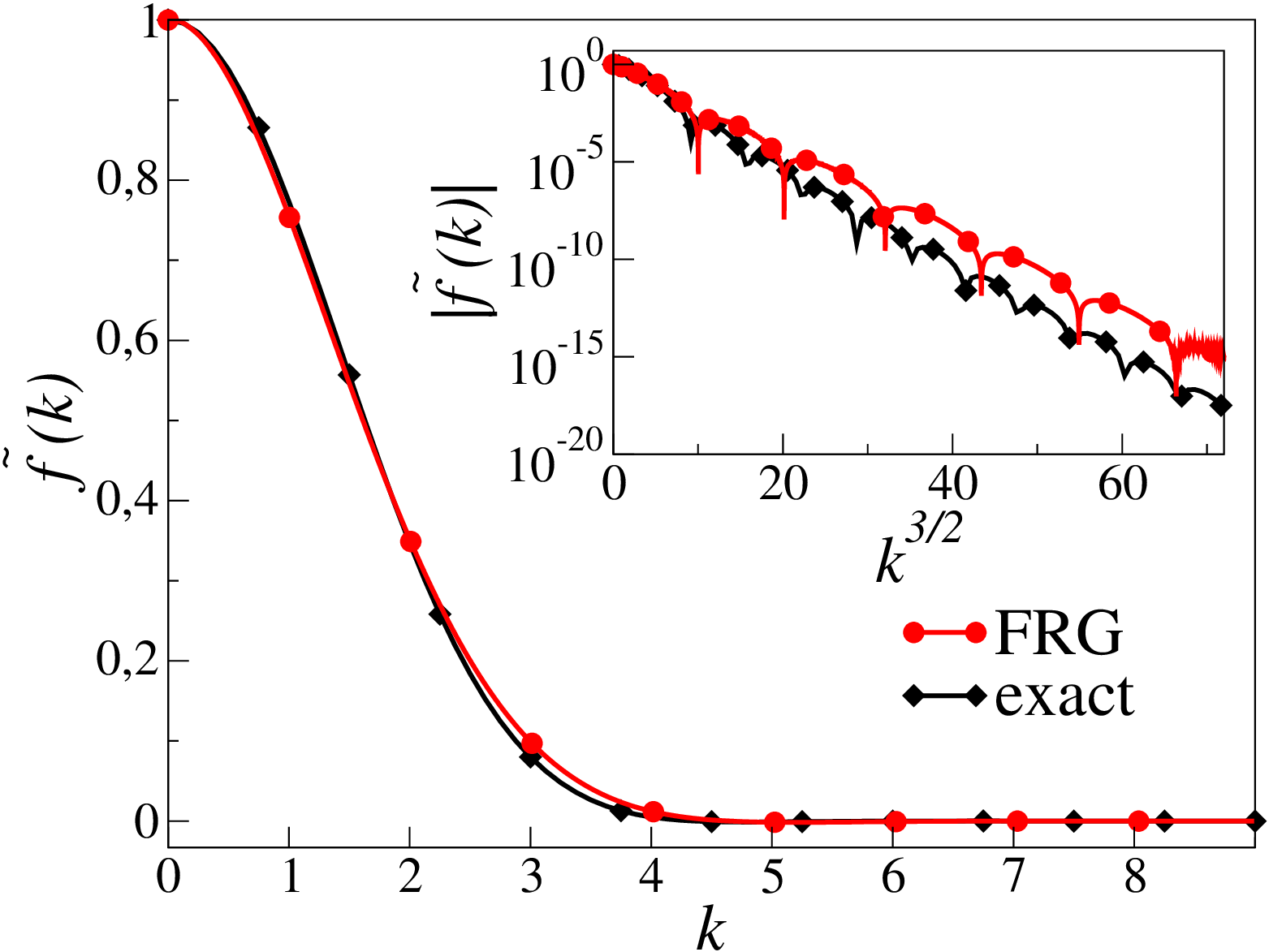}
\caption{Scaling function $f(y)$ ({\it left panel}) and $\tilde f(k)$  ({\it right panel}) for the 1D Burgers-KPZ equation from FRG compared with the exact result from (\cite{Praehofer00}).}
\label{fig:KPZ-scaling}
\end{figure}

In the context of the KPZ equation, it is a very important result. First, this proves the existence of generic scaling, \ie that the  KPZ interface  always becomes critical (rough) in $d=1$, and so that its correlations are indeed described by the universal scaling form \eq{eq:defF}. Second, the associated scaling function have been computed exactly in (\cite{Praehofer00}), and can be compared to the \FRG result. In particular, the exaclty known function, denoted $f(y)$, and its Fourier transform, denoted $\tilde{f}(k)$, are related to the \FRG function $\mathring{F}$ by the integral relations
\begin{equation}
\tilde{f}(k) = \int_0^\infty \frac{d\tau}{\pi} \cos(\tau k^{3/2}) \mathring{F}(\tau) \, ,\quad 
 f(y) =  \int_0^\infty\frac{dk}{\pi} \; \cos(k y )\tilde{f}(k).
\label{eq:defFy}
\end{equation}
The functions $f(y)$ and $\tilde{f}(k)$ computed from the \FRG approach are displayed in \fref{fig:KPZ-scaling} together with the exact results. Let us emphasise that there are no fitting parameters.
The \FRG functions match with extreme accuracy the exact ones. Interestingly,
the function $\tilde{f}(k)$ is studied in detail in (\cite{Praehofer04}), which show that this function first monotonously decreases to vanish at $k_0\simeq 4.4$,  then
exhibits a negative dip, after which it  decays to zero with a stretched
 exponential tail, over which are superimposed tiny oscillations around zero, only apparent on a logarithmic scale (inset of \fref{fig:KPZ-scaling}). The \FRG function reproduces all these features. Regarding the tiny magnitude over which they develop, this agreement is remarkable.
The Burgers equation with other values of $\beta>0$ has been studied in (\cite{Kloss2014a}), which confirms the result \eq{eq:chiBurgers}.
 
\subsection{Results in $d>1$}
\label{sec:BurgersD}

\begin{figure}
	\begin{center}
	\includegraphics[width=0.5\linewidth]{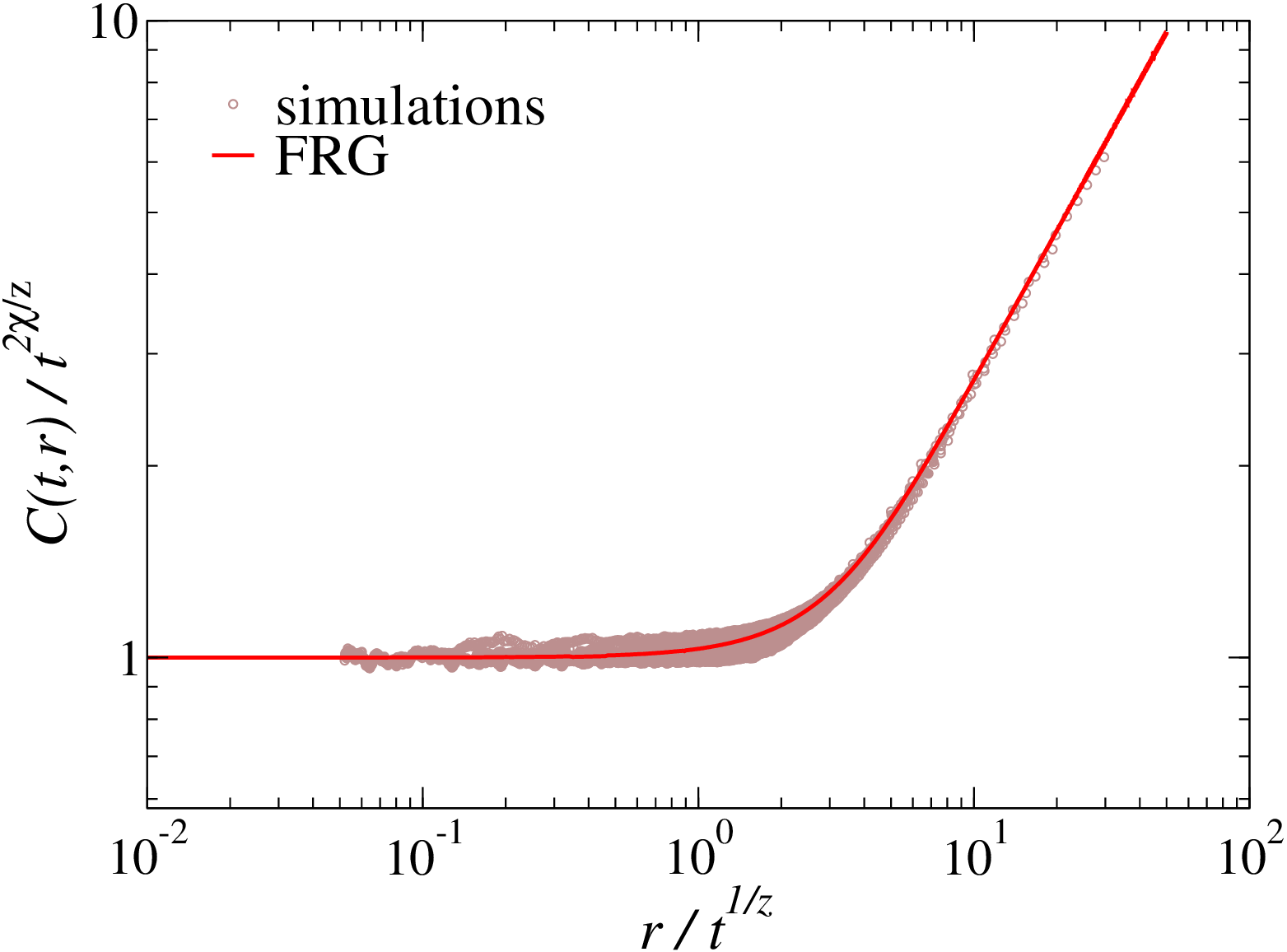}
	\end{center}
	\caption{Scaling function for the 2D Burgers-KPZ equation from FRG compared with numerical simulations from (\cite{Deligiannis2022}).}
	\label{fig:fy-2D}
\end{figure}

In dimensions $d>1$, the Burgers equation has also been studied only for $\beta>0$, which is equivalent to the KPZ equation with long-range noise. As explained at the beginning of the section, for $\beta=2$, the KPZ rough phase is controlled by a strong-coupling fixed-point, and perturbative \RG fails to all orders to find it. Moreover,  none of the mathematical methods used to derive exact results  in $d=1$ can be extended to non-integrable cases, which include $d\neq 1$. In contrast, the \FRG can be straightforwardly applied to any dimension, and thereby offers the only controlled analytical approach to study this problem, and it has been used in several works (\cite{Kloss2012,Kloss2014a,Kloss2014b,Squizzato2019}). We will not review all these results here since they are mainly concern with KPZ. Perhaps the most noteworthy result is that the \FRG provided the critical exponents and scaling functions in the physical dimensions $d=2,3$, as well as other universal quantities, such as universal amplitude ratios. The estimates for these amplitude ratios in $d=2,3$ were later confirmed in large-scale numerical simulations within less than 1\% error in (\cite{Halpin-Healy13,Halpin-Healy13Err}). The whole scaling function in $d=2$ was computed in numerical simulations of the nonequilibrium Gross-Pitaevskii equation, which surprisingly falls in the KPZ universality class for certain parameters (\cite{Deligiannis2022}). The  numerical result agrees very precisely with the \FRG result, as shown in \fref{fig:fy-2D}.

For other values of $\beta$, the results \eq{eq:chiBurgers} were extended to $d>1$, yielding
 \begin{equation}
  \chi = \max\left(\dfrac 1 3(4-d-\beta_t(d), -\dfrac 1 3(4-d-\beta) +1\right) \, ,
 \end{equation}
 where $\beta_t(1)=3/2$ and monotonically decreases with $d$. The complete phase diagram of the problem has also been determined in (\cite{Kloss2014a}).

 The values $\beta>0$ correspond to a forcing mainly exerted at small scales, where it rather plays the role of a microscopic noise, which can be interpreted as a thermal noise for $\beta=2$. This then describes a fluid at rest submitted to thermal fluctuations, as studied in (\cite{Forster77}). The  values $\beta<0$ correspond to a large-scale forcing  relevant for turbulence. This case has not been studied yet using \FRG but it certainly deserves to be addressed in future works. In particular, it would be interesting to investigate whether the decoupling property  breaks down for $\beta<0$, as occurs for \NS, which is presented in the next Section.
 
\section{Fixed-point for Navier-Stokes turbulence}
\label{sec:fixed-point}

As mentioned in \sref{sec:RG}, perturbative \RG approaches to turbulence have been hindered by the need to define a small parameter. The introduction of a forcing with power-law correlations $\propto p^{4-d-\varepsilon}$ leads to a fixed-point with an $\varepsilon$-dependent energy spectrum, which is not easily linked to the expected Kolmogorov one. In this respect, the first achievement of \FRG was to find the fixed-point corresponding to fully developed turbulence generated by a physical forcing concentrated at the integral scale. As explained in \sref{sec:approx}, the crux is that one can devise in the \FRG framework approximation schemes which are not based on a small parameter, thereby circumventing the  difficulties encountered by perturbative \RG.  This fixed point was first obtained in a pioneering work by Tomassini (\cite{Tomassini97}), which has been largely overlooked at the time. He developed an approximation close in spirit to  the \BMW approximation, although it was not yet invented at the time.
 In the meantime, the \FRG framework was successfully developed to study the Burgers-KPZ equation, as stressed in \sref{sec:KPZ}.
 Inspired by the KPZ example, the stochastic \NS equation was revisited using similar \FRG approximations in Refs. (\cite{Monasterio2012, Canet2016}). We now present the resulting fixed-point describing fully developed turbulence.

To show the existence of the fixed point and characterise the associated energy spectrum, one can focus on the two-point correlation functions.  We follow in this Section the same strategy as for the Burgers problem, that is we  resort to an ansatz for the effective average action  $\Gamma_\kappa$  to close the flow equations.  
It is clear that it is an approximation, and we establish the existence of the fixed-point within this approximation. However, as manifest in \sref{sec:KPZ}, it constitutes a very reliable approximation, built and contrained from the symmetries, which  leads to extremely accurate results in the case of the Burgers equation.  It is certainly reliable enough to prove the very existence of the fixed-point. Of course, the quantitative estimates associated with this fixed-point could be systematically improved by implementing  successive orders of the approximation scheme.

\subsection{Ansatz for the effective average action}

 One can first infer the general structure of $\Gamma_\kappa$ for \NS equations stemming from the symmetry constraints analysed in \sref{sec:Ward} and \aref{app:Ward}. On can show that it endows the following form
\begin{equation}
\Gamma_\kappa[\vu,\bar \vu,p,\bar p] = \int_{t,\vx}\Big[ \bar u_\alpha 
\Big(\partial_t u_\alpha+  u_\beta \partial_\beta u_\alpha 
+\frac{\partial_\alpha \spr}{\rho}\Big) +\bar \spr \partial_\alpha u_\alpha\Big] +\tilde \Gamma_\kappa[\vu,\bar \vu]\,,
 \label{eq:anzGammak}
\end{equation}
where  the explicit terms in square brackets are not renormalised, \ie they remain unchanged throughout the \RG flow, and are thus identical to the corresponding terms in the \NS action. The functional $\tilde \Gamma_\kappa$ is renormalised, but it has to be invariant under all the extended symmetries of the \NS action since they are preserved by the \RG flow.

Let us now comment on the choice of the regulator. It should be quadratic in the fields, satisfy the general constraints \eq{eq:cutoff}, and preserve the symmetries. A suitable choice is
\begin{equation}
 \Delta {\cal S}_\kappa[\vu,\bar \vu] =-\int_{t,\vx,\vx'} \Big\{\bar 
u_\alpha(t,\vx) N_{\kappa,\alpha \beta}(|\vx-\vx'|)\bar u_\beta(t,\vx')+\bar u_\alpha(t,\vx) R_{\kappa,{\alpha \beta}}(|\vx-\vx'|) u_\beta(t,\vx') \Big\} \, .
\label{eq:deltaSk}
\end{equation}
The first term is simply the original forcing term, promoted to a regulator by replacing the integral scale $L$ by the \RG scale $\kappa^{-1}$, since it naturally satisfy all the requirements. The forcing thus builds up with the \RG flow, and the physical forcing scale is restored when $\kappa = L^{-1}$.   The additional $R_\kappa$ term can be interpreted as an Eckman friction term in the \NS equation. Its presence is fundamental in $d=2$ to damp the energy transfer towards the large scales. Within the \FRG formalism, the energy pile-up at large scales manifests itself as an \IR divergence in the flow equation when  the $R_\kappa$ term is absent, which is removed by its presence. This term is thus mandatory to properly regularise the \RG flow in $d=2$. Its effect is to introduce an effective energy dissipation at the boundary of the effective volume $\kappa^{-d}$. This form of dissipation  is  negligible in $d=3$ compared to the dissipation at the Kolmogorov scale. 

Introducing a scale-dependent forcing amplitude $D_\kappa$ and  a scale-dependent viscosity $\nu_\kappa$, the two regulator terms can be parametrised as
\begin{equation}
 N_{\kappa,\alpha \beta}(\vq)=\delta_{\alpha\beta} D_ \kappa \hat{n}(\vq/\kappa)\; ,\quad\quad  R_{\kappa,\alpha \beta}(\vq)=\delta_{\alpha\beta} \nu_ \kappa \vq^2\hat{r}(\vq/\kappa)\, .
\end{equation} 
They are chosen diagonal in components without loss of generality since the propagators are transverse due to incompressibility, and thus the component $\propto q_\alpha q_\beta$ plays no role in the flow equations.
 The general structure of the propagator matrix $\bar G_\kappa$, \ie the inverse of the Hessian of $\Gamma_\kappa+{\Delta {\cal S}_\kappa}$ is determined in  \aref{app:floweqLO}. One can show in particular that the propagator in the velocity sector is purely transverse 
\begin{equation}
\bar G_{\kappa,\alpha\beta}(\omega,\vq) = P^\perp_{\alpha\beta}(\vq) \bar G_{\kappa,\perp}(\omega,\vq)\, , \quad\quad  P^\perp_{\alpha\beta}(\vq) = \delta_{\alpha\beta} -\frac{q_\alpha q_\beta}{q^2} \,.
\end{equation}

So far we have just given in \eq{eq:anzGammak} the general structure of $\Gamma_\kappa$ stemming from symmetry constraints.
 Now, one needs to  make an approximation in order to solve the flow equations. For this, we devise an ansatz for $\tilde \Gamma$, which  is very similar 	 to the one considered for the Burgers problem. It is built  to automatically  preserve the Galilean symmetry. For \NS, a lower order approximation than the SO one \eq{eq:anzSO} was implemented. It is called LO (leading order) and was first introduced in the context of Burgers-KPZ in (\cite{Canet2010}). The LO approximation consists in keeping the most general wavevector, but not frequency, dependence for the two-point functions. The corresponding ansatz  reads
\begin{equation}
\tilde \Gamma_\kappa[\vu,\bar \vu] = \int_{t,\vx,\vx'}\Big\{\bar 
u_\alpha(t,\vx) f^{\nu}_{\kappa,\alpha \beta}(\vx-\vx') u_\beta(t,\vx')  -\bar u_\alpha(t,\vx) f^{D}_{\kappa,\alpha \beta}(\vx-\vx') \bar 
u_\beta(t,\vx')\Big\}.
 \label{eq:anzLO}
\end{equation}
Compared to \eq{eq:anzSO}, the functions $f^{\nu}_\kappa$ and $f^{D}_\kappa$ no longer depend on the covariant operator $D_t$, which amounts to neglecting the frequency dependence, but also the dependence in the field $\vu$. For this reason,  all higher-order vertices apart from the one present in the original \NS action vanish. This means that the renormalisation of multi-point interactions is neglected in this ansatz, which is a rather simple approximation. However, it is sufficient for the Burgers-KPZ problem to find the fixed-point in all dimensions. At LO, the flow is  projected onto the two renormalisation functions  $f_\kappa^\nu$ and $f_\kappa^{D}$ which can be interpreted as an effective viscosity and an effective forcing. 

The initial condition of the flow at scale $\kappa=\Lambda$ corresponds to
\begin{equation}
f^{D}_{\Lambda,\alpha \beta}(\vx-\vx')=0 
\quad,\quad\quad
  f^{\nu}_{\Lambda,\alpha \beta}(\vx-\vx')=-\nu 
\delta_{\alpha \beta}\nabla_x^2 \delta^{(d)}(\vx-\vx')
\label{eq:CI}
\end{equation}	
such that one recovers the original \NS action \eq{eq:NSaction} \footnote{Note that the forcing has been incorporated in the regulator part $\Delta {\cal S}_\kappa$ which is why  the initial condition of $f^D_{\Lambda,\alpha \beta}$ is zero.}.
	
The calculation of the two-point functions from the LO ansatz is straightforward. At 
vanishing fields and in Fourier space, one obtains
\begin{align}
\tilde\Gamma_{\kappa,\alpha\beta}^{(1,1)}(\omega,\vp)&=f^{\nu}_{\kappa,\alpha\beta}(\vp)\nonumber\\
\tilde\Gamma_{\kappa,\alpha\beta}^{(0,2)}(\omega,\vp)& =-2f^{D}_{\kappa,\alpha\beta}(\vp).
\label{eq:gam2LO}
\end{align}
Within the LO approximation,  the only non-zero  vertex function is the one present in the \NS action, which reads in  Fourier space
\begin{equation}
\bar \Gamma_{\kappa,\alpha\beta\gamma}^{(2,1)}(\omega_1,\vp_1,\omega_2,\vp_2) = -i 
 (p_2^\alpha \delta_{\beta\gamma} + p_1^\beta \delta_{\alpha\gamma}).
 \label{eq:gamma3L0}
\end{equation}	

One can then compute the  flow equations for the two-point functions  $\tilde\Gamma^{(1,1)}_{\kappa,\alpha\beta}$  and $\tilde\Gamma^{(0,2)}_{\kappa,\alpha\beta}$ from \eq{eq:dkgam2}. They are  purely transverse, and one can deduce from them the flow equations for $f_{\kappa,\perp}^\nu\equiv P_{\alpha\beta}^\perp  f^{\nu}_{\kappa,\alpha\beta}$ and $f_{\kappa,\perp}^{D}\equiv P_{\alpha\beta}^\perp  f^{D}_{\kappa,\alpha\beta}$, see  \aref{app:floweqLO} for details and their explicit expressions. 
 
\subsection{Stationarity}
\label{sec:scalingNS}

Since we are interested in the existence of a fixed point, it is convenient to non-dimensionalise all quantities by the \RG scale as was done for the Burgers equation in \sref{sec:scalingKPZ}.
The two  coefficients $D_\kappa$ and $\nu_\kappa$ are associated with anomalous dimensions $
 \eta_\kappa^D = -\partial_s \ln D_\kappa$ and $\eta^\nu_\kappa = -\partial_s \ln \nu_\kappa$.
 As one expects a power-law behaviour for the coefficient $\nu_\kappa$ beyond a certain scale, \eg the Kolmogorov scale $\eta^{-1}$, one can relate it to the physical viscosity $\nu\equiv \nu_\Lambda$ as
 \begin{equation}
\nu_\kappa = \nu_{\eta^{-1}} ({\kappa}\eta)^{-\eta^\nu} \simeq \nu_{\Lambda} ({\kappa}\eta)^{-\eta^\nu}\, . 
\label{eq:defnuk}
   \end{equation}
  Because of Galilean invariance, the two anomalous dimensions $\eta_\kappa^D$ and $\eta_\kappa^\nu$ are not independent as in the Burgers case and satisfy the same exact relation \eq{eq:expoGal}, which can be used to eliminate for  instance $\eta_\kappa^\nu$ as 
\begin{equation}
\eta_\kappa^\nu = (4-d+\eta_\kappa^D)/3\, .
\label{eq:etanuGal}
\end{equation}

The running anomalous dimension  $\eta^D_\kappa$ should be determined by computing the flow equation for $D_\kappa$, which has to be integrated along with the flow equations for $f_{\kappa,\perp}^\nu(\vp)$ and $f_{\kappa,\perp}^\nu(\vp)$. In the case of fully developed turbulence, the  value of $\eta^D_\kappa$ can be inferred by requiring a stationary state, that is that the mean injection rate balances the mean dissipation rate all along the flow.  The average injected power by unit mass at the scale $\kappa$ can be expressed as (\cite{Canet2016})
\begin{align}
\bar\epsilon =\langle \epsilon_{\rm inj} \rangle &= \big\langle f_\alpha(t,\vx) \sv_\alpha(t,\vx) 
\big\rangle =\lim_{\delta t\to 0^+}\int_{\vx'} N_{\kappa,\alpha\beta}(|\vx-\vx'|)\,  G^{u\bar u}_{\kappa,\alpha\beta}(t+\delta t,\vx;t,\vx')\nonumber\\
&=D_\kappa \kappa^d  \lim_{\delta t\to 0^+} \Bigg[(d-1)
\int_{\hat \omega,\hat \vq} \hat n(\hat \vq) e^{-i\hat \omega \hat {\delta 
t}}\hat G^{u\bar u}_{\kappa,\perp}(\hat \omega,\hat \vq) \Bigg]\nonumber\\
 & \equiv  D_\kappa \kappa^d \hat{\gamma}\, ,
\label{eq:eps-inj}
\end{align} 
where  $\hat{\gamma}$ depends only on the forcing profile since the frequency integral of the response function is one because of causality. It is given by
\begin{equation}
 \hat{\gamma} = (d-1)\int\frac{d^d \hat\vq}{(2\pi)^d} \hat n(\hat \vq) = (d-1)\frac{2 \pi^{d/2}}{(2\pi)^d\Gamma(d/2)}\int_0^\infty d\hat{q}\, \hat{q}^{d-1}\hat n(\hat q)\, ,
 \label{eq:defgamma}
\end{equation}
where the last identity holds for an isotropic forcing.

The  conditions for stationarity depend on the dimension, and in particular on the main form of energy dissipation, which differs in $d=3$ and $d=2$. Let us focus in the following on $d=3$.
One can show that in this dimension, the dissipation at the Kolmogorov scale prevails on the dissipation at the boundary mediated by the regulator $R_\kappa$, and that it does not depend on $\kappa$ (it is given through an integral dominated by its \UV bound $\propto \eta^{-1}$). One concludes that to obtain a steady state, the mean injection should balance the mean dissipation and thus it should not depend on $\kappa$ either. This requires $D_\kappa\sim\kappa^{-d}$, that is 
  $\eta_\kappa^D = d =3 \,.$
  The value of $\eta^\nu$ is then fixed  by \eq{eq:etanuGal} to $\eta^\nu=4/3$.
  
  Thus, one may express the running coefficient $\nu_\kappa$ using \eq{eq:defnuk} and the coefficient $D_\kappa$ in term of the injection rate using \eq{eq:eps-inj} as
 \begin{equation}
 D_\kappa = \frac{\bar\epsilon}{ \hat\gamma} \kappa^{-3}\,, \quad\quad \nu_\kappa = \bar\epsilon^{1/3}\kappa^{-4/3} \, ,
 \end{equation}
 where we used the definition of the Kolmogorov scale as $\nu_\Lambda=\nu =\bar \epsilon^{1/3}\eta^{4/3}$.

Finally, the scaling dimension of any correlation or vertex function can  be deduced from the scaling dimensions of the fields (as determined in \sref{sec:scalingKPZ}). This yields for instance for the two-point correlation function
\begin{equation}
 \bar G^{uu}_{\kappa,\perp}(\omega,\vp) =  \frac{\bar\epsilon^{1/3}}{\hat{\gamma}\kappa^{13/3}} \hat{G}^{uu}_{\kappa,\perp}\left(\hat{\omega} = \frac{\omega}{\bar\epsilon^{1/3} \kappa^{2/3}},\hat{\vp}=\frac{\vp}{\kappa}\right)\, .
\end{equation}	

\subsection{Fixed-point renormalisation functions}	
	
Using the analysis of \sref{sec:scalingNS}, we define the dimensionless functions $\hat h_\kappa^\nu$ and $\hat h_\kappa^D$ as
\begin{equation}
 f_{\kappa,\perp}^\nu(\vp) = \nu_\kappa \,\kappa^2\,\hat p^2\, \hat h_\kappa^\nu(\hat \vp) 
\hspace{0.4cm}\hbox{and} \hspace{0.4cm} f_{\kappa,\perp}^D(\vp) = D_\kappa \, \hat 
p^2\, \hat h_\kappa^D(\hat \vp)\, .
 \end{equation}
 The flow equations of $\hat h_\kappa^\nu$ and $\hat h_\kappa^D$ are  given by
\begin{align}
 \partial_s \hat h_\kappa^\nu(\hat \vp) &=  \eta_\kappa^\nu  \hat h_\kappa^\nu(\hat \vp) +\hat 
\vp \partial_{\hat \vp} \hat h_\kappa^\nu (\hat \vp) +  \frac{\partial_s 
f_{\kappa,\perp}^\nu(\vp)}{\nu_\kappa \kappa^2 \hat p^2} \nonumber \\
\partial_s \hat h_\kappa^D(\hat \vp) &=  (\eta_\kappa^D+2) \hat h_\kappa^D(\hat \vp) 
+\hat \vp \partial_{\hat \vp} \hat h_\kappa^D (\hat \vp) +  
\frac{\partial_s f_{\kappa,\perp}^D(\vp)}{D_\kappa\hat p^2}
\label{eq:flowh}
\end{align}
with the substitutions for dimensionless quantities  in the flow equations 
(\ref{eq:dkfnu}) and (\ref{eq:dkfd}) for $\partial_s f_{\kappa,\perp}^\nu(\vp)$ and $\partial_s f_{\kappa,\perp}^D(\vp)$.  

\begin{figure}
\includegraphics[width=0.48\linewidth]{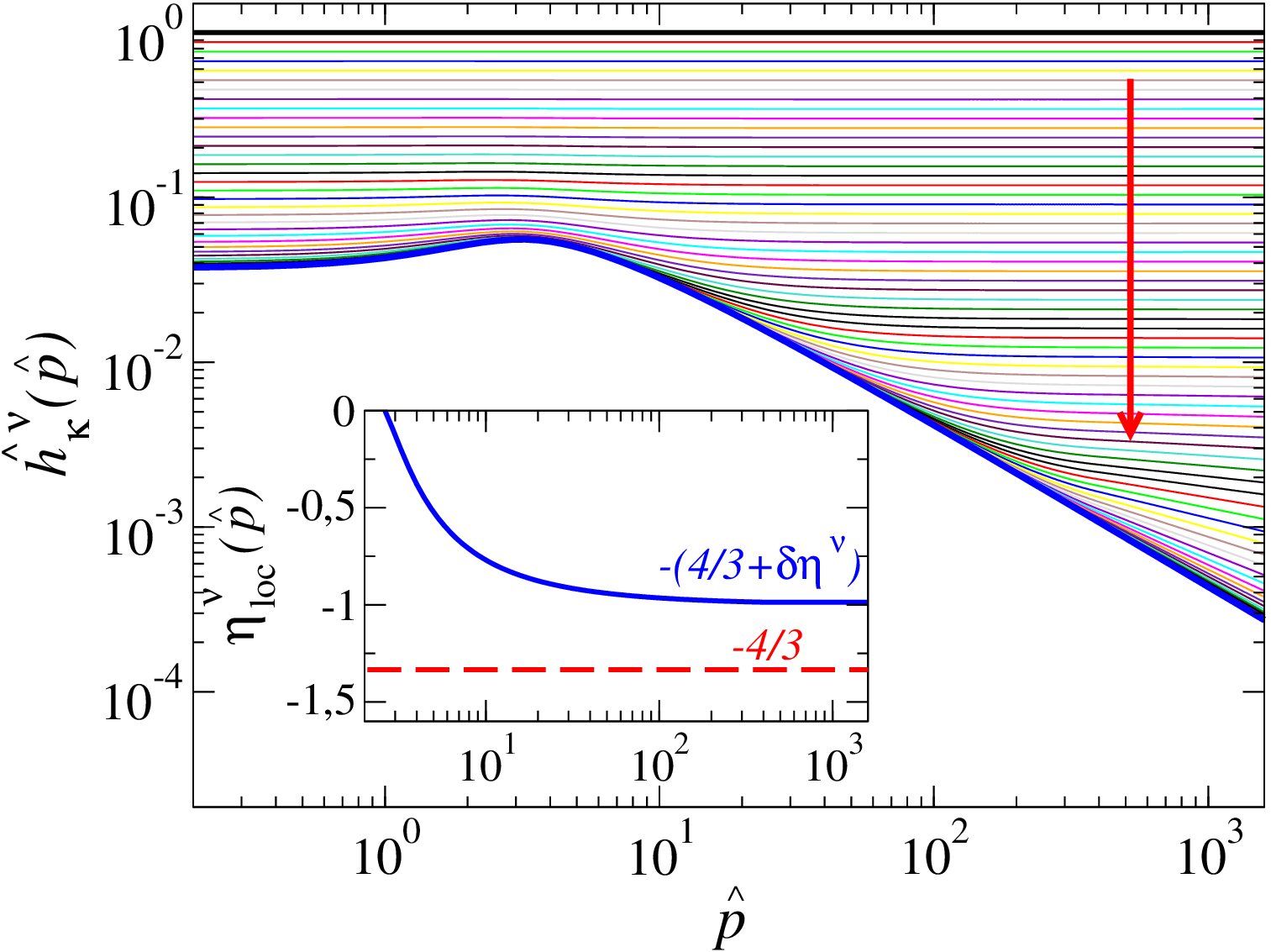}\hspace{0.5cm}
\includegraphics[width=0.49\linewidth]{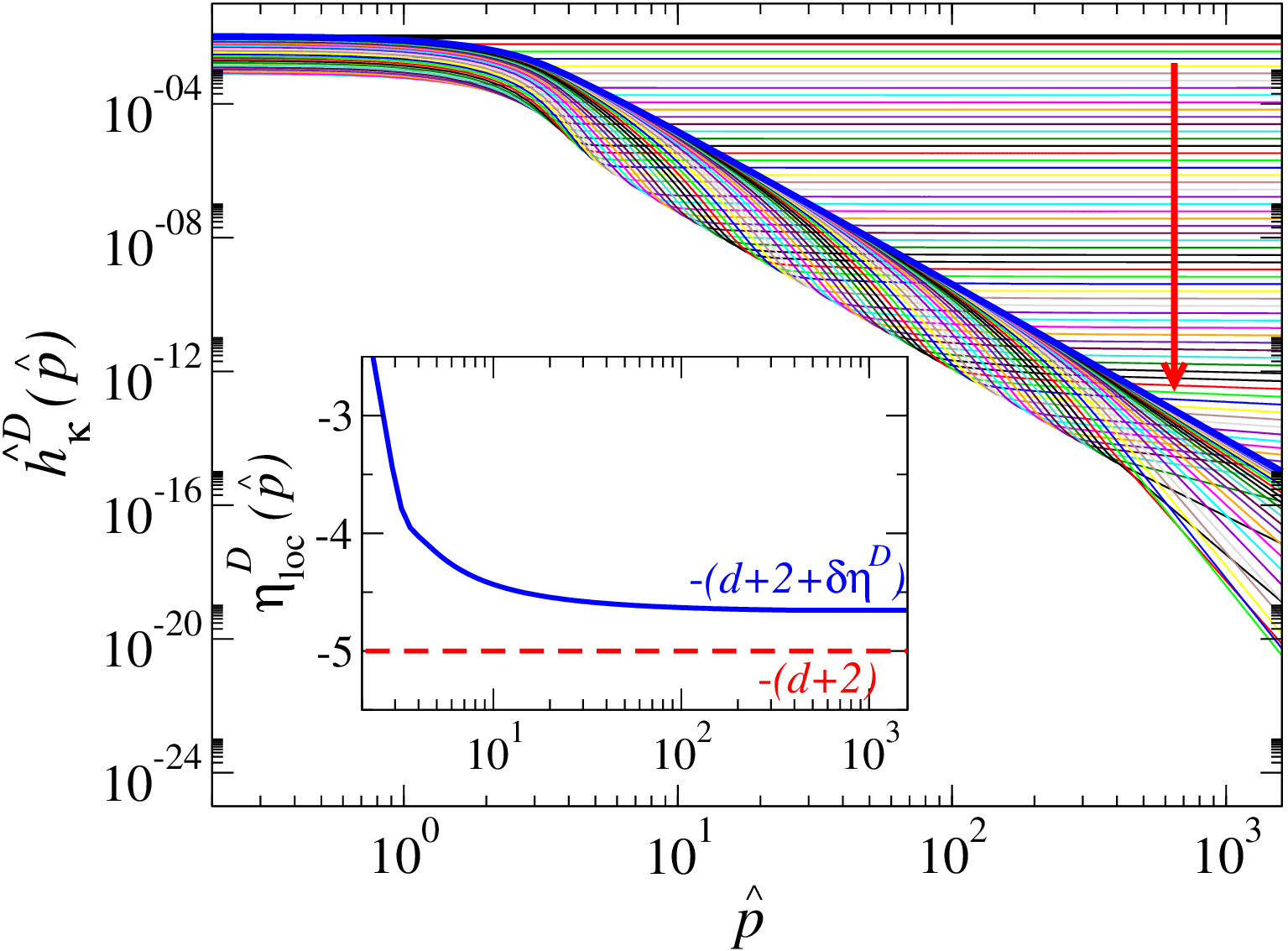}
\caption{\RG evolution of the renormalisation functions $\hat h_\kappa^\nu$ and $\hat h_\kappa^D$ with the \RG scale, from constant initial conditions (black horizontal lines) to their fixed-point shape (bold blue lines). The red arrows indicate the \RG flow, decreasing the \RG time from $s=0$ to $s=-25$. {\it Insets}: local exponents, defined by \eq{eq:locexpo}, at the fixed-point, with the corresponding K41 values indicated as dashed lines.}
\label{fig:evolution}
\end{figure}

These flow equations are first-order differential equation in the \RG scale $\kappa$, they can be integrated numerically from the initial condition \eq{eq:CI}, which corresponds to $\hat h_\kappa^\nu(\hat \vp)=1$ and $\hat h_\kappa^D(\hat \vp)=0$,  down to $\kappa \to 0$.  For this, the wavevectors are discretised on a (modulus, angle) grid. At each \RG time step $s$, the derivatives are computed using 5-point finite differences, the integrals are computed using Gauss-Legendre quadrature, with both interpolation and  extrapolation procedures to evaluate combinations  $\vp+\vq$ outside of the mesh points. One observes that the functions $\hat h_\kappa^\nu$ and $\hat h_\kappa^D$
smoothly deform from their constant initial condition to reach a fixed point where they stop evolving after a typical \RG time $s\lesssim -10$. This is illustrated on \fref{fig:evolution}, where the fixed-point functions recorded at $s=-25$ are highlighted with a thick line. This result shows that the fully developed turbulent state  corresponds to a fixed point of the \RG flow, which means that it is scale invariant. However, this fixed-point exhibits a very peculiar feature.

Indeed, the fixed point functions are found to behave as power laws at large wavenumbers as expected. However, the corresponding exponents differ from their K41 values. The fixed point functions can be described at large $\hat{p}$ as
\begin{equation}
 \hat h_*^\nu(\hat p) \sim \hat p^{-\eta^\nu+\delta \eta^\nu} \quad\hbox{and} \quad \hat 
h_*^D(\hat p) \sim \hat p^{-(\eta^D+2)+\delta \eta^D}  \label{eq:asympLO}
\end{equation}
where $\delta \eta^\nu$ and $\delta\eta^D$ are the deviations from K41 scaling.
The insets of \fref{fig:evolution} show the actual local exponents $\eta^D_{\rm loc}$ and $\eta^\nu_{\rm loc}$ at the fixed-point defined as
\begin{equation}
 \eta^D_{\rm loc} = \frac{d\ln \hat h_*^D(\hat p)}{d\ln\hat p}\quad\quad\eta^\nu_{\rm loc} = \frac{d\ln \hat h_*^\nu(\hat p)}{d\ln\hat p}\,.
 \label{eq:locexpo}
\end{equation}
When a function $f(x)$ behaves as a power law $f(x)\sim x^\alpha$ in some range, the local exponent defined by this logarithmic derivative identifies with $\alpha$ on this range. It is clear that $\eta^{\nu}_{\rm loc}$ differs from its expected K41 value $\eta^\nu=4/3$, and similarly for $\eta^{D}_{\rm loc}$. The deviations are estimated numerically as $\delta\eta^\nu\simeq \delta\eta^D\simeq 0.33$. 

Let us emphasize that it is a very unusual feature, which originates in the breaking of the decoupling property in the \NS flow equations. Non-decoupling means that the loop contributions $\frac{\partial_s f_{\kappa,\perp}^D(\vp)}{D_\kappa\hat p^2}$ and $\frac{\partial_s f_{\kappa,\perp}^\nu(\vp)}{\nu_\kappa \kappa^2 \hat p^2}$ in the flow equations \eq{eq:flowh} do not become negligible in the limit of large wavenumbers compared to the linear terms. Intuitively, a large wavenumber is equivalent to a large mass, and degrees of freedom with a large mass are damped and do not contribute in the dynamics at large (non-microscopic) scale. This means that the \IR (effective) properties are not affected by the \UV (microscopic) details, and this pertains to the mechanism for universality. For turbulence, this is not the case, and the consequences of this breaking, intimately related to sweeping, will be further expounded in   \sref{sec:largep}. Within the simple LO approximation, the signature of this non-decoupling is that the exponent of the power-laws at large $p$ are not fixed by the scaling dimensions $\eta_*^\nu$ and $\eta_*^D$, or in other words, the scaling behaviours in $\kappa$ and in $\vp$ are different. Indeed, the scaling in $\kappa$ is fixed  by \eqref{eq:etanuGal}, but $\eta_\kappa^D$ and $\eta_\kappa^\nu$ do not control the scaling behaviour in $\vp$, given by \eq{eq:asympLO} which exhibit explicit deviations $\delta\eta^\nu$ and $\delta\eta^D$. This is in sharp contrast with what was observed for the Burgers-KPZ case in \sref{sec:KPZ}, where decoupling was satisfied, and the behaviour in $\vp$ and $\omega$, \eg in the scaling form \eq{eq:solf} is indeed controlled by $\eta_\kappa$.
However, it turns out that this peculiar feature plays no role for {\it equal-time} quantities. As we show in the next section, the LO approximation leads to Kolmogorov scaling for the energy spectrum and the structure functions. Hence  the deviations $\delta\eta^\nu$ and $\delta\eta^D$ are a priori  not directly observable.

\subsection{Kinetic energy spectrum}	
\label{sec:spectrum}
 
\begin{figure}
\includegraphics[width=0.48\linewidth]{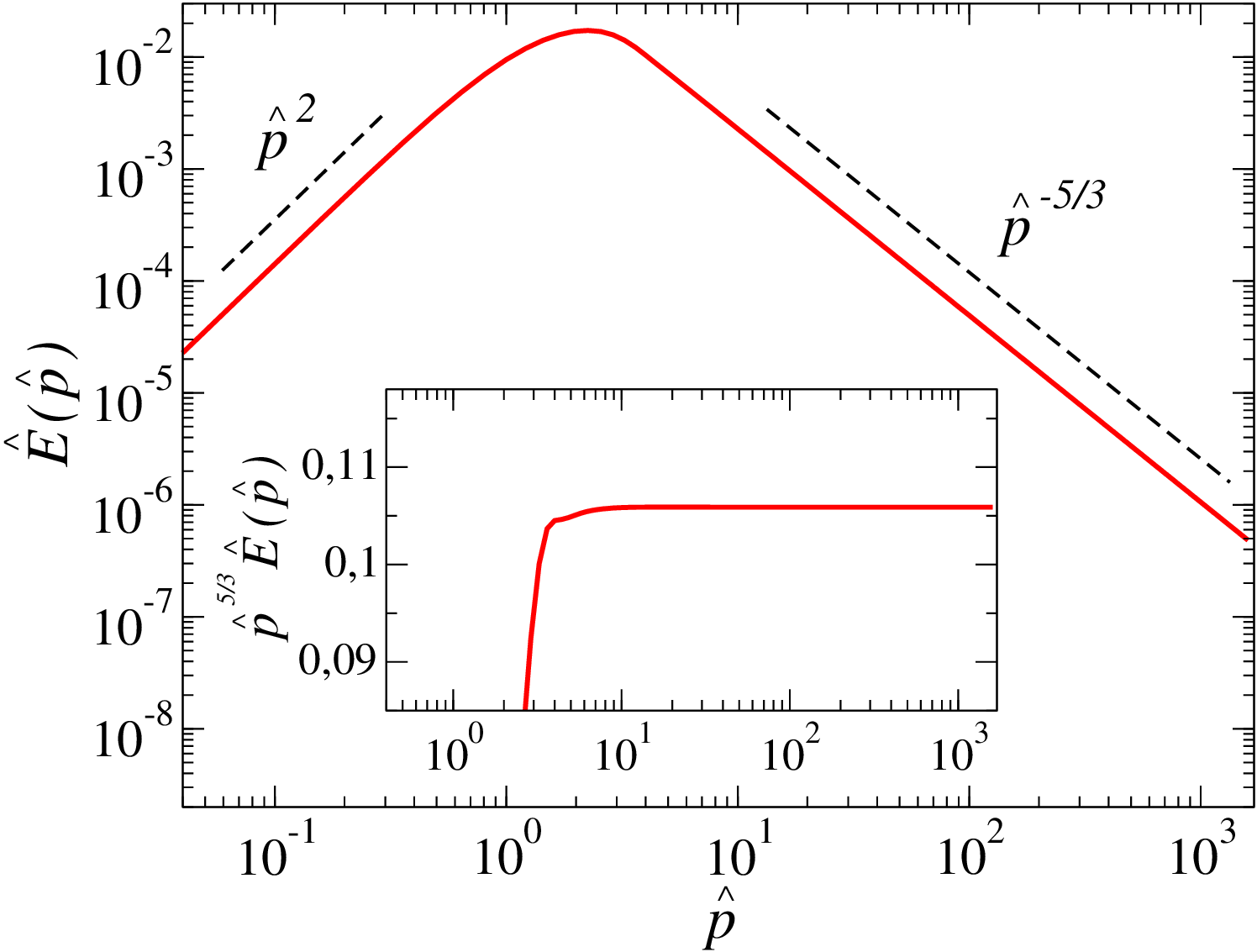}\hspace{0.5cm}
\includegraphics[width=0.49\linewidth]{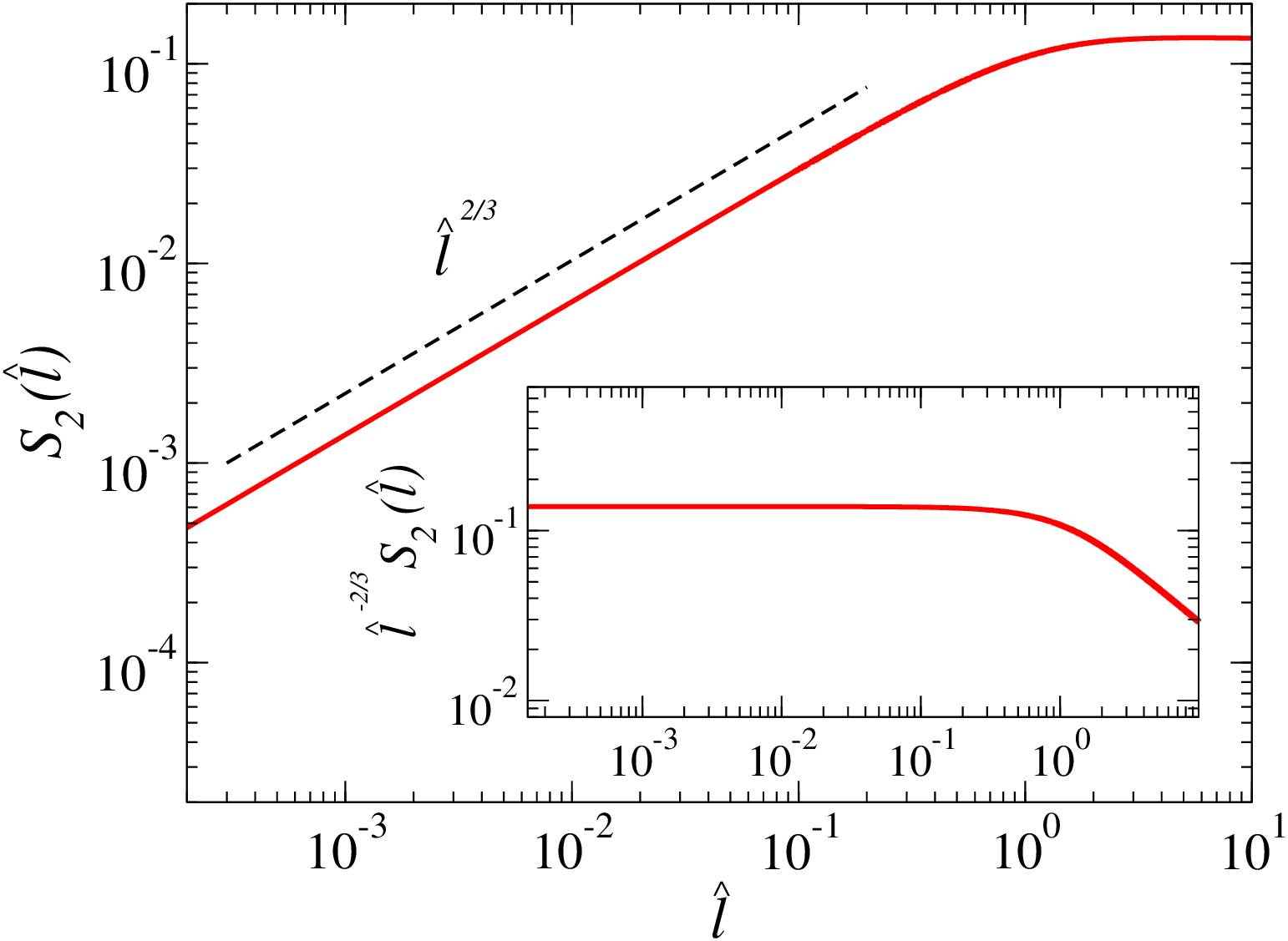}

\hspace{3.5cm}\includegraphics[width=0.49\linewidth]{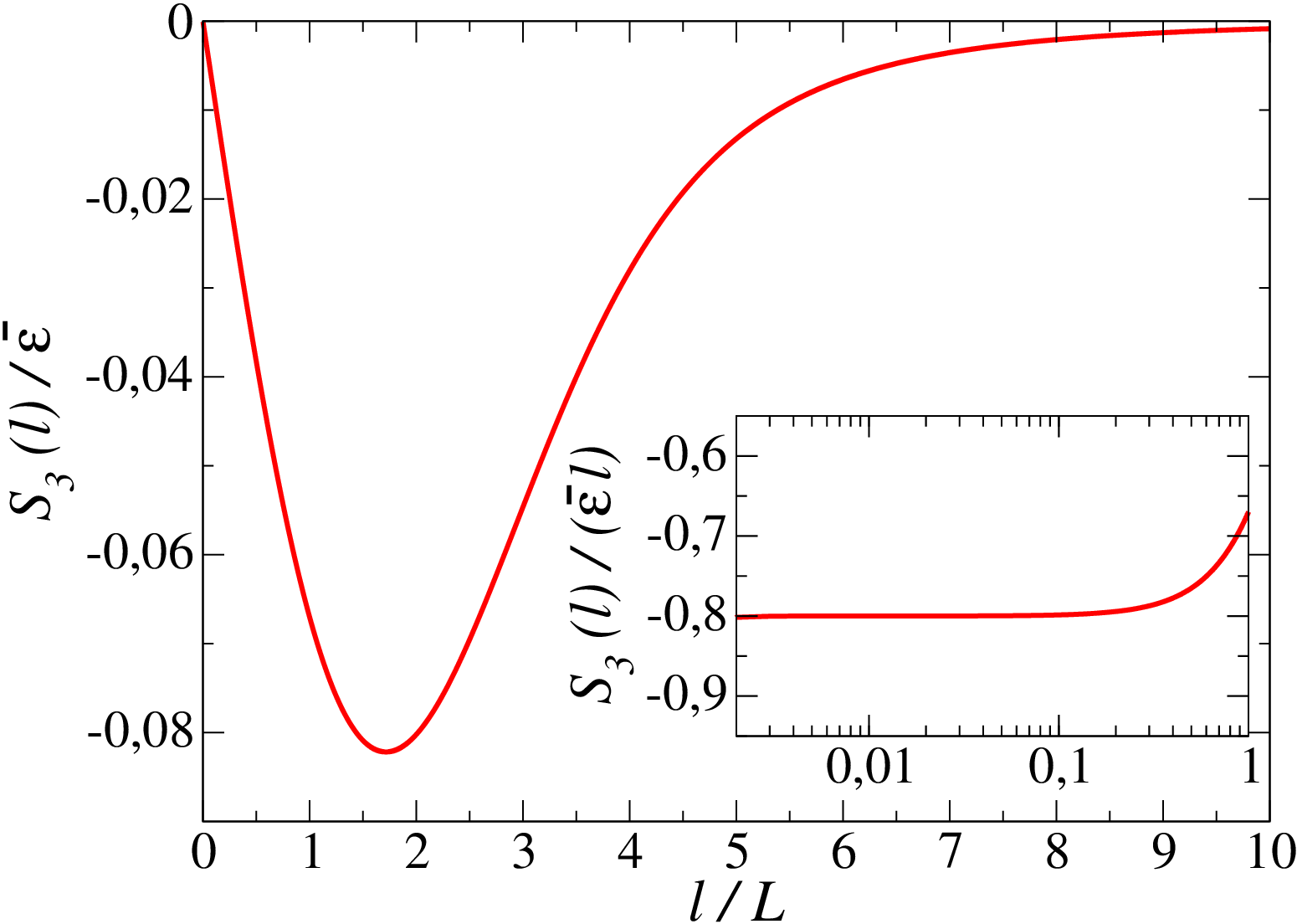}
\caption{{\it Top left panel}: Kinetic energy spectrum ({\it main plot}) and compensated one ({\it inset}); {\it Top right panel}: Second order structure function ({\it main plot}) and compensated one ({\it inset}); {\it Bottom panel}: Third order structure function ({\it main plot}) and compensated one ({\it inset}). All functions are  calculated from the \FRG fixed-point obtained within the LO approximation  in $d=3$.}
\label{fig:spectre}
\end{figure}

 Let us now compute from this fixed point physical observables, starting with the kinetic energy spectrum.
 The mean total energy per unit mass is given by
 \begin{equation}
  \frac 1 2 \langle \vvv(t,\vx)^2\rangle = \frac 1 2 G_{\alpha\alpha}^{uu}(0,0) = \frac 1 2 \int \frac{d^d \vp}{(2\pi)^d} \int \frac{d\omega}{2\pi} \bar G_{\alpha\alpha}^{uu}(\omega,\vp)\, .
  \end{equation}
 The kinetic energy spectrum,  defined as the energy density at wavenumber $p$, is hence 
 given for an isotropic flow by
 \begin{equation}
{\cal E}(p) = \frac 1 2 \frac{2\pi^{d/2}}{(2\pi)^d \Gamma(d/2)} p^{d-1} (d-1) \int \frac{d\omega}{2\pi}   \bar G_{\perp}^{uu}(\omega,p)\, .
\end{equation}
The statistical properties of the system are obtained once all fluctuations have been integrated over, \ie in the limit $\kappa\to 0$. Since the \RG flow reaches a fixed-point, the limit $\kappa\to0$ simply amounts to evaluating at the fixed-point. Within the LO approximation, and in $d=3$, one finds
\begin{equation}
{\cal E}(p) =\displaystyle\frac{\bar\epsilon^{2/3}}{\hat \gamma \kappa^{5/3}} \hat{E}(\hat p) \,, \quad\quad\hbox{with}\quad 
 \hat{E}(\hat p) = \frac{1}{2\pi^2}\, 
\hat p^{2}\,\frac{\hat h_*^D(\hatp)}{\hat h_*^\nu(\hat p) }.
\label{eq:specLO}
\end{equation}
The function $\hat{E}(\hat p)$ is represented in \fref{fig:spectre}, with in the inset
 the compensated spectrum $\hat{p}^{5/3}\hat{E}(\hat p)$.
At small $\hat p$, it behaves as $\hat p^2$, which reflects equipartition of energy.
At large $\hat p$, the two corrections $\delta\eta^\nu$ and $\delta\eta^D$ turn out to   compensate in \eq{eq:specLO}, such that one accurately recovers  the Kolmogorov scaling  $\hat{E}(\hat p) \simeq a \hat  p^{-5/3}$, with $a\simeq 0.106$ a numerical factor, which can be read off from the plateau value of the compensated spectrum $\hat{p}^{5/3}\hat{E}(\hat p)$ shown in the inset of \fref{fig:spectre}.  Thus one obtains for the spectrum
\begin{equation}
 {\cal E}(p) =  \frac{a}{\hat\gamma} \bar\epsilon^{2/3} p^{-5/3} \equiv  C_K \bar\epsilon^{2/3}p^{-5/3}\, .
 \label{eq:defCK}
\end{equation}
 For the present choice of forcing profile, which is  $\hat n(\hat x) = \hat x^2 e^{-\hat x^2}$, one obtains from \eqref{eq:defgamma} $\hat\gamma = \frac{3}{8}\pi^{-3/2}$, and thus $C_K\simeq 1.572$.
 This value is in precise agreement with typical experimental values, as compiled \eg by (\cite{Sreenivasan95}), which yields $C_K\simeq 1.52\pm 0.30$. Note that perturbative \RG approaches yielded the estimate $C_K=1.617$ (\cite{Yakhot86,Dannevik87}) although from an uncontrolled limit. Kraichnan had to resort to an improved DIA (Direct Interaction Approximation) scheme, called Lagrangean history DIA, to obtain the correct K41 spectrum and an estimate of this constant $C_K=1.617$ (\cite{Kraichnan65}). 
 
Let us emphasise that $C_K$ explicitly depends on the forcing profile through $\hat\gamma$. It is thus a priori non-universal. However, it only depends on the integral of this profile, which is required  to vanish at zero,    be peaked around the integral scale and fastly decay after. Thus  one can expect that this integral is not very sensitive to the precise shape of the forcing satisfying these contraints. Indeed, it was shown in (\cite{Tomassini97}) within an approximation similar to LO that the numerical value of $C_K$ varies very mildly, in the range 1.2--1.8, upon changing the forcing  profile.

\subsection{Second and third order structure functions}

The $n$th order structure function is defined as
\begin{equation}
 {\cal S}_n(\ell) = \Big\langle \big[(\vvv(t,\vz +\vell) -\vvv (t,\vz))\cdot 
\hat{\vell} \big]^n  \Big\rangle \, .
\end{equation}
The second order structure function can thus be calculated as
\begin{align}
{\cal S}_2(\ell)&=-2 \hat\ell_\alpha \hat\ell_\beta \Big\langle \sv_\beta(t,\vz+\vell) \sv_\alpha
(t,\vz) - \sv_\beta(t,\vz) \sv_\alpha (t,\vz)\Big\rangle\nonumber\\
&= -2  \displaystyle \int_{\omega,\vq} \, \bar G^{u u}_\perp(\omega,\vq)\,\Big[e^{i \vq\cdot \vell} -1\Big]\,\Big[1-\frac{(\hat \vell\cdot \vq)^2}{q^2} \Big] \, .
\end{align}
Within the LO approximation and in $d=3$, one obtains
\begin{equation}
 {\cal S}_2(\ell) = \frac{\bar\epsilon^{2/3}}{\hat\gamma} \kappa^{-2/3} {\hat S}_2(\hat\ell)\quad\quad \hbox{with}\quad {\hat S}_2(\hat\ell) = -\frac{1}{2\pi^2}\int_0^\infty d\hat q \,\hat q^{2}
\, \frac{\hat h_*^D(\hat q)}{\hat h_*^\nu(\hat q)} I_2(\hat q\hat \ell)\,,
\label{eq:exprS2}
 \end{equation}
and $I_2$ given by the integral
\begin{equation}
 I_2(x) = \displaystyle \frac{4}{x^3}\Big[\sin x -x \cos x -\frac{x^3}{3}\Big]\, .
\end{equation}
The function ${\hat S}_2(\hat\ell)$ is plotted in \fref{fig:spectre}. Again, for small $\hat\ell$, the corrections $\delta\eta^\nu$ and $\delta\eta^D$ precisely compensate such that the second order structure function behaves as the power law ${\hat S}_2(\hat\ell)=b\hat\ell^{2/3}$ with the Kolmogorov exponent 2/3 and $b\simeq 0.139$ (plateau value of the compensated $S_2$). Thus one  obtains
\begin{equation}
{\cal S}_2(\ell) = \bar\epsilon^{2/3} \frac{b}{\hat\gamma} \ell^{2/3}\,\equiv C_2 \bar\epsilon^{2/3}  \ell^{2/3} \,,
\end{equation}
with $C_2\simeq 2.06$, well-within the error bars of the experimental values $C_2\simeq 2.0 \pm 0.4$ (\cite{Sreenivasan95}). 

Of course, the two constants $C_2$ and $C_K$ are not independent.
Using the expression $\frac{1}{2\pi^2}\hat q^2\frac{\hat h_*^D(\hat q)}{\hat h_*^\nu(\hat q) }= \frac{a}{\hat\gamma} \hat q^{-5/3}$ deduced from \eq{eq:specLO} in \eq{eq:exprS2}, one obtains
\begin{equation}
 C_2 = -\displaystyle \frac{a}{\hat\gamma} \int_0^\infty x^{-5/3}\,I_2(x)\, dx\,  = \frac{27}{55}\Gamma(1/3)C_K\, ,
\end{equation}
where $a/\hat\gamma=C_K$ was used according to \eq{eq:defCK}. One thus recovers the well-known relation between these two constants (\cite{Monin2007}). 

Finally, let us compute the third-order structure function. In fact, as we now show, this function is completely fixed in the inertial range  by the spacetime-dependent shift symmetry of the response fields presented in \sref{sec:Karman} (or equivalently by the K\'arm\'an-Howarth relation) (\cite{Tarpin2018thesis}). Thus one recovers the exact result for $S_3$ for any approximation of the form \eq{eq:anzGammak} which automatically preserves the symmetries, and this is the case for the LO approximation. 

The third-order structure function can be expressed, using translational invariance and incompressibility as (\cite{Tarpin2018thesis})
\begin{align}
 S_3(\ell) &= 6\hat \ell_\alpha \hat\ell_\beta\hat\ell_\gamma\big\langle \sv_\alpha(t,0)\sv_\beta(t,0)\sv_\gamma(t,\ell) \big\rangle\nonumber\\
  & = 6 \hat \ell_\alpha \hat\ell_\beta\hat\ell_\gamma \int_{\omega,\vq}  \Big[q^2 (\delta_{\beta\gamma}q_\alpha + \delta_{\alpha\gamma}q_\beta) -2 q_\alpha q_\beta q_\delta \Big]\frac{\bar K(\omega,\vq)}{i q^4} e^{i (\vq\cdot\hat\ell)}\nonumber\\
  &= \frac{6}{(2\pi)^2} \int_0^\infty dq q I_3(q\ell) \int_{-\infty}^\infty\frac{d\omega}{2\pi} \bar K(\omega,\vq)\, ,
\end{align}
where  $I_3$ is given by the integral
\begin{equation}
 I_3(x) = \displaystyle -\frac{8}{x^4}\Big[(x^3-3)\sin x +3 x \cos x\Big]\, .
\end{equation}
The function $\bar K$ is related to the scalar part of the correlation function $\frac{\delta {\cal W}}{\delta L_{\alpha\beta}(t,0)\delta J_\gamma(t,\ell)}$, which can be expressed in terms of the two-point functions. Indeed, taking one functional derivative of the Ward identity \Eq{eq:wardWgaugedshift} with respect to $J_\gamma$, and evaluating  at zero sources, one obtains in Fourier space
\begin{equation}
 \bar K(\omega,\vq) = (i\omega -\nu q^2) \bar G^{u u}_\perp(\omega,\vq)+ 2 D_L\hat n(L \vq)\bar G^{u \bar u}_\perp(-\omega,\vq)\,,
\end{equation}
where $D_L$ is the forcing amplitude at the integral scale,  related to the mean energy injection as $\bar\epsilon = D_L L^{-3} \hat \gamma$ according to  \Eq{eq:eps-inj}. 
The first contribution to $S_3$, proportional to $\bar G^{u u}_\perp$, can be evaluated within the LO approximation. One obtains that it behaves at small $\ell$ as $(\ell/\eta)^{-1/3}$, which implies that it is  negligible in the inertial range. In this range, $S_3$ is hence dominated by the second contribution, proportional to $\bar G^{u \bar u}_\perp$, which writes
\begin{equation}
 S_3(\ell) = -\frac{6}{(2\pi)^2} \frac{\bar\epsilon}{\hat\gamma} L  \int_0^\infty dy \, y\,  I_3\Big(y\frac{\ell}{L}\Big) \hat{n}(y)\label{eq:S3}\, .
\end{equation}
The result is represented in \fref{fig:spectre}, for the choice $\hat n(x) = x^2 e^{-x^2}$. One observes that the 4/5th law is recovered for small scales  in the inertial range. This result can also be demonstrated analytically  from \eq{eq:S3} since in the inertial range, $I_3$ is dominated by the small values  of its argument, and $I_3(x) \sim \frac{8}{15}x$ for $x\to0$.  The remaining integral on $y$ can then be identified with $\hat \gamma$ defined in \Eq{eq:defgamma}, which yields
\begin{equation}
 S_3(\ell) = -\frac{4}{5}\bar \epsilon \ell\, .
\end{equation}
It is clear from this derivation that $C_3=-4/5$ is a universal constant, contrarily to $C_2$ or $C_K$. It has no dependence left on $\hat\gamma$.\\

To conclude this section, the merits of this approach, based on the LO approximation, are to unambiguously establish the existence of the fixed-point  with K41 scaling associated with fully developed turbulence generated by a realistic large-scale forcing.  This has been out of reach of standard perturbative \RG approaches. The approximation used here, in the form of the LO ansatz \eq{eq:anzLO}, is rather simple, since it completely neglects the frequency dependence of the two-point functions, as well as any renormalisation of the vertices. While it is sufficient to recover the exact result for $S_3$, the K41 energy spectrum with a reliable estimate of the Kolmogorov constant, it does not yield an anomalous exponent for $S_2$.
Hence, both  the frequency and field dependences  are bound to be important to capture possible intermittency corrections to K41  exponents.  However, more refined approximations including these aspects (such as the SO one implemented for Burgers-KPZ)  have not been studied yet, and remain a promising but  challenging route to explore in the quest of the computation of intermittency effects from first principles.
 
 In fact, let us emphasise that this program has been recently started  in the context of shell models of turbulence, which offer a much simpler setting, yet exhibiting intermittency. For these models, the \FRG approach provided for the first time to evidence the associated fixed-point, with intermittency corrections to $\zeta_2$ in very good agreement with values from numerical simulations (\cite{Fontaine2022sabra}).

\section{Space-time correlations from FRG}	
\label{sec:largep}

This section stresses the second main achievement stemming from \FRG methods in turbulence,
 which is the general expression for the space-time dependence of any correlation function (for any number of points $n$) of the velocity field in the turbulent stationary state in the limit of large wavenumbers. This expression can be extended to other fields (pressure field, response fields), as well as to correlations of passive scalars transported by a turbulent flow (see \sref{sec:scalar}). Contrarily to the results presented in \sref{sec:KPZ} and \sref{sec:fixed-point}, this result is not based on an ansatz,  it is exact in the limit of infinite wavenumbers.
 
  
In the following, we start with a brief overview of what is known on spatio-temporal correlations in turbulence in \sref{sec:overview}. The principle of the derivation of the \FRG results are  reviewed in \sref{sec:general-flow}, and \sref{sec:general-solution}, before providing a simple heuristic explanation of the large time regime in \sref{sec:heuristic}.  Thorough comparisons with \DNS are reported in \sref{sec:DNS} and \sref{sec:scalar}.

\subsection{Overview of space-time correlations in turbulence}
\label{sec:overview}

One of the earliest insights on the temporal behaviour of correlations was provided by Taylor's celebrated analysis of single particle dispersion in an isotropic turbulent flow (\cite{Taylor22}). The typical time scale for energy transfers in the turbulent energy cascade can be determined based on Kolmogorov original similarity hypothesis. Assuming a constant energy flux throughout the scales forming the inertial range, one deduces from dimensional analysis that the time for energy transfer from an eddy of size $k^{-1}$ to smaller eddies scales as $\tau\sim (\bar\epsilon)^{-1/3}k^{-2/3}$, which corresponds to the local eddy turnover time. This scaling is associated with straining, and  leads in particular to a frequency spectrum $E(\omega)\sim \omega^{-2}$. However, another mechanism, referred to as sweeping, was  identified by Heisenberg (\cite{Heisenberg48}) and consists in the  passive advection of small-scale eddies, without distortion, by the random large-scale eddies, even in the absence of mean flow. This effect introduces a new scale, the root-mean-square velocity $U_{\rm rms}$, related to the large-scale motion, in the dimensional analysis and yields a typical time scale for sweeping $\tau\sim (U_{\rm rms} k)^{-1}$. It then follows that the frequency energy spectrum should scale as $E(\omega)\sim \omega^{-5/3}$. The question  arises as to which of the two mechanisms dominate the Eulerian correlations, and phenomenological arguments (\cite{Tennekes75}) suggest that it is the latter, which means that the small-scale eddies are swept faster than they are distorted by the turbulent energy cascade. This conclusion has been largely confirmed by numerical simulations (\cite{Orszag72,Gotoh93,Kaneda99,He2004,Favier2010}). In particular, the $\omega^{-5/3}$ frequency spectrum is observed in the Eulerian framework, whereas the $\omega^{-2}$  spectrum emerges in the Lagrangian framework, which is devoided of sweeping (\cite{Chevillard2005,Leveque2007}).  
 
 Several theoretical attempts were made  to calculate the effect of sweeping on the Eulerian spatio-temporal correlations. Kraichnan obtained  
 from a simple model of random advection that the two-point velocity correlations should behave as a Gaussian in the variable $kt$, where $k$ is the wavenumber and $t$ the time delay (\cite{Kraichnan64}). In a nutshell, assuming that the velocity field can be decomposed into a large-scale slowly varying component $\vU$ and a small-scale  fluctuating one $\vu$ as $\vvv\simeq \vU + \vu$ with $|\vu|\ll |\vU|$, and assuming that the two components are statistically independent, one arrives at the simplified advection equation  $\p_t \vu +  \vU\cdot \nabla \vu =0$, for scales in the inertial range, \ie neglecting forcing and dissipation. This linear equation can be solved in Fourier space and one obtains $\langle \vu(t,\vk)\cdot \vu(0,-\vk)\rangle\propto \exp(-\frac 1 2 U^2 k^2 t^2)$, which is the anticipated Gaussian, and corroborates the sweeping time scale $\tau\sim (U k)^{-1}$. However, the previous assumptions cannot be justified for the full \NS equation, so it is not clear a priori whether this should be relevant for real turbulent flows. 
 
 Correlation functions were later analyzed  using Taylor expansion in time in \crefs ~(\cite{Kaneda93, Kaneda99}), which yields results compatible with the sweeping time scale $k^{-1}$ for two-point Eulerian correlations, and the straining time scale $k^{-2/3}$ for Lagrangian ones. The random sweeping hypothesis has been used in  several approaches, such as the modelling of space-time correlations within the elliptic model (\cite{He2006,Zhao2009}) or in models of wavenumber-frequency spectra (\cite{Wilczek2012}).
 Beyond phenomenological arguments, more recently, band-pass filtering techniques were applied to the \NS equation, and confirmed  the dominant contribution of sweeping in the Eulerian correlations (\cite{Drivas2017}). Very few works were dedicated to the study of multi-point multi-time correlation functions, and only within the quasi-Lagrangian framework (\cite{Lvov97,Biferale2011}) or within the multifractal approach (\cite{Biferale99}). We refer to (\cite{Wallace2014}) for a general review on space-time correlations in turbulence.

A closely related problem is the nature of space-time correlations of passive scalar fields transported by a turbulent flow. The scalar field can represent temperature or moisture fluctuations,  pollutant or virus concentrations, {\it etc}\dots, which are transported by a turbulent flow. The scalar is termed passive when it does not  affect the carrier flow. This absence of back-reaction  is of course an approximation which only holds at small enough concentration and particle size. Understanding the statistical properties of scalar turbulence plays a crucial role in many domains ranging from  natural processes  to engineering (see \eg \crefs ~(\cite{Shraiman2000,Falkovich2001,Sreenivasan2019}) for reviews).
 The modelling of the scalar spatio-temporal correlations  lies at the basis of many approaches in turbulence diffusion problems (\cite{Mazzino97}). Regarding the relevant time scales, the effect of sweeping for the scalar field was early discussed in \crefs ~(\cite{Chen89}), and its importance was later confirmed in several numerical simulations (\cite{Yeung2002,OGorman2004}) and also experiments of Rayleigh-B\'enard convection (\cite{He2011}).

Of course, one of the major difficulties, which has hindered decisive progress, arises from the non-linear and non-local nature of the hydrodynamical equations. In terms of correlation functions, it implies that the equation for a given $n$-point correlation function -- \ie the correlation of $n$ fields evaluated at $n$ different space-time points -- depends on higher-order correlations, and one faces the salient closure problem of turbulence. In order to close  this hierarchy, one has to devise some approximation, and many different schemes  have been proposed, such as, to mention a few,  the Direct Interaction Approximation (DIA) elaborated by Kraichnan (\cite{Kraichnan64}), approaches based on quasi-normal hypothesis, in particular the eddy damped quasi-normal Markovian (EDQNM) model (\cite{Lesieur2008}), or the local energy transfer (LET) theory (\cite{McComb2017}), we refer to \crefs ~(\cite{Zhou2021}) for a recent review. In this context, field-theoretical  and \RG approaches offer  a systematic framework to study correlation functions, allowing one in principle to  devise controlled approximation schemes. Although this program has been largely impeded within the standard perturbative \RG  framework (see \sref{sec:RG}), it turned out to be fruitful within the \FRG. 

Indeed, the \FRG was shown to provide for homogeneous isotropic turbulence the suitable framework to achieve a controlled  closure for any $n$-point Eulerian correlation function, which becomes asymptotically exact in the limit of large wavenumbers. This closure allows one to establish the explicit analytical expression of the spatio-temporal dependence of any generic $n$-point correlation function in this limit. The main merit of this result is to demonstrate on a rigorous basis, from the \NS equations, that the sweeping mechanism indeed dominates Eulerian correlations, and moreover to provide its exact form at large wavenumbers. Thereby, the phenomenological evidence for sweeping is endowed with a very general and systematic expression, extended to any $n$-point correlation functions. Remarkably, the expression obtained from \FRG also carries more information, and predicts in particular a change of behaviour of the Eulerian correlations at large-time delays. In this regime, another time scale emerges, which scales as $\tau\sim D^{-1}{k^{-2}}$, where $D=U_{\rm rms}^2 \tau_L$ is a turbulent diffusivity with $\tau_L$ the eddy turnover time at the integral scale. The time scale $\tau$ can therefore be called a diffusive time scale. The temporal decay of the correlations hence exhibit a crossover from a fast Gaussian decrease at small time delays to a slower exponential one at large time delays. All these results are largely explained and illustrated throughout this section.

\subsection{Closure in the large wavenumber limit of the FRG flow equations}	
\label{sec:general-flow}

The derivation is based on the large wavenumber expansion, which is inspired by the 
\BMW approximation scheme described in \sref{sec:approx}. Its unique feature for turbulence is that the flow equation for any correlation function can be closed without further approximation than taking the $p_i\to \infty$ limit, and becomes exact in this limit. 

\begin{figure}
\begin{tikzpicture}
 \node[left, scale=1] at (-1,0) {$\partial_{s}$};
 \node[draw,circle,minimum size=1.4cm,pattern=north east lines] (G) at (0,0) {${\cal W}_\kappa^{(n)}$};
     \node[below] at (-130:1.2) {$\varpi_1,\vp_1$};
    \draw (-130:1.2) -- (G);
    \node[below] at (-80:1.2) {$\dots$};
    \draw (-80:1.2) -- (G);
    \draw[-] (-30:1.2) -- (G);
    \draw[-] (20:1.2) -- (G);
    \draw[-] (70:1.2) -- (G);
    \draw[-] (110:1.2) -- (G);
    \draw[-] (160:1.2) -- (G);
\node[left, scale=1] at (2.2,0) {$=\;-\displaystyle\frac{1}{2}$};
 \node[draw,circle,minimum size=1.4cm,pattern=north east lines] (G1) at (3.3,0) {${\cal W}_\kappa^{(n+2)}$};
    \draw (G1) -- ++ (-120:1.2);
    \draw (G1) -- ++ (-70:1.2);
   \draw (G1) -- ++ (-20:1.2);
    \draw (G1) -- ++ (170:1.2);
\draw (G1) -- ++ (52:1.15);
\draw (G1) -- ++ (128:1.15);
\draw (4,0.9) arc (-10:190:0.71);
\node[scale=1] at (2.2,1.) {$\omega,\vq$};
\node[scale=1] at (4.6,1.) {$-\omega,-\vq$};
\draw (3.3,1.71) node {$\boldsymbol{\times}$};
\node[left, scale=1] at (6.1,-0.2) {$+\displaystyle\sum_{{k+\ell=n}}$};
 \node[draw,circle,minimum size=1.3cm,pattern=north east lines] (G2) at (7.3,0) {${\cal W}_\kappa^{(k+1)}$};
 \node[draw,circle,minimum size=1.3cm,pattern=north east lines] (G3) at (10,0) {${\cal W}_\kappa^{(\ell+1)}$}; 
 \draw (8.65,0) node {$\boldsymbol{\times}$};
   \draw (G2) -- (G3);
      \draw (G2) -- ++ (-70:1.2);
   \draw (G2) -- ++ (-120:1.2);
  \draw (G2) -- ++ (70:1.2);
   \draw (G2) -- ++ (120:1.2);
    \draw (G2) -- ++ (170:1.2);
    \draw (G3) -- ++ (120:1.2);
   \draw (G3) -- ++ (70:1.2);
    \draw (G3) -- ++ (-120:1.2);
      \draw (G3) -- ++ (-70:1.2);
      \draw (G3) -- ++ (-20:1.2);
\end{tikzpicture}
  \caption{Diagrammatic representation of the flow of an arbitrary $n$-point generalised correlation function ${\cal W}_\kappa^{(n)}$ \Eq{eq:dkWn}.  The crosses represent the derivative of the regulator $\p_\kappa {\cal R}_\kappa$ and the lines the propagator $\bar G_\kappa$. The correlation ${\cal W}_\kappa^{(n)}$ are represented with hatched disks not to be confused with the vertices $\Gamma^{(n)}_\kappa$. The first diagram involves a loop, along which the internal indices are summed over and  the internal wavevector $\vq$ and frequency $\omega$ are integrated over.}
  \label{fig:diagflowWn}
 \end{figure}
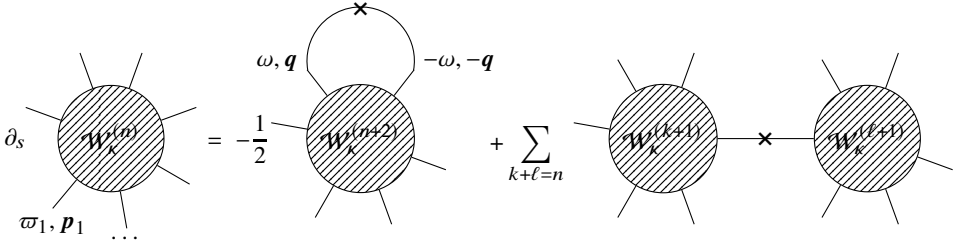

The starting point is the exact flow equation \eqref{eq:flowW} for the generating functional ${\cal W}_\kappa$. A $n$-point correlation function ${\cal W}_\kappa^{(n)}$ is defined according to~\eq{eq:defWn} as the $n$th functional derivative of ${\cal W}_\kappa$ with respect to the corresponding sources ${\cal J}_1,\cdots,{\cal J}_n$. The flow equation for ${\cal W}_\kappa^{(n)}$ is thus obtained by taking  $n$ functional derivatives of~\Eq{eq:flowW}, which yields	
\begin{align}
 \partial_{s}\frac{\delta^{n}{\cal W}_{\kappa}\left[{\cal J}\right]}{\delta {\cal J}_{1}\cdots\delta {\cal J}_{n}} &= 
-\frac{1}{2}\partial_{s}{\cal R}_{\kappa,\alpha\beta}\Biggr[\frac{\delta^{n+2}{\cal W}_{\kappa}\left[{\cal J}\right]}{\delta {\cal J}_{\alpha}\delta {\cal J}_{\beta}\delta {\cal J}_{1}\cdots\delta {\cal J}_{n}} \nonumber \\
& +\sum_{\stackrel{\left(\left\{ a_{k}\right\} ,\left\{ a_{\ell}\right\} \right)}{k+\ell=n}}\frac{\delta^{k+1}{\cal W}_{\kappa}\left[{\cal J}\right]}{\delta {\cal J}_{\alpha}\delta {\cal J}_{a_{1}}\cdots\delta {\cal J}_{a_{k}}}\frac{\delta^{\ell+1}{\cal W}_{\kappa}\left[{\cal J}\right]}{\delta {\cal J}_{\beta}\delta {\cal J}_{a_{k+1}}\cdots\delta {\cal J}_{a_{k+\ell}}}\Biggr]\,,\
\label{eq:dkWn}
\end{align}
where $\left(\left\{ a_{k}\right\} ,\left\{ a_{\ell}\right\} \right)$ indicates all possible bipartitions of the indices $1,\cdots,n$. This equation can be represented diagrammatically as in \fref{fig:diagflowWn}, where the cross stands for $\p_s {\cal R}_\kappa$. It is clear that in the second diagram, the internal line carries a partial sum of external wavevectors $\sum_{i=1}^{k+1} \vp_i$, which enters the derivative of the regulator. If one considers the large wavenumber limit, defined as the limit where all external wavevectors and all their partial sums are large with respect to the \RG scale $\kappa$, then this diagram is exponentially suppressed, and can be neglected in this limit. 	

Let us now consider the first diagram. This diagram involves a loop where the internal wavevector $\vq$ circulates and is integrated over. Because of the presence of the derivative of the regulator (the cross), $\vq$ is effectively cut to values $|\vq|\lesssim\kappa$, and as explained in \sref{sec:approx}, the limit of large wavenumbers is equivalent in the loop to the limit $\vq\to 0$, which we now compute. The wavevector $\vq$ enters two of the legs of ${\cal W}_\kappa^{(n+2)}$. The idea is  to apply the $\vq\to 0$ expansion on these two legs. Technically, within the \FRG framework, approximations are justified and performed on the $\Gamma^{(n)}_\kappa$, not directly on the ${\cal W}_\kappa^{(n)}$, because the flow equation of the formers ensures	 that they remain analytic at any finite scale $\kappa$ in wavevectors and frequencies (both in the \UV and in the \IR) thanks to the presence of the regulator. In particular, they can be Taylor expanded. However, any ${\cal W}_\kappa^{(n)}$ can be expressed in terms of the $\Gamma_\kappa^{(n)}$, as a sum of tree diagrams whose vertices are the $\Gamma_\kappa^{(k)}$, $k\leq n$, and the edges the propagators. Hence, ${\cal W}_\kappa^{(n+2)}$ can be though of, in its $\Gamma$ representation, as a functional of the average field $\Psi$, leading to 
\begin{equation}
\partial_{s}\frac{\delta^{n}{\cal W}_{\kappa}\left[{\cal J}\right]}{\delta {\cal J}_{1}\cdots\delta {\cal J}_{n}} = -\frac{1}{2} \partial_{s}{\cal R}_{\kappa,\alpha\beta} \frac{\delta\Psi_{\gamma}}{\delta {\cal J}_{\alpha}} \frac{\delta\Psi_{\delta}}{\delta {\cal J}_{\beta}} \frac{\delta^2}{\delta\Psi_{\gamma}\delta\Psi_{\delta}} \frac{\delta^{n}{\cal W}_{\kappa}\left[{\cal J}\right]}{\delta {\cal J}_{1}\cdots\delta {\cal J}_{n}}\,,
\label{eq:dsW-uu}
\end{equation}
where $\frac{\delta\Psi_{\gamma}}{\delta {\cal J}_{\alpha}}={\cal W}^{(2)}_{\kappa,\gamma\alpha} \equiv G_{\kappa,\gamma\alpha}$ are simply propagators. The loop wavevector $\vq$ then enters either a propagator $G_\kappa$ or a vertex $\Gamma_\kappa^{(k)}$, which can be evaluated in the limit $\vq \to 0$. At this stage, the symmetries, through the set of associated Ward identities for the $\Gamma_\kappa^{(n)}$, play a crucial role. Indeed, the analysis of \sref{sec:Ward} has established that any vertex $\Gamma_\kappa^{(n)}$ with one wavevector set to zero can be expressed exactly in a very simple way: if the $\vq=0$ is carried by a velocity field, then it is given in terms of lower-order vertices $\Gamma_\kappa^{(n-1)}$ through the ${\cal D}_\alpha$ operator \Eq{eq:wardGalN}, or it vanishes if it is carried by any other fields. This has a stringent consequence, which is that only contributions where $\Psi_\gamma$ and $\Psi_\delta$ are velocity fields survive in \Eq{eq:dsW-uu}, yielding  
\begin{equation}
\partial_{s} {\cal W}^{\left(n \right)}_{\kappa,\alpha_1\cdots\alpha_n}   =  -\frac{1}{2}\int_{\omega,\vq}\left(\bar{G}_{u_{\gamma}u_{\alpha}}\partial_{s}{\cal R}_{\kappa,u_{\alpha}u_{\beta}}\bar{G}_{u_{\beta}u_{\delta}}\right)\left(\omega,\vq\right)
\frac{\delta}{\delta u_{\gamma}\left(-\omega,-\vq\right)}\frac{\delta}{\delta u_{\delta}\left(\omega,\vq\right)}{\cal W}^{\left(n \right)}_{\kappa,\alpha_1\cdots\alpha_n} \,,
\end{equation}
where the $\alpha_k$'s denote both the field indices and their space components.
One can then prove, after some algebra, the following identity
\begin{align}
 \frac{\delta}{\delta u_{\gamma}\left(-\omega,-\vq\right)}\frac{\delta}{\delta u_{\delta}\left(\omega,\vq\right)}{\cal W}_{\kappa,\alpha_1\cdots\alpha_n}^{\left(n\right)}&\left(\omega_{1},\vp_{1},\cdots\right)\Bigr|_{\vq=0} 
  =\nonumber\\ 
&{\cal D}_{\gamma}\left(-\omega\right){\cal D}_{\delta}\left(\omega\right){\cal W}_{\kappa,\alpha_1\cdots\alpha_n}^{\left(n\right)}\left(\omega_{1},\vp_{1},\cdots\right)\,.
\end{align}
The proof relies on expressing the ${\cal W}_{\kappa}^{\left(n\right)}$ in terms of the $\Gamma_\kappa^{(n)}$ for which the wavenumber expansions are justified and then on using the Ward identities. We refer the interested reader to (\cite{Tarpin2018,Tarpin2018thesis}) for the complete proof. One thus obtains the explicit flow equation
\begin{align}
 \partial_{s}{\cal W}_{\kappa,\alpha_1\cdots\alpha_n}^{(n)}&\left(\cdots,\varpi_{i},\vp_{i},\cdots\right) 
=
\frac{1}{2}\int_{\omega, \vq}\left(\bar{G}_{u_{\gamma}u_{\alpha}}\partial_{s}{\cal R}_{\kappa,u_{\alpha}u_{\beta}}\bar{G}_{u_{\beta}u_{\delta}}\right)\left(\omega,\vq\right)
\qquad
\nonumber \\
& 
\sum_{k,\ell=1}^{n}\frac{p_{k}^{\gamma}p_{\ell}^{\delta}}{\omega^{2}}{\cal W}_{\kappa,\alpha_1\cdots\alpha_n}^{\left(n\right)}\left(\cdots,\varpi_{k}+\omega,\vp_{k},\cdots,\varpi_{\ell}-\omega,\vp_{\ell},\cdots\right)\quad +\quad{\cal O}(p_{\rm max})\, , 
\end{align}
which constitutes one of the most important results obtained with \FRG methods. The flow equation for an arbitrary $n$-point connected correlation  function ${\cal W}_{\kappa}^{(n)}$ (of any fields, velocity, pressure, response fields, \dots) is closed in the large wavenumber limit, in the sense that it no longer involves higher-order vertices. Remarkably, it does not depend on  the whole hierarchy of ${\cal W}_{\kappa}^{(m)}$ with $m\leq n$, but only on the $n$-point function for which the flow is expressed and on propagators (\ie two-point functions). Moreover, it is a linear equation in  ${\cal W}_{\kappa}^{(n)}$. Once the two-point functions are determined (within  some approximation scheme), any $n$-point correlations can be simply obtained. 
 The term on the r.h.s. is calculated exactly, \ie it is the exact leading term in the large wavenumber expansion. Subleading corrections are at most of order $p_{\rm max}$ where $p_{\rm max}$ indicates the largest external wavevector at which the function is evaluated.

This equation can be further simplified, by noting that ${\cal W}_{\kappa}^{(n)}$ in the r.h.s. does not enter the integral over $\vq$, but only the integral over the internal frequency $\omega$. Moreover, ${\cal W}_{\kappa}^{(n)}$ depends on the latter only through local shifts of the external frequencies. Hence, let us denote by $H_{\kappa,\gamma\delta}$ the term containing the propagators and the regulator
\begin{equation}
 H_{\kappa,\gamma\delta}(\omega,\vq) = \left(\bar{G}_{u_{\gamma}u_{\alpha}}\partial_{s}{\cal R}_{\kappa,u_{\alpha}u_{\beta}}\bar{G}_{u_{\beta}u_{\delta}}\right)\left(\omega,\vq\right) \, . 
 \label{eq:exprH}
\end{equation}
On can show that because of incompressibility, the propagators are transverse, hence $H_{\kappa,\gamma\delta}(\omega,\vq)=P^\perp_{\gamma\delta}(\vq) H_{\kappa,\perp}(\omega,\vq)$. Finally, by Fourier transforming back in time variables, the shifts in frequency can be transferred to the Fourier exponentials, such that the function ${\cal W}_{\kappa}^{(n)}$ in the mixed time-wavevector coordinates $(t_i,\vp_i)$ can be pulled out of the frequency integral also, yielding
\begin{align}
 \partial_{s}{\cal W}_{\kappa,\alpha_1\cdots\alpha_n}^{(n)}&\left(\cdots,t_{i},\vp_{i},\cdots\right) 
=\frac{d-1}{2d}{\cal W}_{\kappa,\alpha_1\cdots\alpha_n}^{(n)}\left(\cdots,t_{i},\vp_{i},\cdots\right) \nonumber\\
&\times  \sum_{k,\ell=1}^{n}{\vp_{k}\cdot \vp_{\ell}} \int_{\omega, \vq} H_{\kappa,\perp}(\omega,\vq)
\frac{e^{i\omega(t_k-t_\ell)}-e^{i\omega t_k} - e^{-i\omega t_\ell}+1}{\omega^{2}} \quad+\quad{\cal O}(p_{\rm max})\, . 
\label{eq:flowWn}
\end{align}
 Thus, one is left with a linear ordinary differential equation, which can be solved explicitly, as detailed in \sref{sec:general-solution}. Again, the detailed proof of this expression can be found in (\cite{Tarpin2018,Tarpin2018thesis}). We discuss in \aref{app:nexttoleading} the next-to-leading order in the large wavenumber expansion, which has been computed for 2D turbulence in (\cite{Tarpin2019}).

\subsection{General expression of the time-dependence of $n$-point correlation functions}
\label{sec:general-solution}

Let us now discuss the form of the general solution of the flow equation \eq{eq:flowWn} for a generalised correlation function ${\cal W}_\kappa^{(n)}$. As emphasised in \sref{sec:general-flow}, \Eq{eq:flowWn} is very special in many respects: i) it is closed as it depends only on  ${\cal W}_\kappa^{(n)}$ (besides the propagators) and not on higher-order correlation functions,  ii) it is linear in ${\cal W}_\kappa^{(n)}$, iii) the leading order in the large $p$ expansion is exact, and iv) it does not decouple, which implies that the large wavenumber limit contains non-trivial information. This information is related to the breaking of standard scale invariance, and reflects the fact that the small-scale properties of turbulence are affected by the large scales.

We have shown in \sref{sec:fixed-point} that the \FRG flow for \NS equation, in some reasonable approximation, with a physical large-scale forcing, reaches a fixed point.  This means that the flow essentially stops when $\kappa$ crosses the integral scale $L^{-1}$. However, because of the non-decoupling, the large scale remains imprinted in the solution. It is instructive to uncover the underlying mechanism. At a fixed-point of a \RG flow, the system does not depend any longer on the \RG scale $\kappa$ by definition, which leads to scale invariance. The usual way to evidence such a behaviour is to introduce as in  \sref{sec:fixed-point} non-dimensionalised quantities, denoted with a hat symbol. Wavevectors are measured in units of $\kappa$, \eg $\hat{\vp} = \vp/\kappa$, frequencies in units of $\nu_\kappa \kappa^2$, \eg $\hat{\omega}=\omega/(\nu_\kappa\kappa^2)$. Denoting generically $d_n$ the scaling dimension of a given correlation function ${\cal W}^{(n)}_{\kappa}$, one also defines $\hat{{\cal W}}^{(n)}_{\kappa,\alpha_1\cdots\alpha_n}(\cdots,\hat{t}_i,\hat{\vp}_i,\cdots) = \frac{{\cal W}^{(n)}_{\kappa,\alpha_1\cdots\alpha_n}}{\kappa^{d_n}}(\cdots,\nu_\kappa \kappa^2 t_i,\vp_i/\kappa,\cdots)$. For a generalised correlation funtion of $m$ velocity fields and $\bar{m}$ response fields with $m+\bar{m}=n$ in mixed time-wavevector coordinates, one has  $d_n = 3(m-1)+(m-{\bar{m}/3})$ in $d=3$. The flow equation for the dimensionless correlation function then writes
\begin{equation}
 \Big\{\p_s -d_n - \hat{\vp}_i \cdot \p_{\hat{\vp}_i} + z \hat t_i \p_{\hat t_i}\Big\}
 \hat{{\cal W}}^{(n)}_{\kappa,\alpha_1\cdots\alpha_n}(\cdots,{\hat t_i,\hat{\vp}_i}, \cdots) = \hat{{\cal F}}_{\rm loop}(\cdots,{\hat t_i,\hat{\vp}_i}, \cdots) \, ,
 \label{eq:flowhatWn}
\end{equation}
where $\hat{{\cal F}}_{\rm loop}$ denotes the non-linear part of the flow, corresponding to the contribution of the loop \eq{eq:flowWn}, expressed in dimensionless quantities.

Scale invariance, as encountered in usual critical phenomena (say at a second order phase transition), emerges if two conditions are fulfilled: existence of a fixed-point, and decoupling. Indeed, as shown in \sref{sec:KPZ} for the Burgers-KPZ equation, these two conditions yield that the two-point correlation function takes a scaling form \eq{eq:defF} -- which is the hallmark  of scale invariance. This derivation can be generalised for any correlation function. Let us assume that the flow reaches a fixed-point, which means $\p_s\hat{{\cal W}}^{(n)}_{\kappa}=0$ and that decoupling occurs, which means that the loop contribution of the flow is negligible in the limit of large wavenumber, \ie
\begin{equation}
 \frac{\hat{{\cal F}}_{\rm loop}}{\hat{{\cal W}}^{(n)}_{\kappa}} \stackrel{\hat{\vp}_i\to\infty}{\longrightarrow} 0\, .\label{eq:decoupling}
\end{equation}
One can easily show, using some elementary changes of variables, that the general form of the solution to the remaining homogeneous equation is  a scaling form
\begin{equation}
 \hat{{\cal W}}_{*,\alpha_1\cdots\alpha_n}^{(n)}(\cdots,{\hat t_i,\hat{\vp}_i}, \cdots)= \frac{1}{\hat{p}_{\rm tot}^{d_n}}\hat{w}_{\alpha_1\cdots\alpha_n}^{(n)}\left(\cdots,\hat{t}_i \hat{p}_i^z,\frac{\hat{\vp}_i}{\hat{p}_i},\cdots\right)\, ,
 \label{eq:solWnK41}
\end{equation}
where $\hat{\vp}_{\rm tot}=\sum_{i=1}^{n-1}\hat{\vp}_i$ and $\hat{w}^{(n)}$ is a universal scaling function which form is not known, and could be determined by explicitly integrating the flow equation.

If there is no decoupling, \ie if the condition \eq{eq:decoupling} does not hold, the solution to \eq{eq:flowhatWn} is no longer a scaling form, but is modified by the loop contribution. The explicit expression of this solution can be obtained as an integral expression, which can be simplified in two limits, the limits of small and large time delays, that we study separately in the following.

\subsubsection{Small time regime}

The limit of small time delays corresponds to the limit where all $\hat t_i \to 0$. In this limit, one can expand the exponential in the second line of \eq{eq:flowWn} 
\begin{equation}
\int_{\omega, \vq} H_{\kappa,\perp}(\omega,\vq)
\frac{e^{i\omega(t_k-t_\ell)}-e^{i\omega t_k} - e^{-i\omega t_\ell}+1}{\omega^{2}} \stackrel{t_i \to 0}{\simeq}
 I_\kappa t_k t_\ell\quad\quad\hbox{with}\quad I_\kappa \equiv \int_{\omega,\vq} H_{\kappa,\perp}(\omega,\vq)\, .
\end{equation}
Note that the explicit expression of $H_\kappa$ is not needed in the following, one only demands that the frequency integral converges.
The flow equation for the dimensionless $\hat{{\cal W}}^{(n)}_{\kappa}$ then simplifies to 
\begin{equation}
 \Big\{\p_s -d_n - \hat{\vp}_i \cdot \p_{\hat{\vp}_i} + z \hat t_i \p_{\hat t_i}\Big\}
 \hat{{\cal W}}^{(n)}_{\kappa,\alpha_1\cdots\alpha_n}(\cdots,{\hat t_i,\hat{\vp}_i}, \cdots) =  \frac{\hat I_\kappa}{3}\big|\hat t_\ell \hat{\vp}_\ell \big|^2\hat{{\cal W}}^{(n)}_{\kappa,\alpha_1\cdots\alpha_n}(\cdots,{\hat t_i,\hat{\vp}_i}, \cdots) \, .
 \label{eq:flowhatWn-small-t}
\end{equation}
Since the flow reaches a fixed-point, one can focus on the fixed-point equation as previously ($\p_s \cdot = 0$). At the fixed-point, $\hat I_\kappa \to \hat I_*$ which is just a number.
 To solve the fixed-point equation, one can introduce a $(n-1)\times(n-1)$ rotation matrix ${\cal R}$ in wavevector space such that
${\cal R}_{i1} = \frac{t_i}{\sqrt{t_\ell t_\ell}}$ and define new variables $\vrho_k$ such that $\vp_i={\cal R}_{ij} \vrho_j$.
 This transforms the sum  $|t_\ell \vp_\ell |^2$ into $t_\ell t_\ell|\vrho_1|^2$, and allows one to obtain the explicit solution,  which reads in physical units (\cite{Tarpin2018})
 \begin{align}
 &\log \Big[\bar\epsilon^{\frac{\bar{m}-m}{3}}L^{-d_n}{\cal W}^{(m,\bar{m})}_{\alpha_1\dots\alpha_{n}}({t_1,\vp_1}, \cdots, {t_{n-1},\vp_{n-1}})\Big] = 
 -\alpha_0\bar\epsilon^{2/3}L^{2/3} \,  t_\ell t_\ell\, \rho_1^2\nonumber\\
 - d_n \log (\rho_1 L) &+ {{w_0}^{(m,\bar{m})}}_{\alpha_1\dots\alpha_{n}}\left(\rho_1^{2/3}\bar\epsilon^{1/3} t_1,\frac{\vrho_{1}}{\rho_1}, \cdots,
 \rho_1^{2/3}\bar\epsilon^{1/3}t_{n-1}, \frac{\vrho_{n-1}}{\rho_1}\right) + {\cal O}(p_{\rm max} L)\, ,
\label{eq:solWn-smallt}
\end{align}
where ${{w_0}^{(m,\bar{m})}}$ is a universal scaling function, which explicit form is not given by the fixed point equation only, but can be computed numerically by integrating the flow equation in some approximation. The constant $\alpha_0 = \hat \gamma\hat I_*/2$ is  non-universal since it depends on the forcing profile through $\hat \gamma$, but it does not depend on the order $n$ of the correlation, nor on the specific fields involved (velocity, response velocity). It is thus the same number for all correlation functions of a given flow.

All the terms in the second line of \eq{eq:solWn-smallt} correspond to the K41 scaling form obtained in \eq{eq:solWnK41}. However, these terms are subdominant, that is of the same order as the indicated neglected terms ${\cal O}(p_{\rm max} L)$ in the flow equation. Hence, they should be consistently discarded (included in ${\cal O}(p_{\rm max} L)$), and were kept here  only for the sake of the discussion. As they are of the same order as the neglected terms, it means that they could receive corrections, \ie be modified by these terms. In particular, at equal time, the leading term in the first line of \eq{eq:solWn-smallt} vanishes, and one is left with only these subdominant contributions, which are of the form K41 plus possible corrections. In  other words, the \FRG calculation says nothing at this order about intermittency corrections on the scaling exponents of the structure functions, which are equal-time quantities.

However, at unequal times, the important part of the result is the term in the first line of \eq{eq:solWn-smallt}, which is the exact leading term at large wavenumber. For this reason, the most meaningful way to write this result is
under the form
\begin{equation}
 {\cal W}^{(m,\bar{m})}_{\alpha_1\dots\alpha_{n}}({t_1,\vp_1}, \cdots, {t_{n-1},\vp_{n-1}}) \propto \exp\Big[ -\alpha_0 (L/\tau)^2 |t_\ell \vp_\ell |^2 + {\cal O}(p_{\rm max} L) \Big]\, ,
\label{eq:solWngauss}
 \end{equation}
where $\tau = (L^2/\bar\epsilon^{1/3})$ is the eddy turnover time at the integral scale, and $(L/\tau)$ identifies with $U_{\rm rms}$,  yielding  the sweeping time scale $(U_{\rm rms}p)^{-1}$ expected from phenomenological arguments.
 The salient feature of this result is that it breaks standard scale invariance, which directly originates in the non-decoupling. As stressed before, if there were decoupling, \ie if the fixed point conformed with a standard critical point, a time variable could only appear through the scaling combination $p^z t$ with $z=2/3$ in K41 theory. Instead, time enters in \eq{eq:solWngauss}
 through the combination $pt$, which can be interpreted as an effective dynamical exponent $z=1$. This is a large correction, which can be attributed to the effect of random sweeping.
 As a consequence, the argument in the exponential  explicitly depends on a scale, the integral scale $L$. This residual dependence in the integral scale of the statistical properties of turbulence in the inertial range is well-known, and crucially distinguishes turbulence from standard critical phenomena.

Let us mention that, interestingly, the non-decoupling also occurs for the passive scalars in the Kraichnan model (see \sref{sec:scalar}). In this context, the breaking of scale invariance was associated to the existence of zero modes (\cite{Falkovich2001}). It would be very interesting to investigate the link between the existence of zero modes and non-decoupling, which could give new insights for the case of \NS turbulence.

\subsubsection{Large time regime}
Let us now consider the large time limit of \eq{eq:flowWn}, corresponding to $t_i\to \infty$.
One can show that in this limit
\begin{align}
&\int_{\omega,\vq} H_{\kappa,\perp}(\omega,\vq)
 \frac{e^{i\omega(t_k-t_\ell)}-e^{i\omega t_k} - e^{-i\omega t_\ell}+1}{\omega^{2}} \stackrel{t_i \to \infty}{\simeq}\frac{J_\kappa}{2} \Big(|t_k|+|t_\ell|-|t_\ell-t_k|\Big)\nonumber\\
 & \quad\quad\hbox{with}\quad J_\kappa = \int_{\vq} H_{\kappa,\perp}(0,\vq)\, ,
\end{align}
and thus the dimensionless flow equation for $\hat{{\cal W}}^{(n)}_{\kappa}$ becomes in this limit
\begin{align}
 \Big\{\p_s -d_n - \hat{\vp}_i \cdot \p_{\hat{\vp}_i} + z \hat t_i \p_{\hat t_i}\Big\}&
 \hat{{\cal W}}^{(n)}_{\kappa,\alpha_1\cdots\alpha_n}(\cdots,{\hat t_i,\hat{\vp}_i}, \cdots) = \nonumber\\
 &\frac{\hat J_\kappa}{6}\sum_{k,\ell} \hat\vp_k\cdot\hat\vp_\ell \Big(|\hat t_k|+|\hat t_\ell|-|\hat t_\ell-\hat t_k|\Big)\hat{{\cal W}}^{(n)}_{\kappa,\alpha_1\cdots\alpha_n}(\cdots,{\hat t_i,\hat{\vp}_i}, \cdots) \, .
 \label{eq:flowhatWn-large-t}
\end{align}
This equation can be solved at the fixed-point in an analogous way as for the small time limit, introducing another rotation matrix such that in the new wavevector variables $\vvvarrho_1 = \sum_k\vp_k$. Specifying to equal time lags $t_i=t$ for simplicity, the solution reads (\cite{Tarpin2018})
\begin{align}
 &\log\Big(\bar\epsilon^{\frac{\bar{m}-m}{3}}L^{-d_n}{\cal W}^{(m,\bar{m})}_{\alpha_1\dots\alpha_{n}}(t,{\vp_1}, \cdots, {\vp_{n-1}})\Big)=  
 -\alpha_\infty\bar\epsilon^{1/3} L^{4/3} \,  |t|\, \varrho_1^2 \nonumber\\
 & -d_n \log (\varrho_1 L)+ {{w_\infty}^{(m,\bar{m})}}_{\alpha_1\dots\alpha_{n}}\left(\varrho_1^{2/3}\bar\epsilon^{1/3} t,\frac{\vvvarrho_{1}}{\varrho_1}, \cdots, \frac{\vvvarrho_{n-1}}{\varrho_1}\right)
 + {\cal O}(p_{\rm max} L)\, .
 \label{eq:solWn-larget}
\end{align}
The non-universal constant $\alpha_\infty= \hat \gamma\hat J_*/4$  is again the same for any generalised correlation function, irrespective of its order or fields content.
Moreover, the same comments apply for the subleading terms in the second line of \eq{eq:solWn-larget}: they merely represent K41 scaling, with possible corrections induced by the neglected terms. The main feature of this solution is thus captured by the expression
\begin{equation}
 {\cal W}^{(m,\bar{m})}_{\alpha_1\dots\alpha_{n}}({t_1,\vp_1}, \cdots, {t_{n-1},\vp_{n-1}})\propto \exp\Big( -\alpha_\infty (L^2/\tau)  \Big|\sum_\ell \vp_\ell \Big|^2 \, |t| + {\cal O}(p_{\rm max} L) \Big)\, .
 \label{eq:solWnexp}
\end{equation}
At large time delays, the temporal decay of correlation functions is also sensitive to the large scale and thus explicitly breaks scale invariance as in the small time regime. However, the decay drastically slows down, since it crosses over for all wavenumbers from a Gaussian  at small time delays to an exponential at large ones.  Note that the dependence in the wavenumbers  is quadratic in both regime.  This large time regime had not been predicted before. However, it was already in germ in early studies of sweeping, as was noted in (\cite{Gorbunova2021}). This allows one to provide a simple physical interpretation of these two regimes, see \sref{sec:heuristic}. Let us first summarise for later comparisons the explicit forms of the two-point  correlation function of the velocity fields:
\begin{align}
 C_{\alpha\beta}^{(2)}(t,\vk) &\equiv \text{FT} \Big[ \Big\langle  \sv_\alpha(t_0, \vr_0) \sv_\beta(t_0 + t, \vr_0 + \vr) \Big\rangle\Big]  \nonumber \\  
  &= C^{(2)}_{\alpha\beta}(0,\vk)  \left\{\begin{array}{l l} \exp \big(-\alpha_0 (L/\tau)^{2} t^2 k^2 \big) \quad & t\ll \tau\\ 
  \exp \big(-\alpha_\infty (L^2/\tau) |t| k^2 \big) \quad & t\gg \tau
  \end{array} \right.\label{eq:C2-FRG}
\end{align}
and of the three-point correlation function at small time delays $t_1,t_2 \ll \tau$
\begin{align}
	C^{(3)}_{\alpha \beta \gamma}  (t_1, \vk_1, t_2, \vk_2) &\equiv    
	 \text{FT} \Big[ \Big\langle  \sv_\alpha(t_0 + t_1, \vr_0 + \vr_1) \sv_\beta(t_0 + t_2, \vr_0 + \vr_2) \sv_\gamma(t_0, \vr_0) \Big\rangle \Big]  \nonumber\\ 
	&= C^{(3)}_{\alpha \beta \gamma}  (0, \vk_1, 0, \vk_2) \exp\Big(- \alpha_0(L/\tau)^{2}  \left|\vk_1 t_1 + \vk_2 t_2 \right|^2 \Big)\, .
	\label{eq:C3-FRG}
\end{align}

\subsection{Intuitive physical interpretation}
\label{sec:heuristic}

Let us now provide an heuristic derivation of these results, proposed in (\cite{Gorbunova2021}).
The early analysis of Eulerian sweeping effects by Kraichnan (\cite{Kraichnan64}) was based on
 the Lagrangian expression for the Eulerian velocity field as
\begin{equation}
  u_i(t,\vr) = \exp_{\to}[\vxi(t|\vr,t_0)\cdot\vnabla] u_i(t_0, \vr) +\int_{t_0}^t ds\  \exp_{\to}[\vxi(t|\vr,s)\cdot\vnabla] \left[ \nu \nabla^2 u_i(s, \vr)-\nabla_ip(s,\vr)\right]\, . 
 \label{eq:convecteq}
 \end{equation}
 In this equation, $\vxi(t|\vr,t_0)=\vX(t|\vr,t_0)-\vr$ is the Lagrangian displacement vector, where $\vX(t|\vr,t_0)$ denotes the position at time $t$ of the Lagrangian fluid particle located at position $\vr$ at time $t_0$, which satisfies 
 \begin{equation}
 \frac{d}{ds}\vX(t|\vr,s)=\vu(s,\vX(t|\vr,s)), \quad \vX(t|\vr,t)=\vr\, .
 \end{equation}
 One can write using \eq{eq:convecteq} the expression of the two-point correlation function. 
 This expression can be simplified under some general assumptions, namely slow variations in space and time of the displacement field, and  its statistical independence from the initial velocities at sufficiently long time $t$. Moreover, since the displacements are dominated by the large scales of the flow, whose statistics are nearly Gaussian (\cite{Corrsin59,Corrsin62}), one may plausibly further assume a normal distribution for $\xi$, resulting in the following expression
\begin{equation}
C^{(2)}(t,\vk) = \exp\left[-\frac{1}{2}\langle | \vxi(t|\vr,0)|^2\rangle k^2 \right] C^{(2)}(0,\vk)  \Big\{1+{\cal O}(\nu k^2|t|,k U_{rms}(k)|t|)\Big\}.
 \label{eq:Kr2pcorr} 
  \end{equation}
According to these arguments, the two-point velocity correlation undergoes a rapid decay in the time-difference $t$ which arises from an average over rapid oscillations in the phases of Fourier modes due to sweeping, or ``convective dephasing'' (\cite{Kraichnan64}).
 
The connection with the \FRG results \eq{eq:C2-FRG} stems from exploiting 
 the results of the classical study by Taylor (\cite{Taylor22}) on one-particle turbulent 
 dispersion. In this study, the Lagrangian displacement field was shown to exhibit two regimes   
\begin{equation}
\langle |\vxi(t|\vr,0)|^2\rangle \sim \left\{ \begin{array}{ll} U_{\rm rms}^2  t^2 &  |t|\ll \tau\cr
2D|t| &   |t|\gg \tau 
\end{array} \right. 
\label{eq:xivar} \end{equation}
where the early-time regime corresponds to ballistic motion with the rms velocity $U_{\rm  rms}$ and 
the long-time regime corresponds to diffusion with a turbulent diffusivity $D\propto U_{\rm rms}^2\tau.$
 Inserting these results in \eq{eq:Kr2pcorr} then yields the \FRG expressions \eq{eq:C2-FRG}. This argument can be generalised in principle to multi-point correlation functions, which, under the same kind of assumptions, would also lead to similar expressions as the ones derived from \FRG.
 Of course, the \FRG derivation is far more systematic and rigorous than these heuristic arguments, but the latters provide an intuitive physical interpretation of these results.

The \FRG results have been tested in extensive \DNS, focused on the temporal dependence of two- and three-point correlation functions. We report in \sref{sec:DNS} and \sref{sec:scalar} the main results.

\section{Comparison with Direct Numerical Simulations}
\label{sec:DNS}

Extensive \DNS were performed on high-performance computational clusters to quantitatively test the \FRG results. The \DNS consist of direct numerical integration of the \NS equation with a stochastic forcing using standard pseudo-spectral methods and Runge-Kutta scheme for time advancement, with 
 typical spatial resolution conforming the standard criterion $k_{max}\eta\simeq 1.5$, and Taylor based Reynolds numbers $R_\lambda$ from 40 to 250 for corresponding grid sizes from $64^3$ to $1024^3$.  Details can be found, \eg in (\cite{Gorbunova2021thesis}). The analysis focused on the two-point and three-point correlation function of the velocity field, which were computed in the stationary state, using averages over spherical shells in wavenumber and over successive time windows, typically
 \begin{equation}
 C^{(2)}(t,\vk) = \frac{1}{N_t}\sum_{j=1}^{N_t} \frac{1}{M_n}\sum_{\vk\in  S_n}\Re\big[u_i(t_{0j},\vk)u_i^*(t_{0j}+t,\vk)\big]\, ,
 \label{eq:C2-DNS}
\end{equation}
where $N_t$ is the number of time windows in the simulation, $M_n$ is the number of modes in the spectral spherical shell $S_n$, and $k=n\Delta k$, $n\in \mathbb{Z}$.

\subsection{Small time regime and sweeping}
\label{sec:sol-small-t}

\begin{figure}
	\includegraphics[width=0.48\linewidth]{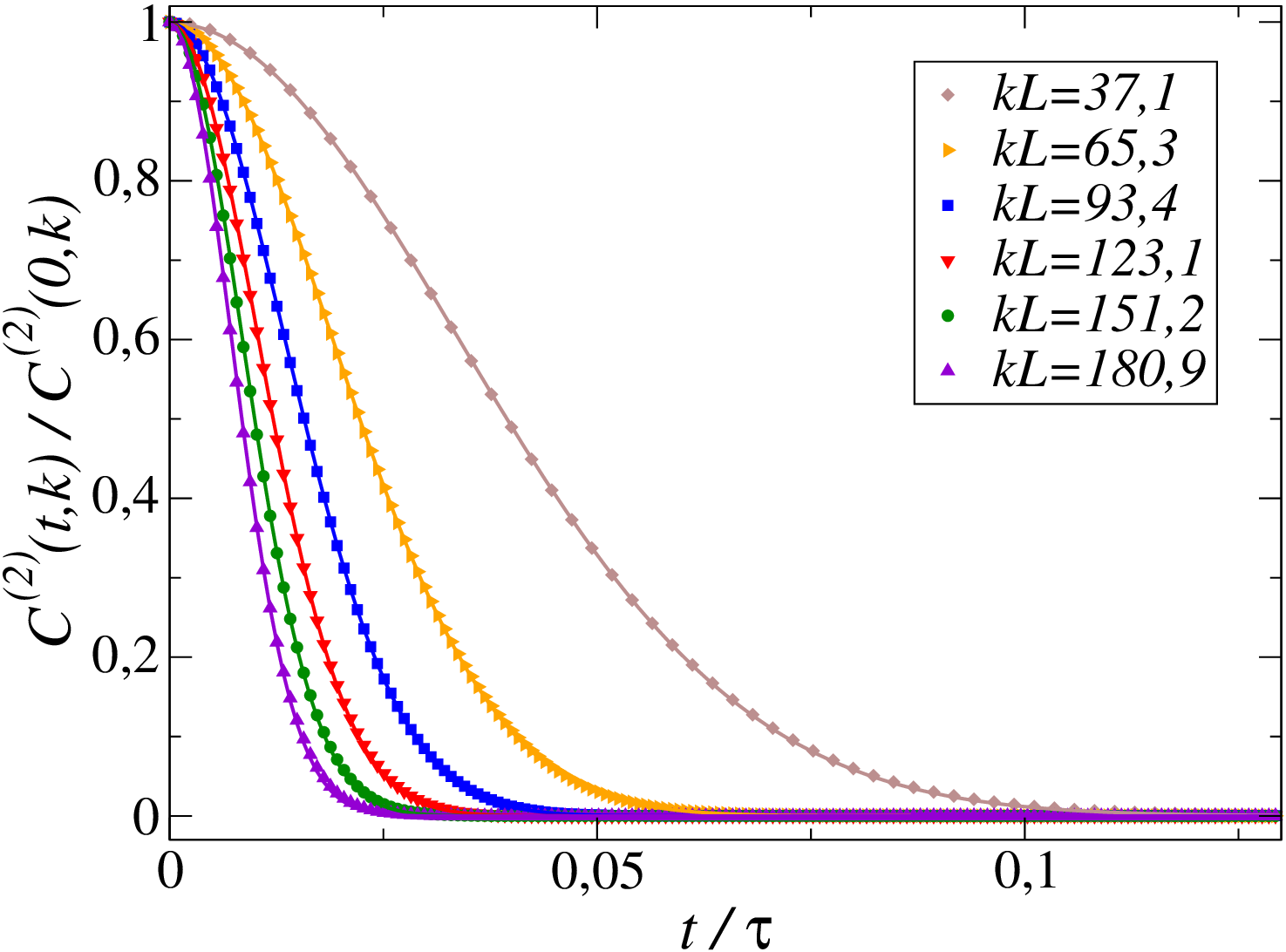}\hspace{0.5cm}
	\includegraphics[width=0.49\linewidth]{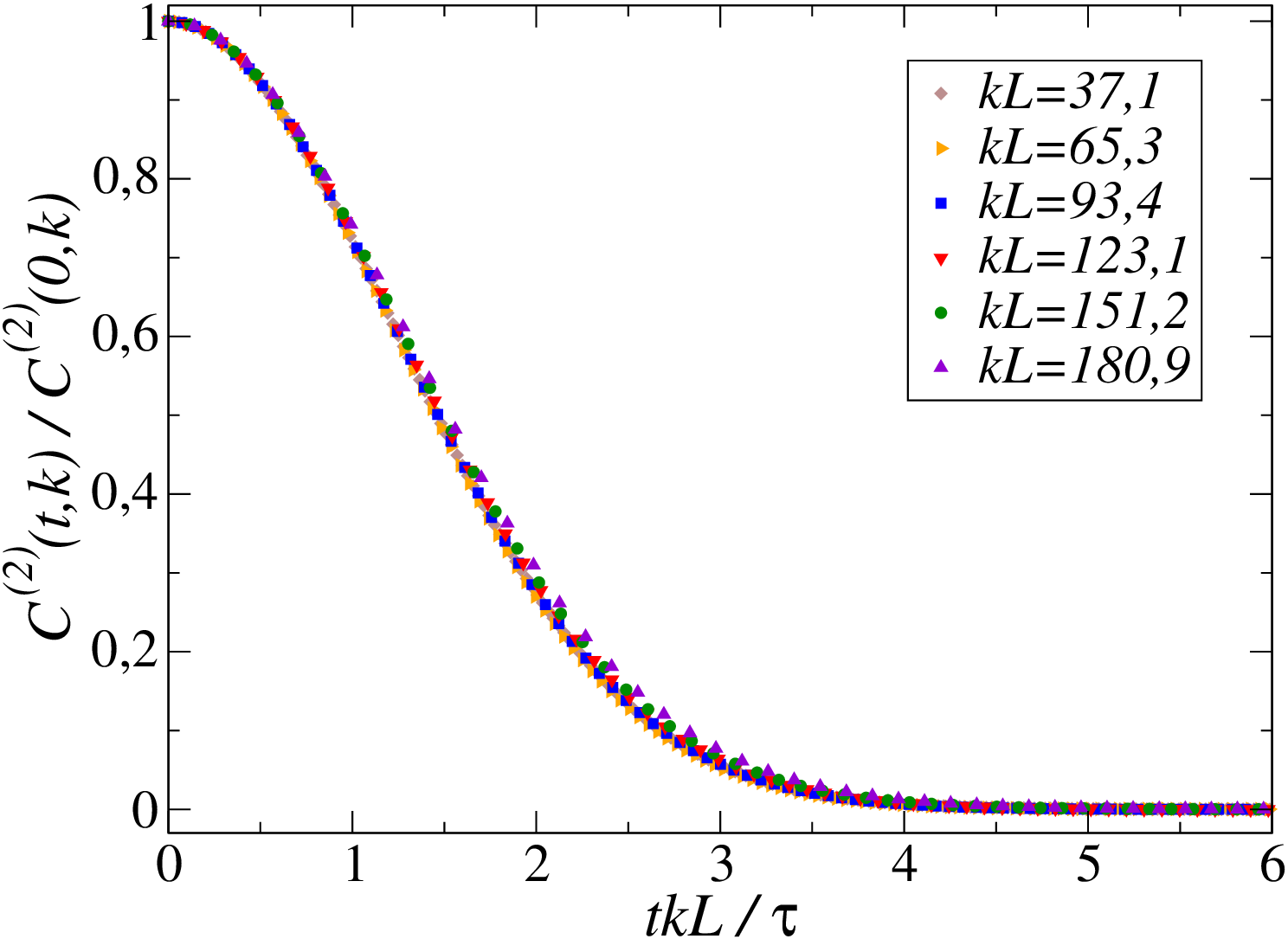}
	\caption{
{\it Left panel}: Time dependence of the normalised two-point correlation function \(C^{(2)} (t,k)\) 
	at different wavenumbers \(k\) for a simulation at \(R_\lambda = 90\). Data from the simulation are represented with symbols, and their Gaussian fits with plain lines. {\it Right panel:} Same data plotted as a function of $kt$, which leads to their collapse as expected from the \FRG result \eq{eq:C2-FRG}. }
	\label{fig:C2-smallt}
\end{figure}

We first present the result for the two-point correlation function at small time delays,
 whose theoretical expression is given by \Eq{eq:C2-FRG}. The time dependence of \(C^{(2)} (t,k)\) is displayed in the left panel of \fref{fig:C2-smallt} for a sample of different wavenumbers $k$. The larger ones exhibit a faster decorrelation as expected. For each $k$, the normalised curve  \(C^{(2)} (t,k)/C^{(2)} (0,k)\) is fitted with a two-parameter Gaussian function $f_{\rm Gaus}(t) = c\exp(- (t/\tau_0)^2)$, which provides a very accurate model for all curves, as illustrated in the figure. Moreover, all curves collapse into a single Gaussian function when plotted as a function of $kt$, as expected from the theory \eq{eq:C2-FRG}. The fine quality of the collapse is shown in the right panel of \fref{fig:C2-smallt}. 
 
\begin{figure}
\includegraphics[width=0.48\linewidth]{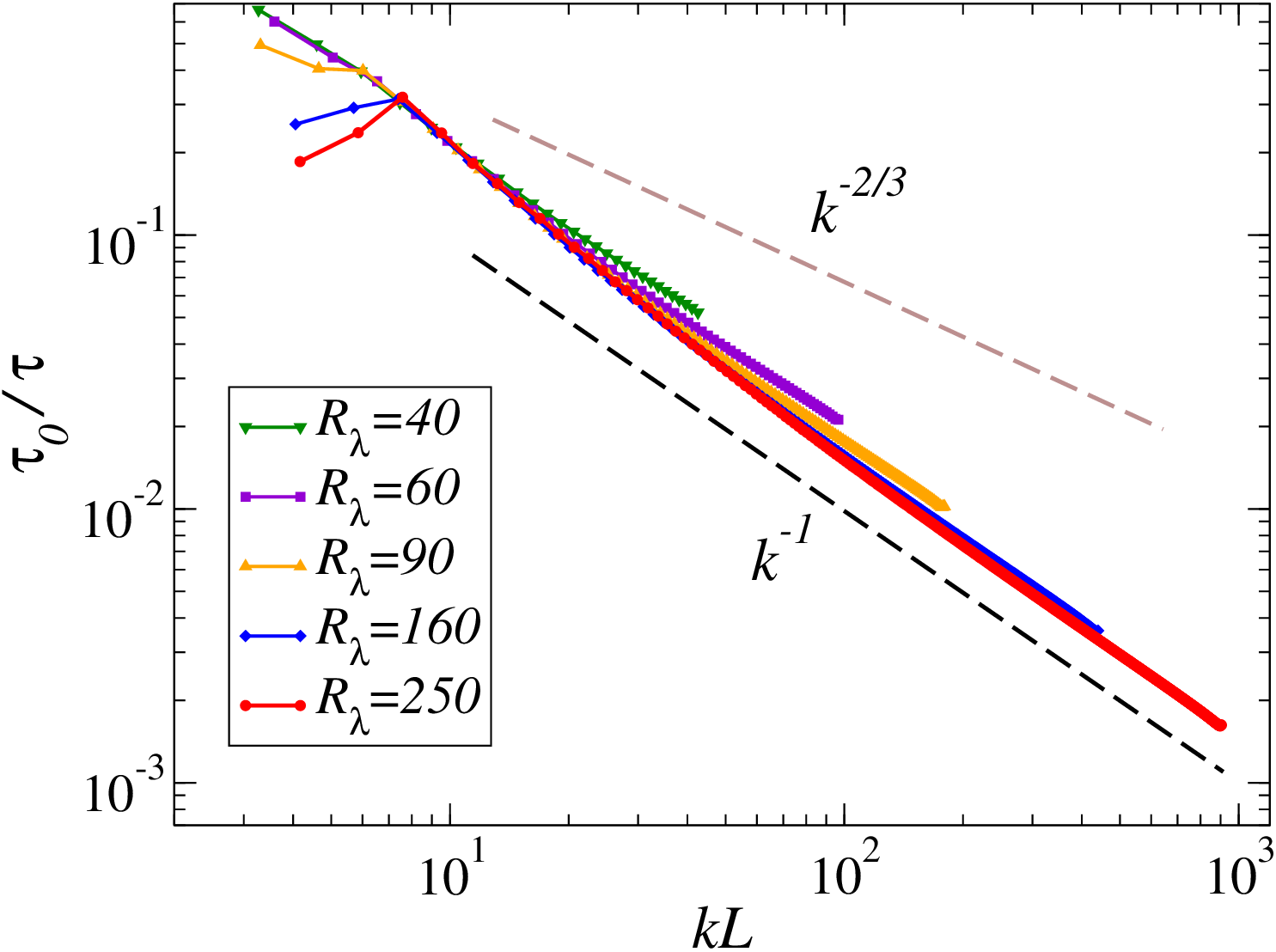}\hspace{0.5cm}
\includegraphics[width=0.49\linewidth]{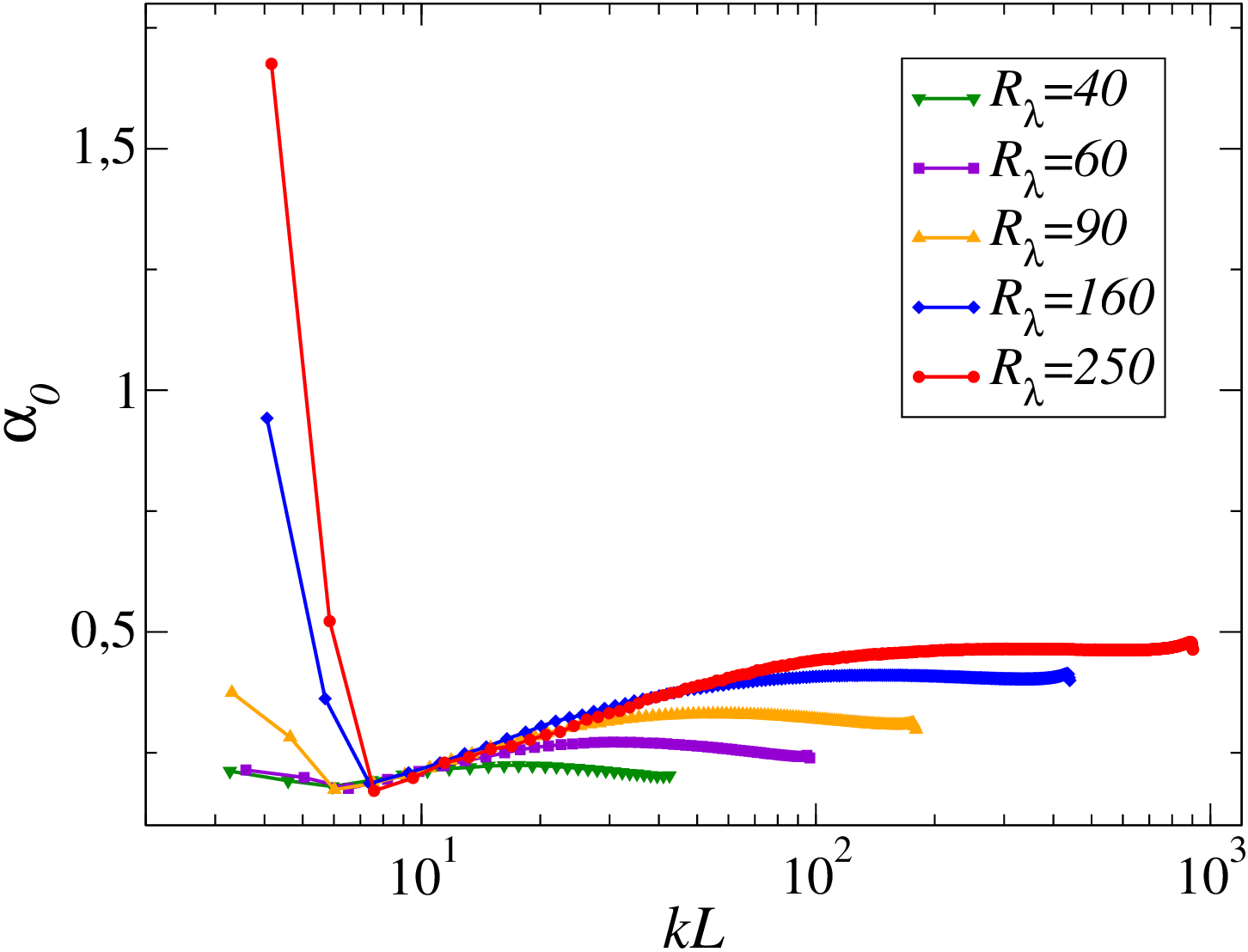}
	\caption{{\it Left panel}: Decorrelation time $\tau_0$ extracted from the Gaussian fit as a function of the wavenumber for various $R_\lambda$. {\it Right panel}: Non-universal constant $\alpha_0$ obtained as $\alpha_0 = \tau^2/(\tau_0 kL)^2$. It reaches a plateau at large wavenumber, which extends with increasing $R_\lambda$.}
	\label{fig:tauS}
\end{figure}

 The fitting parameter $\tau_0$ is plotted as a function of $kL$ in the left panel of \fref{fig:tauS}, which clearly confirms that it is proportional to $k^{-1}$ at large enough $k$, and not to the K41 time scale $k^{-2/3}$, in plain agreement with the \FRG result. Similar results were also obtained in \crefs ~(\cite{Sanada92, Favier2010}), where the characteristic decorrelation time is estimated by integrating the correlation function, as well as in the work of (\cite{Kaneda99}), where the characteristic time is measured through the second derivative of the correlation function. 

The non-universal constant $\alpha_0$ can be deduced from $\tau_0$ as $\alpha_0 = \tau^2/(\tau_0 kL)^2$. As can be observed in the right panel of  \fref{fig:tauS}, this quantity reaches a plateau, which value corresponds to the theoretical $\alpha_0$ at large $k$.  The physical origin of the departure from the plateau at intermediate and small wavenumbers can be clearly identified in the simulations as a ``contamination'' from the forcing (\cite{Gorbunova2021}). Indeed, by analyzing the modal energy transfers, one observes that the plateau is reached when the energy transfer to a mode $k$ is local, \ie mediated by neighbouring modes only through the cascade process. The modes at intermediate and small $k$ also receive energy via direct non-local transfers from the forcing range in the simulations. This means that the large wavenumber limit underlying the \FRG derivation corresponds to the modes $k$ in the inertial range with negligible direct energy transfer from the forcing range. Of course, the ``large wavenumber'' domain extends with the Reynolds number, as visible in \fref{fig:tauS}. 
One can also observe that the value of the plateau in the simulations depends on the $R_\lambda$.
 Although only few points are available, they suggest a saturation at high $R_\lambda$ beyond the values accessible in the \DNS. Moreover, although the forcing location and its width are fixed for the different $R_\lambda$, its amplitude varies. The theoretical expression for  $\alpha_0=\hat\gamma \hat I_*/2$ does depend in the precise forcing profile, which could also explain the $R_\lambda$ dependence. This point was not quantitatively investigated.

\subsection{Three-point correlation function}

Let us now present the results for the three-point correlation function. The general expression of this correlation \eq{eq:C3-FRG} involves a product of velocities which is not local in $\vk$. This is unpracticable in \DNS heavily relying on parallel computation and memory distribution. To overcome this difficulty, one can focus on a particular configuration, which corresponds to an advection-velocity correlation.
This correlation is defined as
 \begin{eqnarray}
	&{T}(t, \vk) \equiv \Big\langle {N}_\ell(t_0 + t, \vk) {u}^*_\ell(t_0, \vk) \Big\rangle \label{eq:defT}\\
	&{N}_\ell(t, \vk) = -i k_n P^\perp_{\ell m}(\vk) \sum_{\vk^\prime} \Big\langle {u}_m (t, \vk^\prime) {u}_n(t, \vk -\vk^\prime) \Big\rangle \nonumber\, ,
\end{eqnarray}
where $N$ is simply the Fourier transform of the advection and pressure term appearing in the \NS equation expressed in spectral space
\begin{equation}
	\partial_t {u}_\ell (t, \vk) = {N}_\ell (t, \vk) - \nu k^2  {u}_\ell (t, \vk) + {f}_\ell  (t, \vk)\, .
	\label{eq:ns_spec}
\end{equation}
At equal times, $T(0,\vk)$ is thus  the usual energy transfer, and \eq{eq:defT} represents its time-dependent generalisation. Let us note that $T(t,\vk)$  is still a statistical moment of order three, although evaluated in a specific configuration involving only two space-time points. It is proportional to a linear combination of three-point correlation function $C^{(3)}$ as
\begin{equation}
    {T}(t, \vk) \sim  \sum_{\vk^\prime} C^{(3)}_{m n \ell} (t, \vk^\prime, t, \vk - \vk^\prime)  \stackrel{{\rm all} \;|\vk|\gg L^{-1}}{\sim}  \exp\left(- \alpha_0 (L/\tau)^{2} k^2 t^2 \right)
    \label{eq:T-C3}
\end{equation}
and according to the \FRG result, it should behave in the limit where all wavenumbers $|\vk|$, $|\vk'|$ and $|\vk-\vk'|$ are large compared to $L^{-1}$,  as the indicated Gaussian in $kt$. 

However, the sum in \eq{eq:T-C3} involves all possible wavenumber $\vk'$, and not only large ones. Hence, in order to compare it with the \FRG prediction, one needs to perform a scale decomposition, as defined \eg in Refs. (\cite{Frisch95,
Verma2019}). Each velocity field $\vu(\vk)$ can be decomposed into a small-scale component $\vu^S(\vk)$ for $|\vk|> K_c$  and a large-scale one $\vu^L(\vk)$ for $|\vk|\leq K_c$ as $\vu(\vk) = \vu^S(\vk)+\vu^L(\vk)$, where $K_c$ is a  cutoff wavenumber. This cutoff is chosen  such that the direct energy transfers from the forcing range to the modes $|\vk|>K_c$ are negligible, which coincides with the regime of validity of the large wavenumber limit of the \FRG identified on the basis of the two-point correlation function. 
This scale decomposition thus leads to terms of the form  ${T}^{XYZ} (t,\vk) = -{[{u}_i^X]^*}(\vk,t_0) \ \text{FT}[u_j^Y \partial_j u_i^Z](\vk,t_0 + t)$ where $X,Y,Z$ stand for $S \ \text{or}\ L$. Using the terminology of \crefs ~(\cite{Verma2019}) for instantaneous energy transfers (\ie for $t=0$), the first
superscript of ${T}^{XYZ}$ is related to the mode receiving energy in
a triadic interaction process, 
 the intermediate
superscript denotes the mediator mode and the last superscript is related to
the giver mode that sends the energy to the receiver mode. The mediator mode
does not loose nor receive energy in the interaction, it corresponds to the advecting
velocity field, which comes as prefactor of the operator $\nabla$ in the non-linear
term of the Navier-Stokes equation.

We are interested in the large wavenumbers $\vk$, which corresponds to short scales in the previous terminology. Thus, fixing $X=S$, the possible contributions  for $T$ are given by the following expression:
\begin{equation}
	{T}(t,\vk) = \left[{T}^{SSS} + {T}^{SLS} + {T}^{SSL} + {T}^{SLL}\right](t,\vk) \,.
	\label{eq:tdecompos}
\end{equation}
We focus on the term ${T}^{SSS}$, which gathers all triadic interactions where the three modes belong to the small scales (all wavenumbers are large). This term constitutes with $T^{SLS}$ the  turbulent energy cascade.

\begin{figure}
	\includegraphics[width=0.48\linewidth]{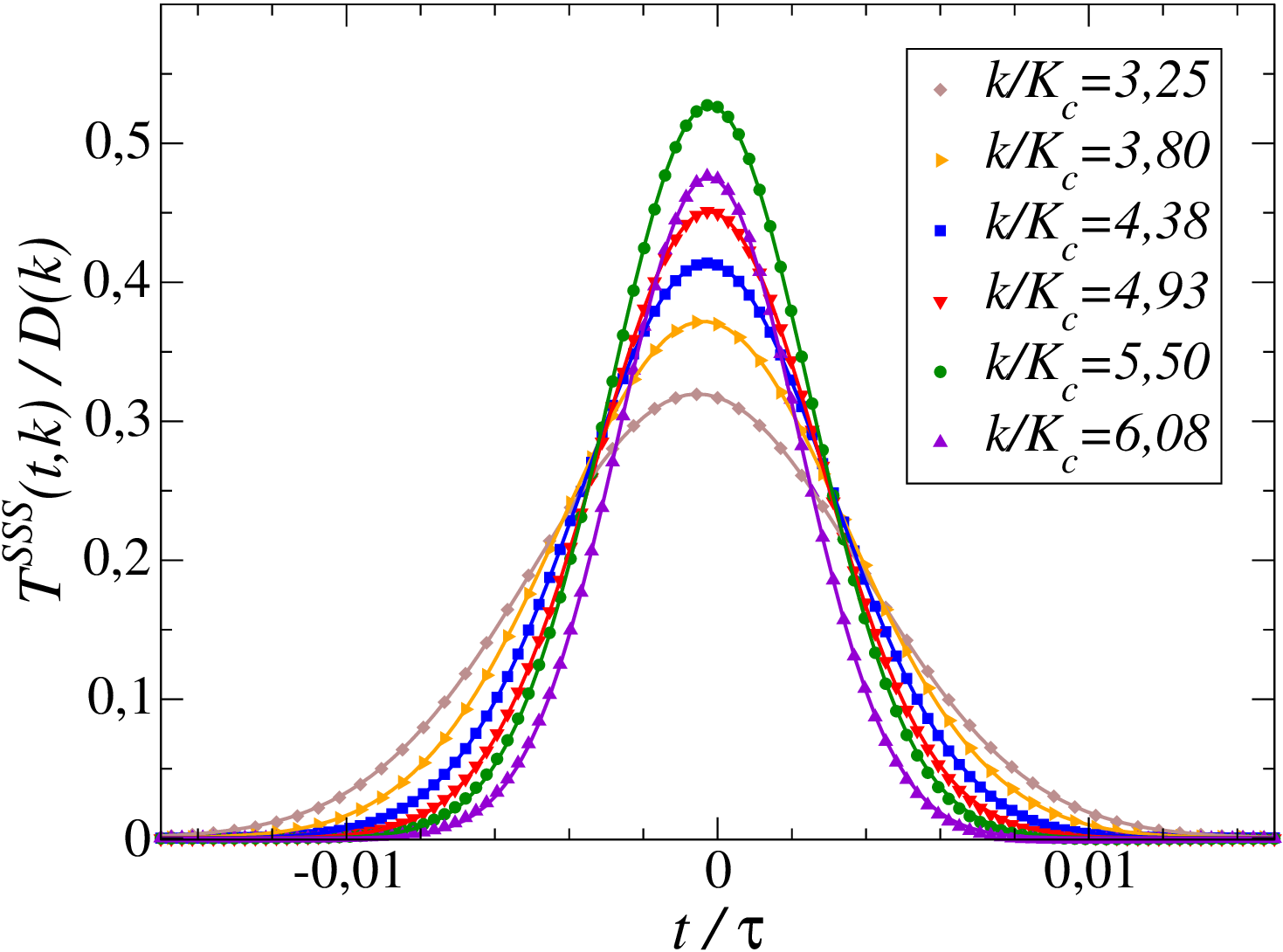}\hspace{0.5cm}
	\includegraphics[width=0.49\linewidth]{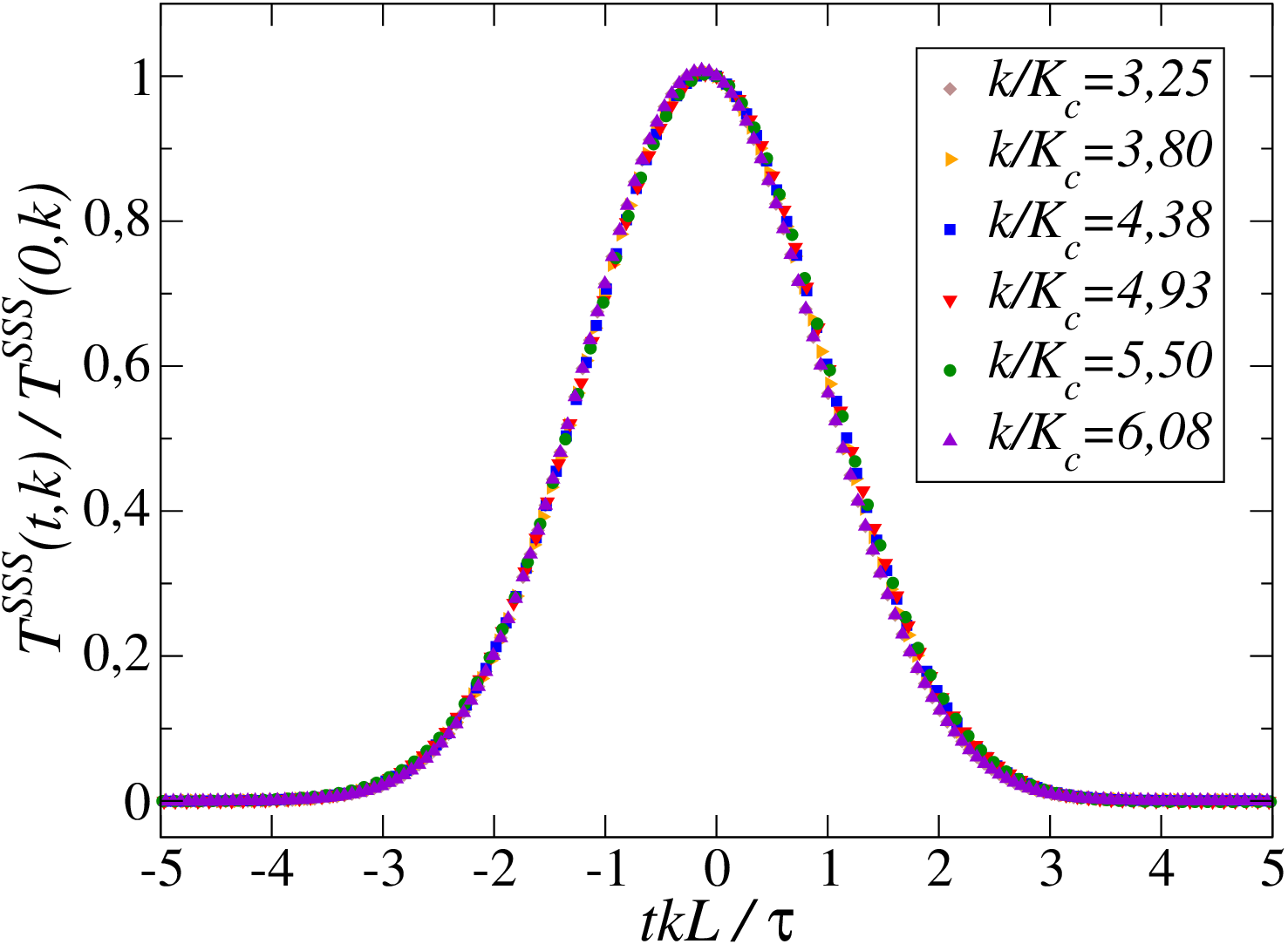}
	\caption{{\it Left panel:} Advection-velocity spatio-temporal correlation function \({T}^{SSS}(t,{k})\) from \DNS at $R_\lambda =90$, normalised by the dissipation spectrum \(\hat{D}(k) = \nu k^2 C^{(2)}(0,k)\). Data from the simulation are represented with symbols, and their Gaussian fits with plain lines. {\it Right panel:} same curves \({T}^{SSS}(t,{k})\) normalised by $T^{SSS}(0,\vk)$ and plotted as a function of $k t$, which induces their collapse.}
	\label{fig:C3}
\end{figure}

We show in \fref{fig:C3} the results for $T^{SSS}(t,\vk)$ computed from \DNS. 
 All the curves for $T(t,\vk)$ and $T^{SSS}(t,\vk)$ for the different wavenumbers $k$ are fitted with the function $g_{\rm Gaus}(t)=c(1-t/\tau_l)\exp(-(t/\tau_0)^2)$, where $c$, $\tau_l$ and $\tau_0$ are the fit parameters. This function provides a very accurate modellisation of the data both for $T$ and $T^{SSS}$ for all $k$. However, only the curves $T^{SSS}$ are well approximated by pure Gaussians ($\tau_l\ll \tau_0$), whereas without scale decomposition the curves for $T$ are in general not symmetric and can take negative values (see (\cite{Gorbunova2021}) for details). 
 
\begin{figure}
 \includegraphics[width=0.48\linewidth]{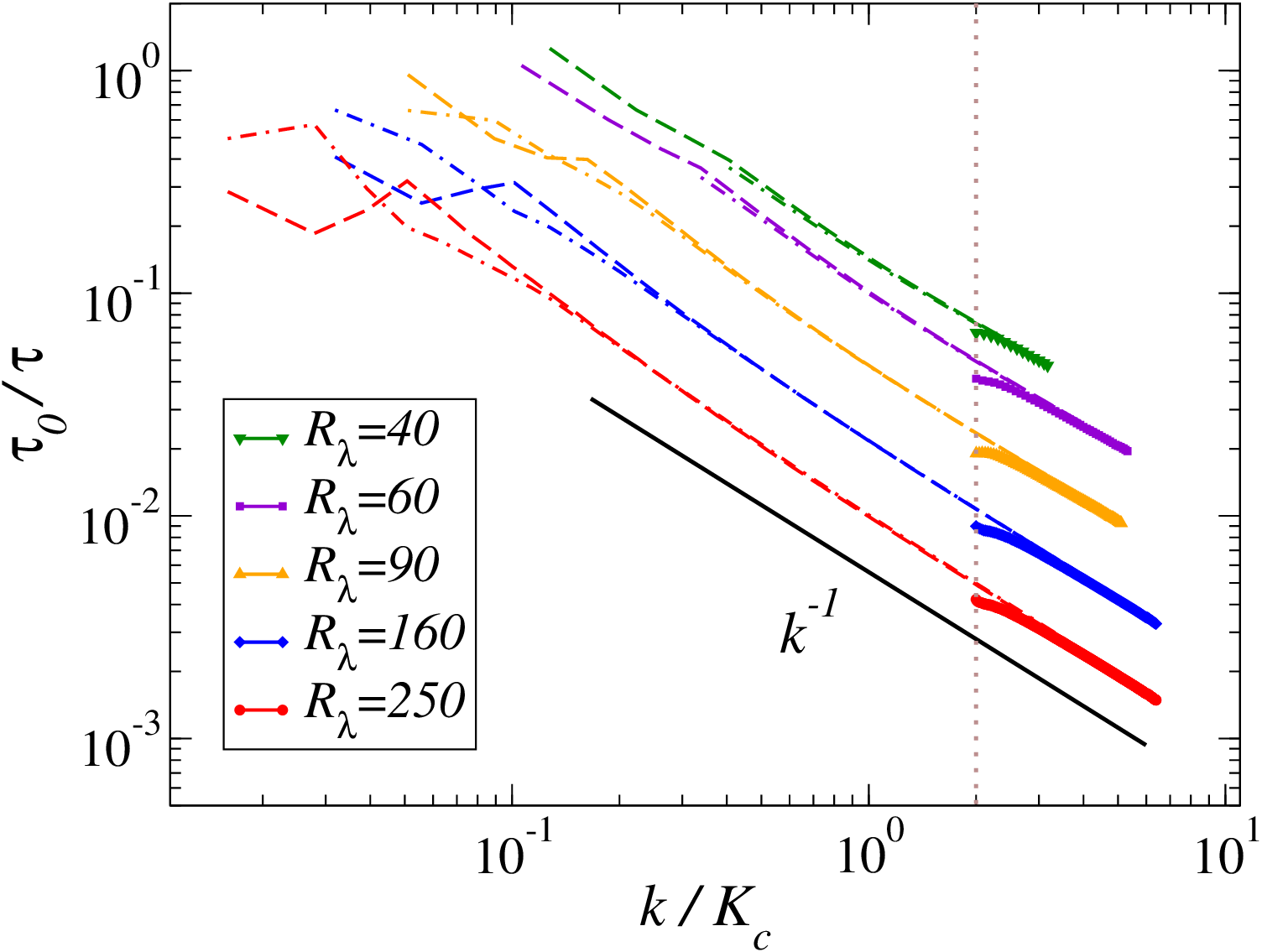}\hspace{0.5cm}
	\includegraphics[width=0.49\linewidth]{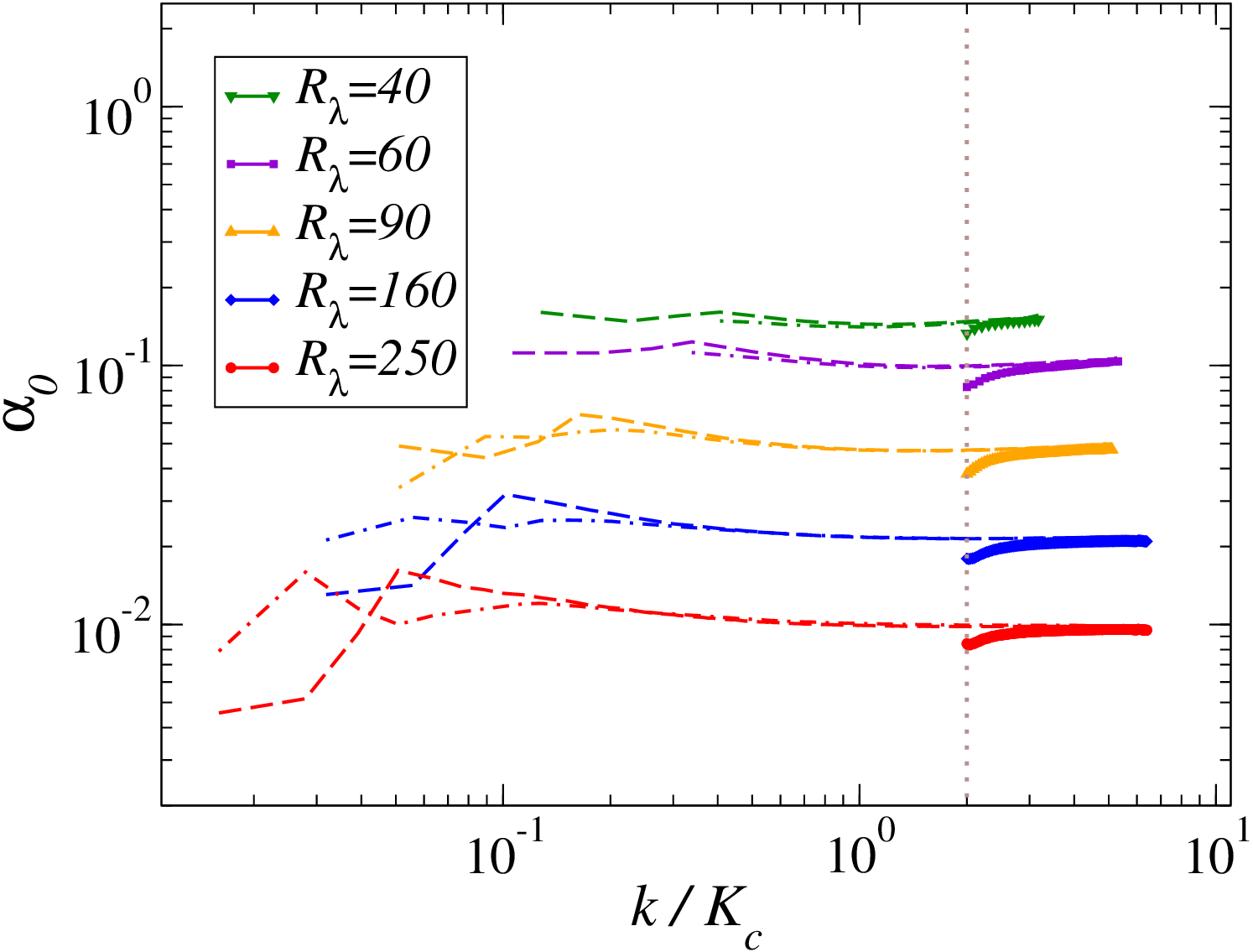}
	\caption{{\it Left panel}: Decorrelation time $\tau_0$ extracted from the Gaussian fits  of the purely small-scale advection-velocity correlation $T^{SSS}$ (plain lines), the two-point correlation function $C^{(2)}$ (dashed lines) and the full advection-velocity correlation $T$ (dashed-dotted lines).  {\it Right panel}: non-universal parameter $\alpha_0$ obtained as $\alpha_0 = \tau^2/(\tau_0 kL)^2$. The data for the different R$_\lambda$ in both panels have been shifted vertically for clarity.}
	\label{fig:taua}
\end{figure}

 The Gaussian form of $T^{SSS}$ is in agreement with the \FRG result \eq{eq:T-C3}. As shown in  \fref{fig:C3}, all the curves for different wavenumbers collapse onto a single Gaussian when plotted as a function of the variable $kt$, as expected from \Eq{eq:T-C3}. Moreover, let us emphasise  that the non-universal parameter $\alpha_0$ in this equation is predicted to be the same as the one for the two-point correlation function in \Eq{eq:C2-FRG}. The decorrelation time  $\tau_0$ extracted from the fit with $g_{\rm Gaus}$ is displayed in \fref{fig:taua},  together with the value obtained from $C^{(2)}$ and also the one from the full correlation $T$. They show as expected a $k^{-1}$ dependence. The coefficient $\alpha_0$ can be obtained as $\alpha_0 = \tau^2/(\tau_0 kL)^2$, and is shown in \fref{fig:taua}. At sufficiently large wavenumbers, corresponding to the regime of validity of \FRG, the three values coincide with very good precision. At smaller wavenumbers, they are also in good agreement for $C^{(2)}$ and for $T$, although in this regime $\tau_l$ (parameter of the linear part of the fitting function $g_{\rm Gaus}$) can be large.    
  The detailed analysis of these results can be found in \crefs ~(\cite{Gorbunova2021}).
  
 To summarise this part, the results for the spatio-temporal dependence of the two-point and the specific configuration of the three-point correlation function studied in \DNS hence confirm with accuracy the \FRG prediction in the small-time regime, including the equality of the non-universal prefactors in the exponentials. This regime corresponds to a fast decorrelation on a time scale $\tau\sim k^{-1}$ related to sweeping. Because of this fast initial decay, the observation in the \DNS of the crossover to another slower regime at large-time  is very challenging for \NS turbulence, as we now briefly  discuss.

\subsection{Large time regime}
\label{sec:sol-large-t}

The initial Gaussian decay of the two-point correlation function leads to a fast decrease of the amplitude of the signal within a short time interval. In all the simulations of \NS equation, the correlation function falls to a level comparable with the numerical noise while it is still in the Gaussian regime, which prevents from detecting the crossover to an exponential and resolving the large time regime.  
 
\begin{figure}
	\begin{center}
	\includegraphics[width=0.5\linewidth]{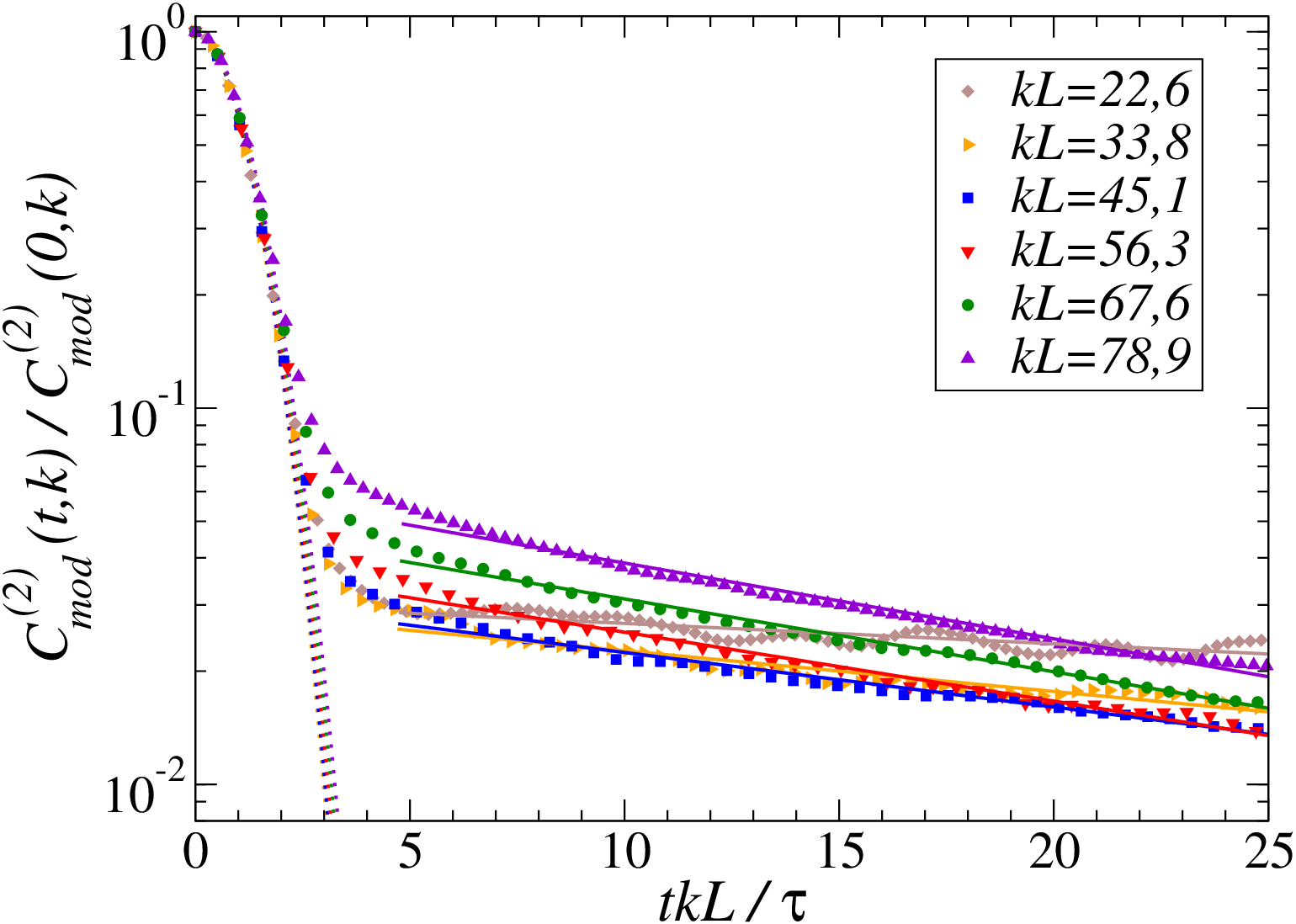}
	\end{center}
	\caption{
Time dependence of the normalised two-point correlation function of the velocity modulus \({C}_{mod}^{(2)} (t,k)\) at $R_\lambda = 60$ for different wavenumbers \(k\). 
The numerical data is represented with dots,  the dashed lines correspond to Gaussian fits  and the plain lines to exponential fits.}
	\label{fig:C2-larget}
\end{figure}

 Interestingly, although this crossover could not be accessed for the real part of the correlation function, it was observed for a different quantity, namely the correlation function of the modulus of the velocity, defined as 
 \begin{equation}
  {C}_{mod}^{(2)} (t,k) = \big \langle \parallel\hat{\vu}(t_0,\vk)\parallel \parallel \hat{\vu}(t_0+t,\vk)\parallel \big\rangle - \big \langle \parallel\hat{\vu}(t_0,\vk)\parallel \big\rangle\big \langle\parallel \hat{\vu}(t_0+t,\vk)\parallel  \big\rangle\, .
 \end{equation}
The result obtained in \DNS for this quantity is displayed in \fref{fig:C2-larget}, which clearly shows the existence of a Gaussian regime at short time followed by  an exponential regime at large time, analogous to the one expected for the real part of the two-point correlation function. Remarkably, very comparable results were obtained in air jet experiments where a similar correlation is measured (\cite{Poulain2006}).
Of course, ${C}_{mod}^{(2)}$ is a quite different object from $C^{(2)}$, and there is no theoretical understanding so far of its behaviour, neither from \FRG nor from a heuristic argument. Indeed, in the latter argument, the decorrelation stems for a rapid random dephasing of the Fourier modes, while the phases obviously cannot play a role for the decorrelation of the modulus. We mention it here as an interesting puzzle, which deserves further work to be explained.

Although the large time regime appears difficult to access in \DNS of \NS equation, it can be unraveled in the case of scalar turbulence, which we present in \sref{sec:scalar}.
Prior to this, let us briefly mention another result which concerns the kinetic energy spectrum in the near-dissipation range.

\subsection{Kinetic energy spectrum in the near-dissipation range}
\label{sec:dissipative}

The expression for the two-point correlation function of the velocity obtained from \FRG can also be used to study the kinetic energy spectrum beyond the inertial range, that is beyond the Kolmogorov scale $\eta = \nu^{3/4}\bar\epsilon^{-1/3}$. Based on Kolmogorov hypotheses and dimensional analysis, the kinetic energy spectrum is expected to endow the universal form
\begin{equation}
	E(k) = \bar\epsilon^{2/3} k^{-5/3} F(\eta k)\, 
\end{equation}		
where $F(x) \to C_K$ for $x\lesssim 1$ since viscous effects are negligible in the inertial range, and $F(x)$ fastly decays for $x\gtrsim 1$ in the dissipation range.
Although $F$ is expected to be universal, its analytical expression is not known. Many empirical expressions of the form $F(x)\sim x^{-\beta} \exp(-\mu x^\gamma)$	with different values for $\gamma$ ranging from 1/2 to 2 were proposed (\cite{Monin2007}).  We refer to \eg (\cite{Khurshid2018, Gorbunova2020, Buaria2020}) for recent overviews on the various predictions. Despite the absence of consensus on the precise form of $F(x)$,  the following general features emerge. There exist two successive ranges: 
 \begin{itemize}
 \item the near-dissipation range for $0.2\lesssim k\eta \lesssim4$ where the logarithmic derivative of the spectrum is not linear and its curvature clearly indicates $\gamma < 1$,
 \item the far-dissipation range for $k\eta \gtrsim 4$ where the spectrum is well described by a pure exponential decay, which corresponds to $\gamma=1$,
 \end{itemize} 
 irrespective of the value for $\beta$.
These observations hold for the \NS equation in the absence of thermal noise. Of course, at large wavenumbers, the thermal fluctuations become non-negligible and  drastically affect the shape of the spectrum, leading to the equilibrium  $k^2$ spectrum reflecting equipartition of energy (\cite{Bandak2021,Bell2022,McMullen2022}). This crossover was shown to occur well above the mean-free path, at scales within the dissipation range. Thus, depending on the system actual scales for thermal and turbulent fluctuations, the pure exponential decay associated with far-dissipation range may be unobservable in real turbulence and superseded by the thermal spectrum.

\begin{figure}
	\includegraphics[width=0.48\linewidth]{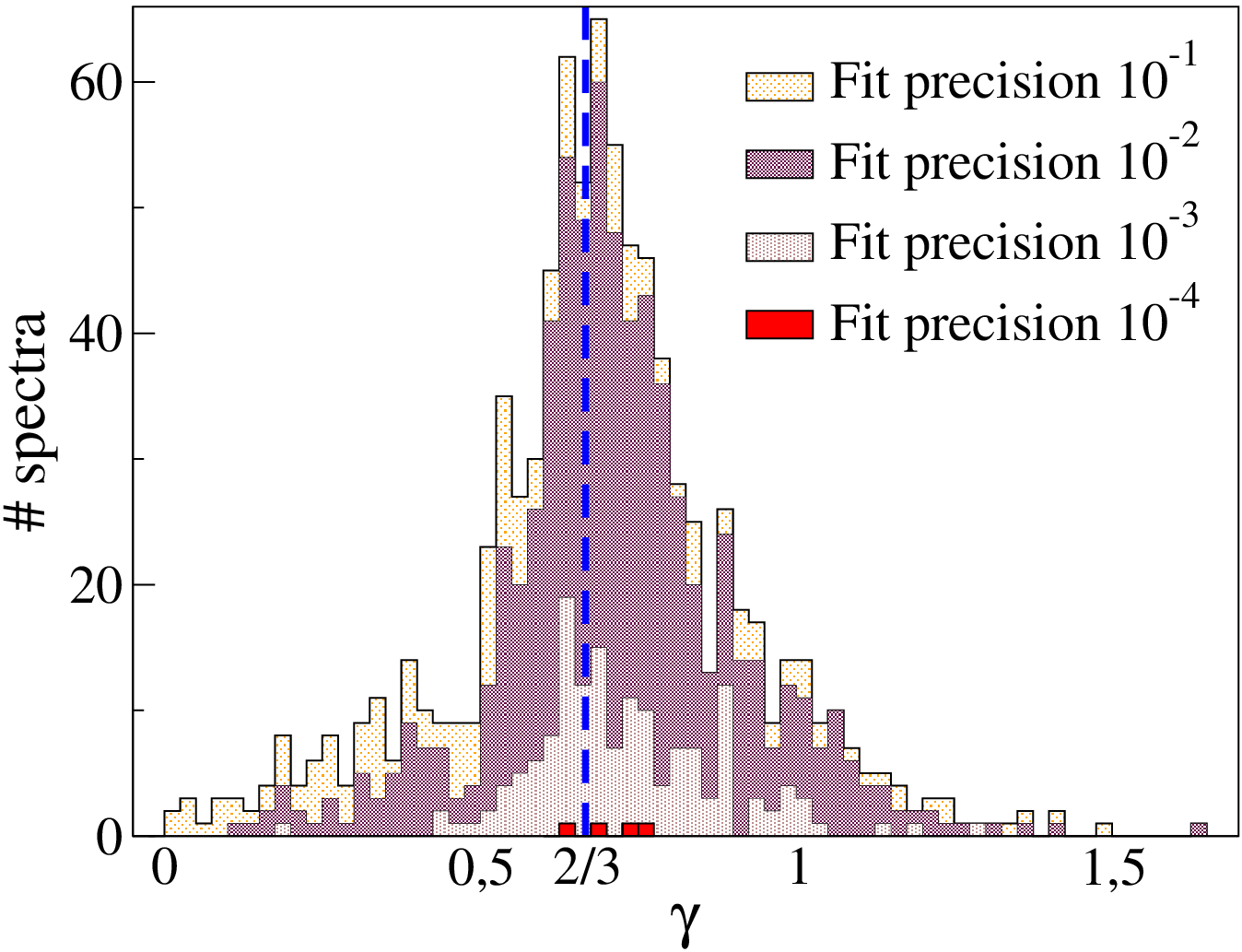}\hspace{0.5cm}
	\includegraphics[width=0.49\linewidth]{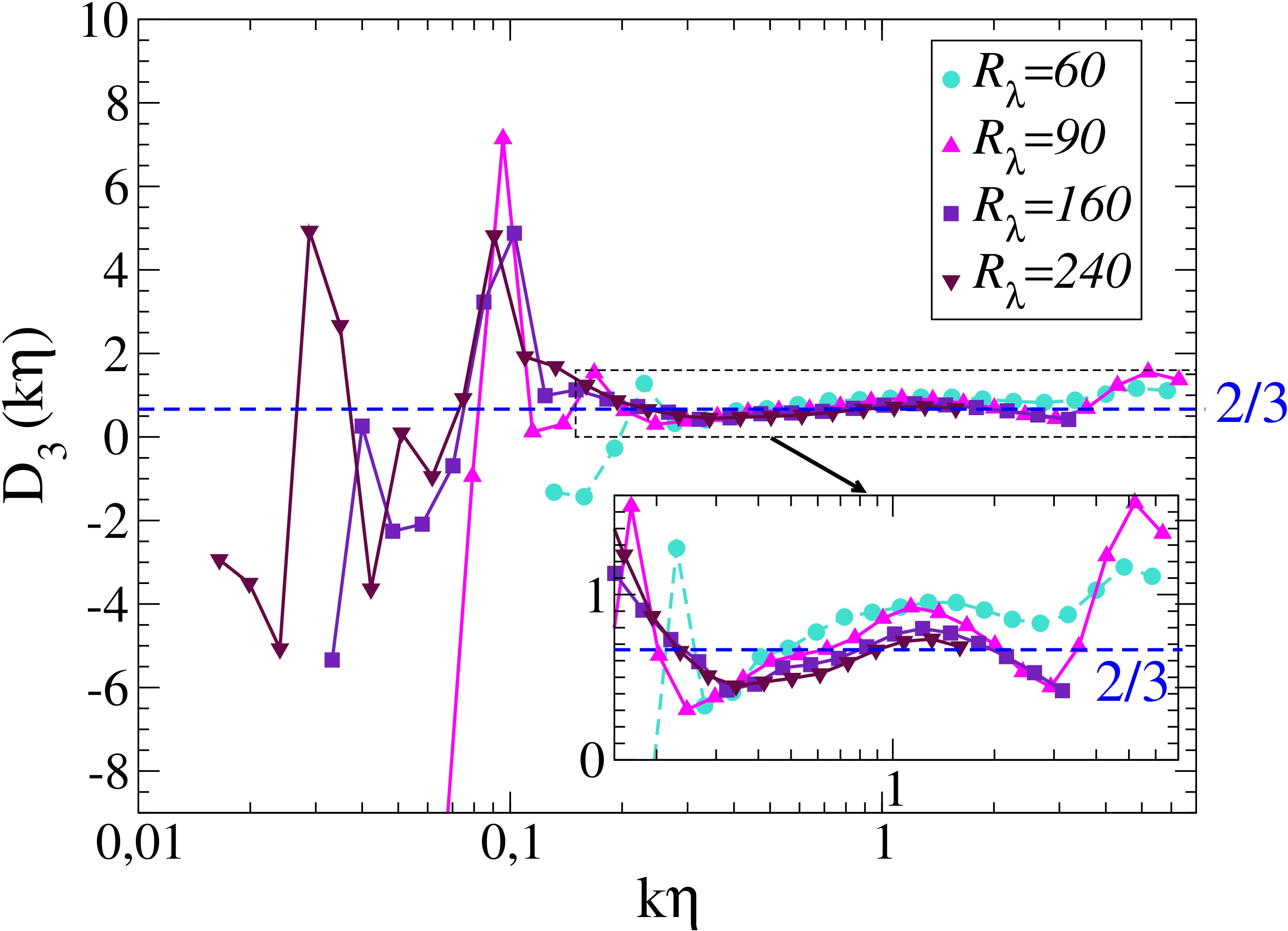}
	\caption{{\it Left panel:} distribution  of the exponent $\gamma$ of the stretched exponential for over 1600 spectra collected in experimental data from Modane wind tunnel, with different precision criteria for the fit of the spectrum in the dissipation range (see (\cite{Gorbunova2020}) for details). {\it Right panel:} Third logarithmic derivative $D_3$ of $E(k)$  as a function of the wavenumber from \DNS at different Taylor Reynolds number. The spectrum in the near-dissipation range is magnified in the inset. The local exponent approaches the blue dashed line, which indicates the \FRG prediction $\gamma=2/3$, as $R_\lambda$ increases.}
	\label{fig:dissipative}
\end{figure}

Leaving aside this issue, the form of the spectrum just beyond the inertial range can be deduced from \FRG by taking the appropriate $t\to 0$ limit in the expression for the two-point correlation function at large wavenumbers.  If one assumes that the scaling variable $t k^{2/3}$ saturates when $t$ approaches the Kolmogorov  time-scale $\tau_K=\sqrt{\nu/\bar\epsilon}$ and $k$ reaches $L^{-1}$, one obtains   for the energy spectrum
\begin{equation}
E(k) = \lim_{t\to 0} 4\pi k^2 C^{(2)}(t,\vk) = A \bar\epsilon^{2/3} (k\eta)^{-\beta}\exp(-\mu(k\eta)^\gamma)
\label{eq:spec-diss}
\end{equation}
with $\gamma=2/3$ and $\beta = 5/3$. The expression for $C^{(2)}$ is obtained in the large wavenumber expansion and at the fixed-point, thus this behaviour is expected to be valid at wavenumbers large $k\gg L^{-1}$ but still controlled by the fixed-point, which corresponds to the near-dissipation range. Let us emphasise that the result \eq{eq:spec-diss} does not have the same status as the expression for $C^{(2)}$ \eq{eq:C2-FRG}. It clearly relies on an additional assumption, the saturation of the scaling variable, which is reasonable but not rigorous. In particular, if one assumes the existence of different successive scalings, associated with the emergence of quasi-singularities (\cite{Dubrulle2019}), one would obtain a different result.
In fact, it would be interesting to investigate whether the distribution of $\gamma$ displayed in \fref{fig:dissipative} is compatible with a multi-fractal description.

However, it seems reasonable to assume that the first scaling to dominate close to the inertial range is the Komogorov one, which leads to the stretched exponential \eq{eq:spec-diss} with the $\gamma=2/3$ exponent. This prediction has been verified in high-resolution \DNS (\cite{Canet2017,Gorbunova2020,Buaria2020}). In the simulations,  
the exponent $\gamma$ was determined through the local exponent defined as
\begin{equation}
D_3(k) = \dfrac{d\ln}{d\ln(k\eta)}\Big[\dfrac{d\ln}{d\ln(k\eta)}\Big[-\dfrac{d\ln E(k)}{d\ln(k\eta)}]\Big]
\end{equation}
Indeed, if $E(k)$ is described by \Eq{eq:spec-diss} on a certain range of $k$, one obtains $D_3(k) = \gamma$  on this range (\cite{Gorbunova2020}), irrespective of the value for $\beta$. In the \DNS, significant averaging was used in order to obtain smooth enough data to allow for the numerical evaluation of the three successive derivatives by finite differences. 

It has also been observed in experimental data from von K\'arm\'an swirling flows (\cite{Debue2018}) and from the Modane wind tunnel (\cite{Gorbunova2020}). The latter yields the estimate $\gamma\simeq 0.68\pm 0.19$, which is in very good agreement with the \FRG prediction. 
Two examples of determination of $\gamma$ are shown on \fref{fig:dissipative}.

\section{Time-dependence of correlation functions in passive scalar turbulence}
\label{sec:scalar}
 
We now consider a passive scalar field $\theta(t,\vx)$ which dynamics is governed by the advection-diffusion equation \eq{eq:passive-scalar}. In this section, we consider three cases for the carrier flow: i) the advecting field $\vvv(t,\vx)$ is a turbulent velocity field solution of the incompressible \NS equation \eq{eq:NSeq}, ii) $\vvv(t,\vx)$ is a random vector field with a Gaussian distribution characterised by zero mean and covariance
\begin{equation}
\bigr\langle \sv_\alpha \left(t,\vx\right) \sv_\beta \left(t^{\prime},\vy \right) \bigr\rangle  
= 
\delta\left(t-t^{\prime}\right) D_{0}\int_{\vp} \frac{e^{i \vp\cdot\left(\vx-\vy\right)}P^\perp_{\alpha\beta}(\vp)}{\left(p^{2}+m^{2}\right)^{\frac{d}{2}+\frac{\varepsilon}{2}}}
 \label{eq:cov-kraichan} 
\end{equation}
where $P^\perp_{\alpha\beta}(\vp)$ is the transverse projector which ensures incompressibility, and 
iii) the same random vector field with finite time correlations instead of $\delta(t-t')$.
 The model ii) was proposed by Kraichnan~(\cite{Kraichnan68,Kraichnan94}), and has been thoroughly studied (we refer to (\cite{Falkovich2001,Antonov2006}) for reviews). In this model, the parameter $ 0<\varepsilon/2<1$ corresponds to the H{\"o}lder exponent, describing the velocity roughness from very rough for $\varepsilon\to 0$ to smooth for $\varepsilon\to 2$, and $m$ acts as an \IR cutoff. 
 
 The spatio-temporal correlations  of the passive scalar in the three models for the carrier flow has been investigated within the \FRG formalism in \crefs ~ (\cite{Pagani2021}). We first summarise the main results, and then present comparison with \DNS for the three cases.

\subsection{FRG results for the correlation functions of passive scalars}
 
 \subsubsection{Scalar field in Navier-Stokes flow}
 
 We start with the model i), \ie a scalar field transported by a \NS turbulent flow.
 We focus on the inertial-convective range, in which the energy spectrum of the scalar 
 decays as $E_\theta \left(\vp\right)\sim p^{-5/3}$ as established by Obukhov and Corrsin  in their seminal works (\cite{Obukhov49,Corrsin51}.
  As shown in~\sref{sec:Ward}, the action for the passive scalar possesses similar extended symmetries as  the \NS action. As a consequence, the structure of the related Ward identities for the vertices is the same: they all vanish upon setting one wavevector to zero, unless it is carried by a velocity field in which case it can be expressed using the ${\cal D}_\alpha$ operator \eq{eq:wardGalNtheta}. Hence, the derivation of \sref{sec:general-flow}
 can be reproduced identically in the presence of the scalar fields.  We refer to (\cite{Pagani2021}) for details.
 This yields the general expression for the time-dependence of any $n$-point generalised correlation function of the scalar in the limit of large wavenumbers.
 
 In this section, we only consider two-point correlation functions, so we drop the superscript $^{(2)}$ on $C$ in the following. The  correlation function of the scalar, defined in time-wavevector coordinates as $C_{\theta, {\rm NS}}(t,\vp)\equiv \langle \theta\left(t,\vp\right)\theta\left(0,-\vp \right) \rangle$, is given in the \FRG formalism by
 \begin{equation}
C_{\theta, {\rm NS}}\left(t,\vp\right) = \frac{\bar\epsilon_\theta \bar\epsilon^{-1/3}}{ p^{11/3}} \left\{
\begin{array}{l l}
 C_{0}  \exp\left({-\alpha_{0} (L/\tau)^2 p^2 t^2}\right)\,,\quad &t\ll \tau \\ 
 C_{\infty} \exp\left({-\alpha_{\infty} (L^2/\tau) p^2 |t|}\right)\, \quad  &t\gg \tau \, ,
 \end{array}
 \right.
\label{eq:C2-scalNS}
\end{equation}
where $\bar\epsilon$ and $\bar\epsilon_\theta$
 denote the  mean energy dissipation rates of the velocity and scalar fields respectively,
  $\tau\equiv \left(L^2/\bar\epsilon \right)^{-1/3}$  the eddy-turnover time at the 
energy injection scale, and $C_{0,\infty}$ and $\alpha_{0,\infty}$ are constants. In fact, $\alpha_{0,\infty}$ are the same constants as the ones in the \NS velocity correlation function \Eq{eq:C2-FRG}. They can be calculated using \FRG by integrating the flow equations, for example within the LO approximation presented in \sref{sec:fixed-point}, 
for the forcing profile under consideration (or a reasonable model of it).
Hence, the temporal decay of the scalar correlations is determined  in the inertial-convective range by the  one of the carrier fluid. In particular, these constants depend on the profile of the forcing exerted on the velocity, but not on the one exerted on the scalar field.

\subsubsection{Scalar field in white-in-time synthetic velocity flow}

Let us now focus on the  model ii) proposed by Kraichnan  (\cite{Kraichnan68}), in which the \NS velocity field is replaced by a random vector field with white-in-time Gaussian statistics. The remarkable feature of this model is that despite its extreme simplication compared to real scalar turbulence, it still retains its essential features, and in particular universal anomalous scaling exponents of the structure functions. 
 Moreover, it is simple enough to allow for analytical calculation in suitable limits
(\cite{Chertkov96,Chertkov95,Gawedzki95,Bernard96,Bernard98,Adzhemyan98,Adzhemyan2001,Kupiainen2007,Pagani2015}),
 see  (\cite{Falkovich2001,Antonov2006}) for reviews. The temporal dependence of the scalar correlation function was analyzed in (\cite{Mitra2005, Sankar2008}).
 Within the \FRG framework, the large wavenumber expansion  also allows in this case for the closure of the flow equations for the scalar correlation functions. Indeed, for Kraichnan model, the action for the velocity field ${\cal S}_{\rm NS}[\vvv,\bar\vvv,p,\bar p]$ is simply replaced by 
 \begin{equation}
  {\cal S}_{\rm K}[\vvv] = \int_{\vx} \sv_\alpha(t,\vx)\frac{(-\p^2 +m^2)^{\frac{d+\varepsilon}{2}}}{2 D_0} \sv_\alpha(t,\vx) \,.
 \end{equation}
Remarkably, the total action ${\cal S}_{\rm K} + {\cal S}_\theta$ still possesses the same extended symmetries, but for the time-dependent shifts of the response fields \eq{eq:shiftjauge} and \eq{eq:shiftjaugetheta}, which reduce to $\bar\theta(t,\vx) = \bar\theta(t,\vx)+\bar\epsilon(t)$.
The resulting Ward identities endow an essentially identical structure, such that the derivation of the large wavenumber expansion proceeds in the same way. The essential difference arises when computing the loop integral in the second line of \Eq{eq:flowWn}, in which the velocity propagator in \eq{eq:exprH} is replaced by
\begin{equation}
 \bar{G}_{\sv_\alpha \sv_\beta}(\omega,\vq) = 2 D_0 \frac{P^\perp_{\alpha\beta}(\vq)}{(q^2 +m^2)^{\frac{d +\varepsilon}{2}}+{\cal R}_{\kappa, vv}(\vq)} \, .
 \label{eq:GK}
\end{equation}
Because of  the white-in-time (frequency-independent) nature of the Kraichnan propagator, the loop integral can be computed explicitly for any time delay, and yields a linear dependence in $t$.
 This implies that the decay of the scalar correlation is always exponential in time, \eg for the two-point function 
 \begin{equation}
 C_{\theta, \rm{K}}\left(t,\vp\right) = F\left(p\right) e^{-2\kappa_{\rm ren} p^2 |t|}\, ,
 \label{eq:C2-scalK}
 \end{equation}
 hence there is no Gaussian regime at small times.
Moreover, one obtains an explicit expression for the non-universal constant $\kappa_{\rm ren}$ 
\begin{eqnarray}
\kappa_{\rm ren} = \kappa_\theta +
\frac{d-1}{2d}
\int_{\vp} \frac{D_{0}}{\left(p^{2}+m^{2}\right)^{\frac{d}{2}+\frac{\varepsilon}{2}}}\,.
\label{eq:def-renormalised-mol-visc}
\end{eqnarray}
The second term in the renormalised diffusivity $\kappa_{\rm ren}$  embodies the effect of fluctuations and can therefore be interpreted as an eddy diffusivity.
  The prefactor $F\left(p\right)$ in \eq{eq:C2-scalK} is given by an  integral, which behaves 
 in the inertial range as $F\left(p\right) \sim p^{-d-2+\varepsilon}$
while in the weakly non-linear regime (\ie when the convective term is perturbative), 
 it behaves as $F\left(p\right) \sim p^{-d-2-\varepsilon}$ (\cite{Frisch96,Pagani2021}).
For the Kraichnan model, there is no anomalous correction to the second-order structure function (\cite{Falkovich2001,Antonov2006}). The dimensional (analogous to K41) scaling  of $F(p)$ is in this case an exact result. 

\subsubsection{Scalar field in time-correlated synthetic velocity flow}

Let us introduce a slight extension of the Kraichnan model, model iii), in which the white-in-time covariance \eq{eq:cov-kraichan} of the synthetic velocity field in replaced by one with tunable time-correlations $D_{T_e}(t-t')$, where $T_e$ represents the typical decorrelation time of the synthetic velocity.  The pure Kraichnan model is simply recovered in the limit $T_e\to 0$ where  $D_{0}(t-t')\equiv \delta(t-t')$. The calculation of the scalar correlation functions within the \FRG large wavenumber expansion is striclty identical to the pure Kraichnan case, except for the loop integral  in \Eq{eq:flowWn}. Indeed, as soon as some  time dependence is introduced in the velocity covariance, hence in the velocity propagator \eq{eq:GK}, one restores the two time regimes. Let us focus for simplicity on the two-point scalar correlation function $C_{\theta, {\rm K_{T_e}}}(t,\vp)$. For small time delays $t\ll T_e$,  the Fourier exponentials in the loop integral can be expanded and one finds the Gaussian decay in $p t$, whereas for large time delays $t\gg T_e$, $T_e$ can be replaced by zero in the integral and one recovers the Kraichnan exponential decay in $p^2 |t|$. 

The great advantage of this model is that the time-scale of the correlation function of the velocity field is adjustable and can be varied independently of the scalar dynamics, in particular in \DNS, such that it allows one to access the crossover between the two time regimes, as shown in \sref{sec:KraichnanTe-DNS}.

\subsection{Passive scalars in Navier-Stokes flow}
\label{sec:scalar-DNS}

The spatio-temporal behaviour of the correlation function of the scalar transported by \NS flow has been studied in \DNS in (\cite{Gorbunova2021scalar}), using  similar numerical methods as for the \NS case. The Schmidt number of the scalar, defined as the ratio of the fluid viscosity to the scalar diffusivity $Sc=\nu/\kappa_\theta$ was varied from 0.7 to 36, and the two-point correlations of both the scalar and the velocity fields were recorded during the runs with similar averages as in \Eq{eq:C2-DNS}.

\begin{figure}
	\includegraphics[width=0.48\linewidth]{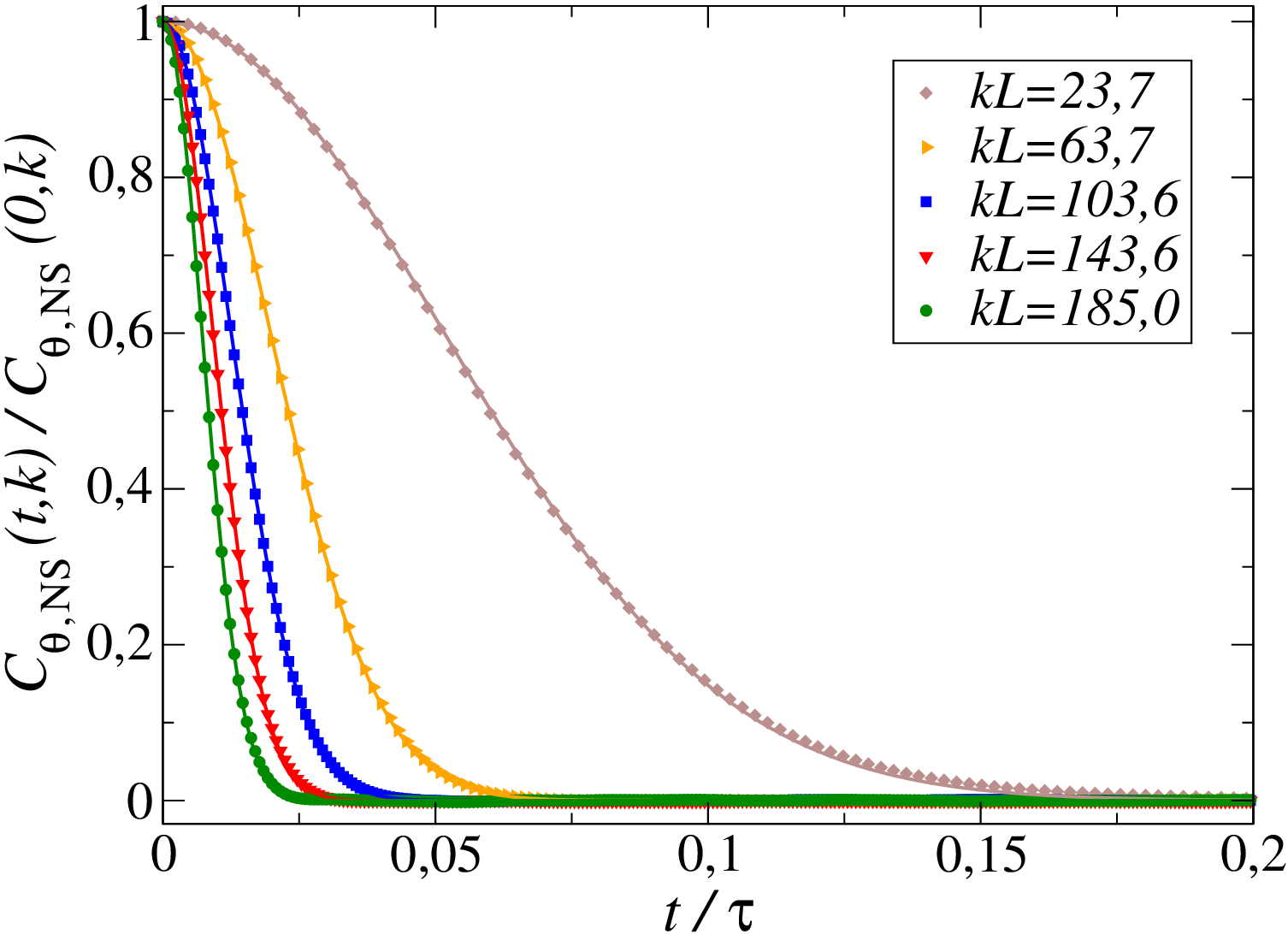}\hspace{0.5cm}
	\includegraphics[width=0.49\linewidth]{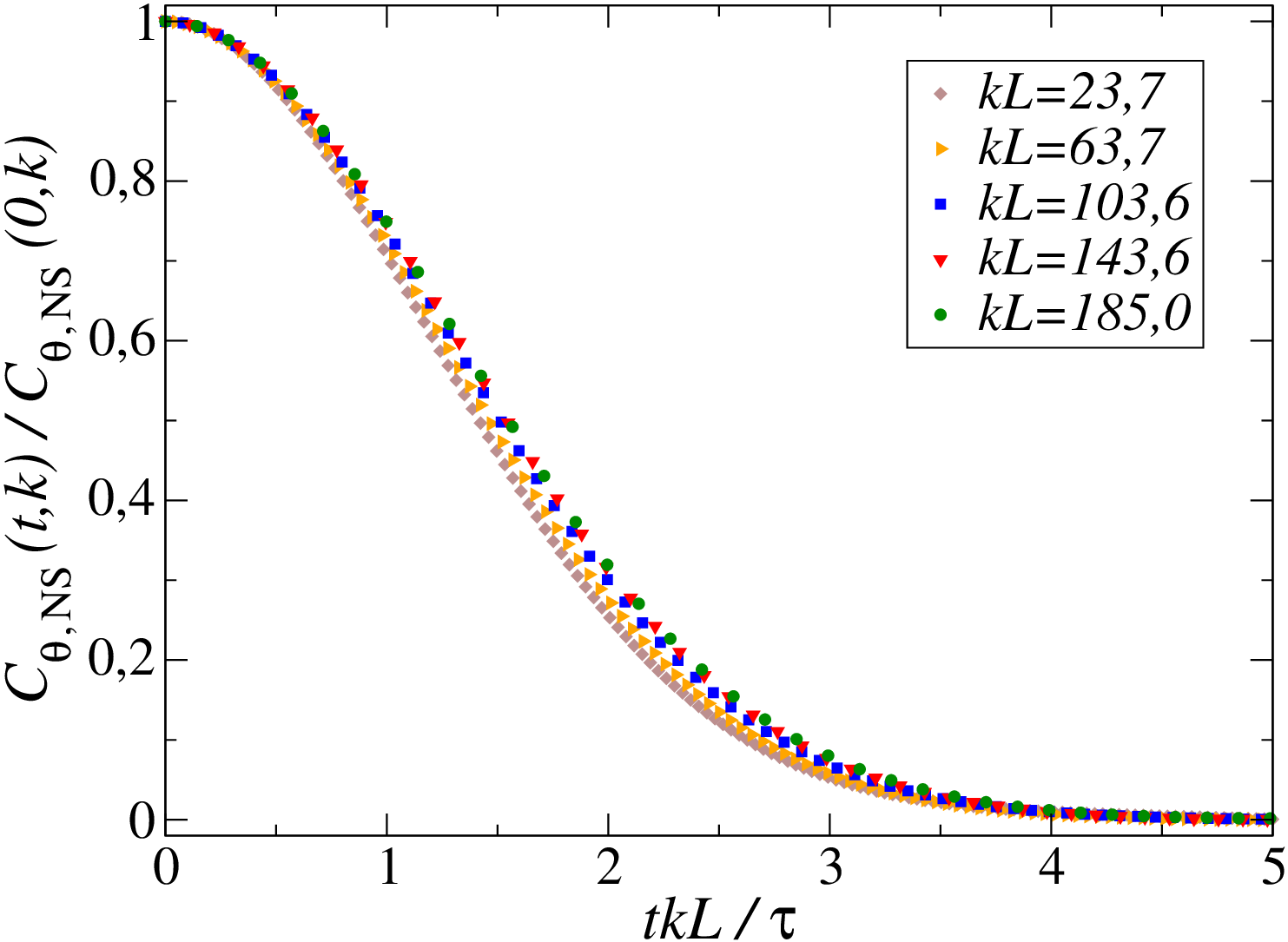}
	\caption{
 Time dependence of the normalised two-point correlation function \(C_{\theta,{\rm NS}} (t,k)\) of the scalar  in the \NS flow
	at different wavenumbers \(k\) for \(R_\lambda = 90\), as a function of $t$. ({\it left panel}) and of $kt$ ({\it right panel}), which results in their collapse.  \(L\) is the integral length scale, \(\tau\) is the  eddy-turnover time scale at the integral scale.}
	\label{fig:C2-scalNS}
\end{figure}

The results for $C_{\theta,{\rm NS}}(t,\vk)$ are presented in \fref{fig:C2-scalNS}. Each curve for a fixed wavenumber is accurately fitted by a Gaussian $f_{\rm Gaus}(t) = c\exp(-(t/\tau_0)^2)$. Moreover, all curves for different wavenumbers collapse when plotted as a function of the variable $kt$, as expected from the \FRG result \eq{eq:C2-scalNS}. This behaviour is completely similar to the one of the \NS velocity field presented in \fref{fig:C2-smallt}, as anticipated. The  transported scalars behave in the inertial-convective range as the carrier fluid particles.

\begin{figure}
	\includegraphics[width=0.48\linewidth]{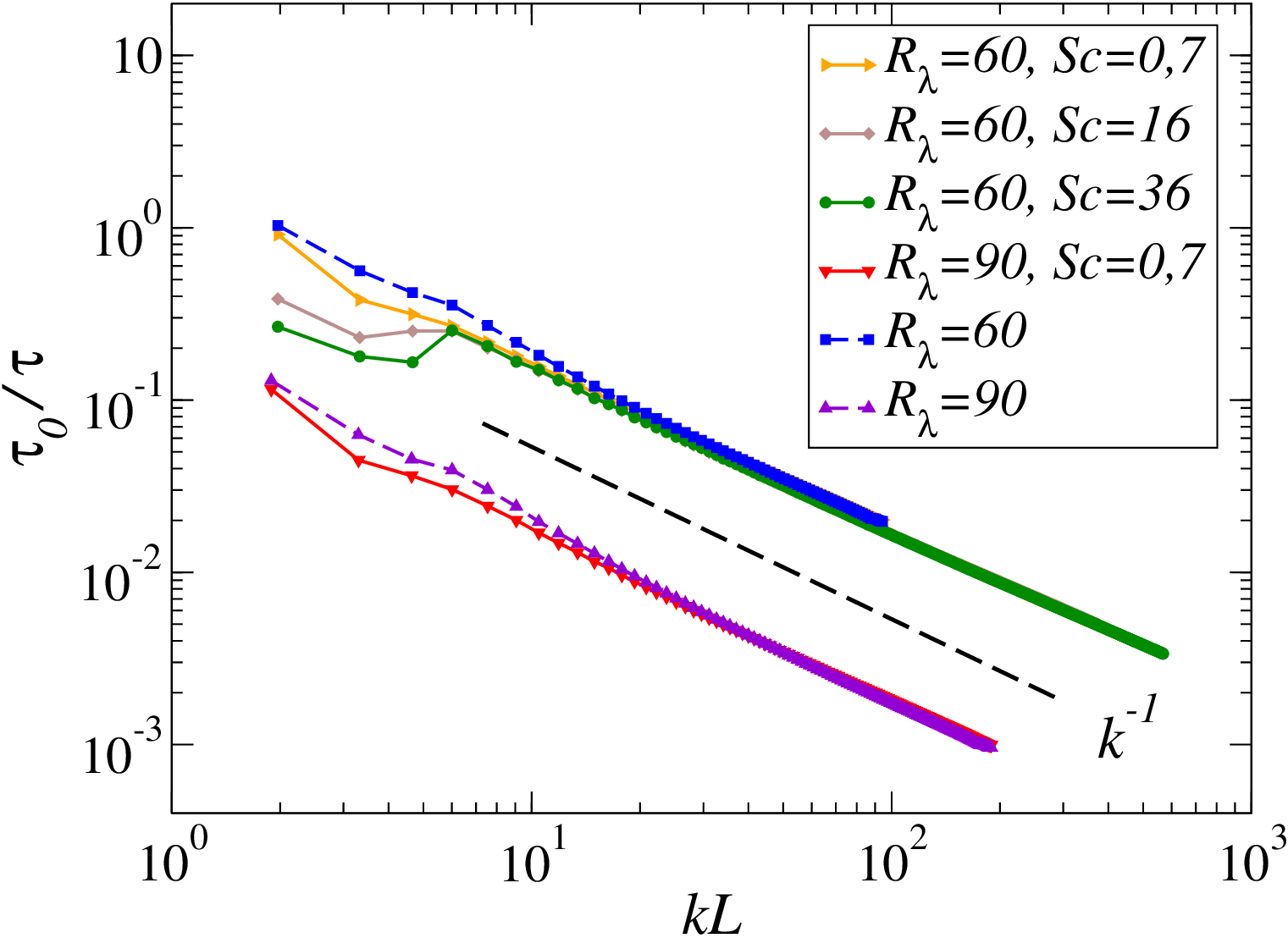}\hspace{0.5cm}
	\includegraphics[width=0.49\linewidth]{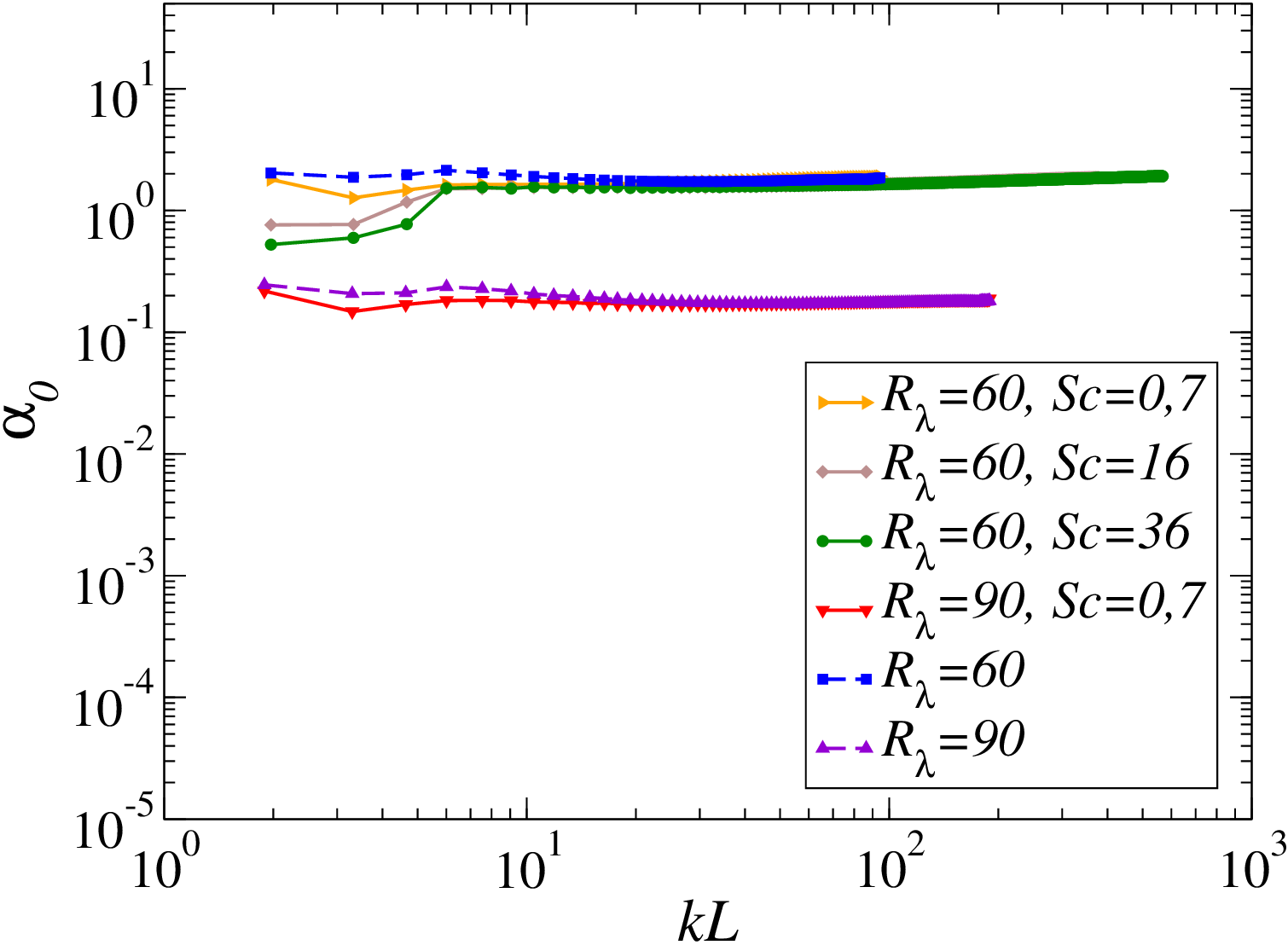}
	\caption{{\it Left panel}: Decorrelation time $\tau_0$ extracted from the Gaussian fit of the scalar correlations (plain lines) and of the velocity correlations (dashed lines) as a function of the wavenumber for various $R_\lambda$ and $Sc$. {\it Right panel}: non-universal parameter $\alpha_0$ obtained as $\alpha_0 = \tau^2/(\tau_0 kL)^2$.  In both panels, the data for $R_\lambda=90$ are shifted downward by a factor 10 for visibility.}
	\label{fig:tauS-scal}
\end{figure}

The decorrelation time $\tau_0$ extracted from the Gaussian fit is displayed in \fref{fig:tauS-scal}. According to \eq{eq:C2-scalNS}, it is related to $\alpha_0$ as $\tau_0/\tau = (\sqrt{\alpha_0} kL)^{-1}$, and it is found to precisely conform to the expected $k^{-1}$ decay. Moreover, beyond this behaviour, the \FRG analysis yields that the prefactor $\alpha_0$ is uniquely fixed by the properties of the carrier fluid, and is therefore equal for the velocity and the transported scalars. The numerical data for $\alpha_0$ shown in \fref{fig:tauS-scal} confirms this result, since the values of $\alpha_0$ for the velocity correlations and for the scalar correlations are in very close agreement, and almost independent of the scalar properties. Indeed, the variation of $Sc$ from 0.7 to 36 does not lead 
to significant changes of $\alpha_0$. Hence the dynamics of the scalar field is dominated by the random advection and the  sweeping effect.

At large time delays, a crossover from the Gaussian in $kt$ to an exponential in $k^2t$ is predicted by the \FRG result \eq{eq:C2-scalNS}. However, the Gaussian decay at small time induces a fast decrease of the correlations, such that the signal approaches zero and becomes oscillatory is the simulations before the crossover can occur. This prevents from its detection, as for the  \NS flow, which is not surprising since in this inertial-convective range the behaviour of the scalar closely follows the one of the velocity field. It would be very interesting to study other regimes of the scalar in order to find a more favorable situation to observe the large-time regime.
Meanwhile, the  crossover can be evidenced in the case of synthetic  flows, as we now show. 

\subsection{Passive scalars in the Kraichnan model}
\label{sec:Kraichnan-DNS}

The correlation function of the scalar field transported in a white-in-time synthetic velocity field has also been studied in \crefs ~(\cite{Gorbunova2021scalar}). A random vector field, isotropic and divergenceless, conforming to a Gaussian distribution with the prescribed covariance \eq{eq:cov-kraichan} is generated. The advection-diffusion equation \eq{eq:passive-scalar} with this synthetic field is then solved using \DNS. The details of the procedure and parameters can be found in (\cite{Gorbunova2021scalar}).
In order to achieve a precision test of the \FRG result \eq{eq:C2-scalK} and \eq{eq:def-renormalised-mol-visc}, 24 different sets of simulations were analyzed, varying the different parameters: H\"older exponent $\varepsilon$,  amplitude $D_0$ of the velocity covariance, and   diffusivity $\kappa_\theta$ of the scalar.  

\begin{figure}
	\includegraphics[width=0.48\linewidth]{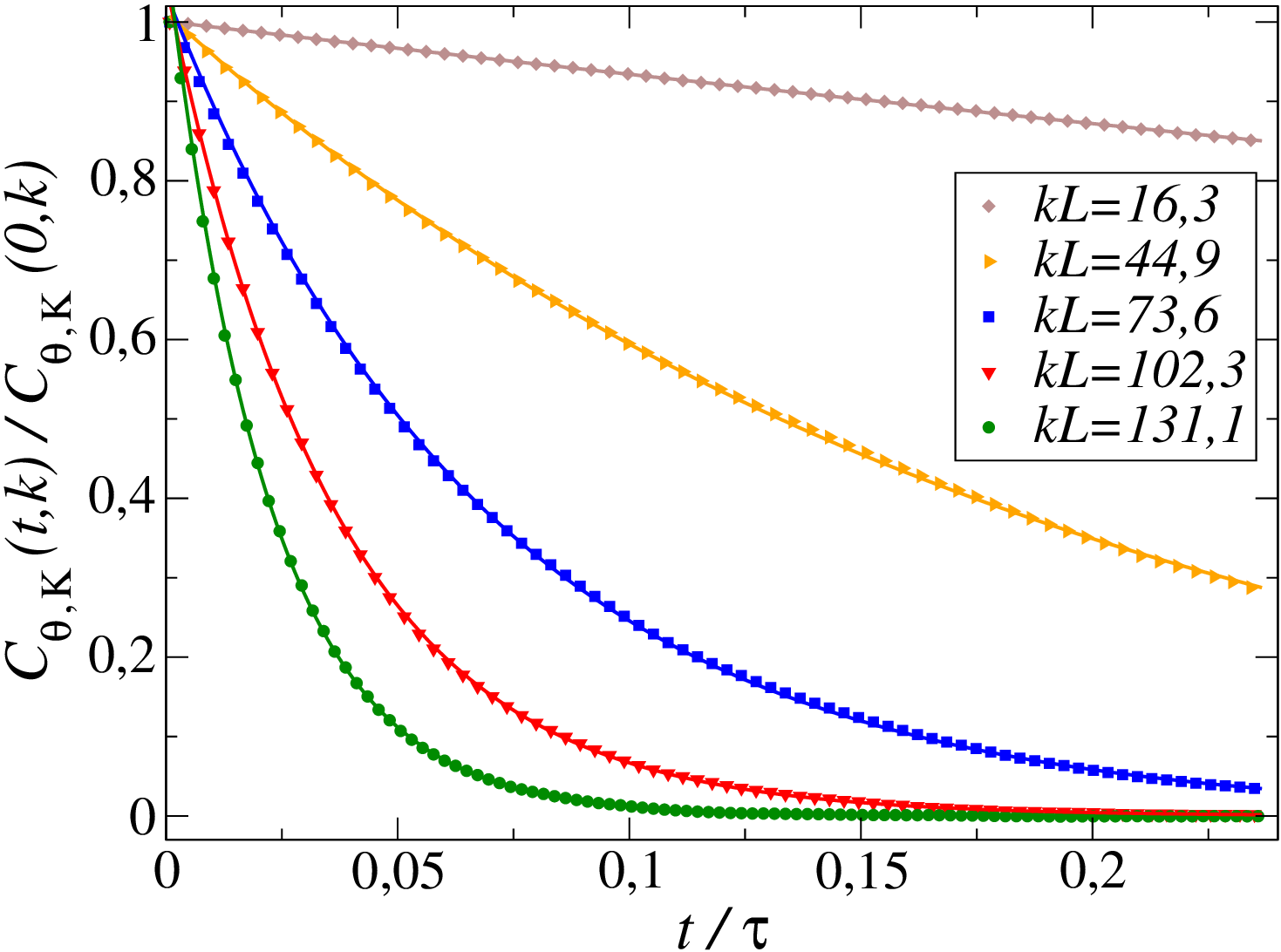}\hspace{0.5cm}
	\includegraphics[width=0.49\linewidth]{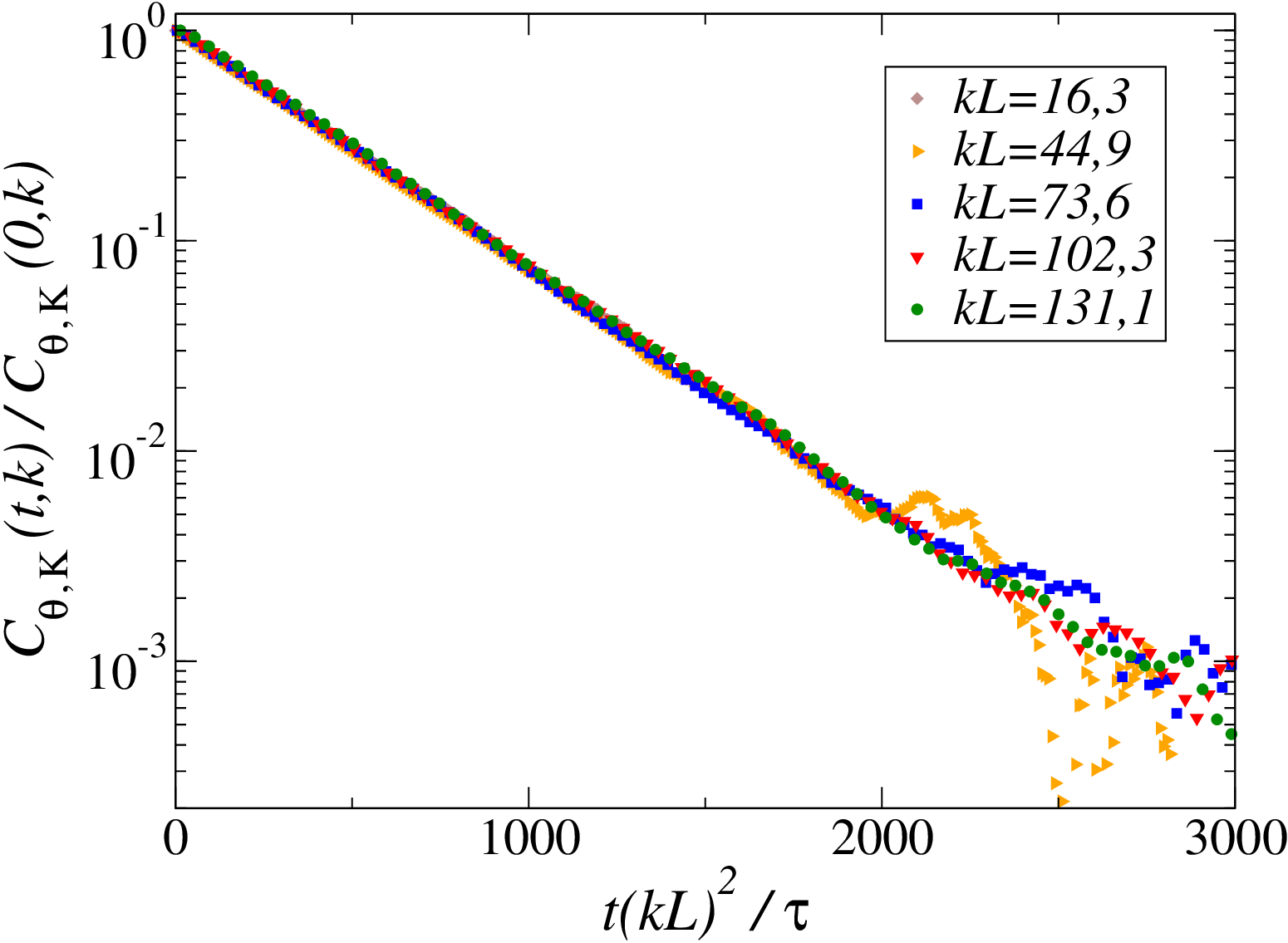}
	\caption{
Time dependence of the normalised two-point correlation function of the scalar \(C_{\theta, {\rm K}} (t,k)\) in Kraichnan model at different wavenumbers, as a function of time ({\it right panel}) and of the variable $k^2 t$ ({\it left panel}) which leads to their collapse.
	Data from the numerical simulation, for $\varepsilon=1$ in this example, are denoted with dots and their exponential fits with continuous lines.}
	\label{fig:C2-Kraichnan}
\end{figure}

\begin{figure}
	\includegraphics[width=0.48\linewidth]{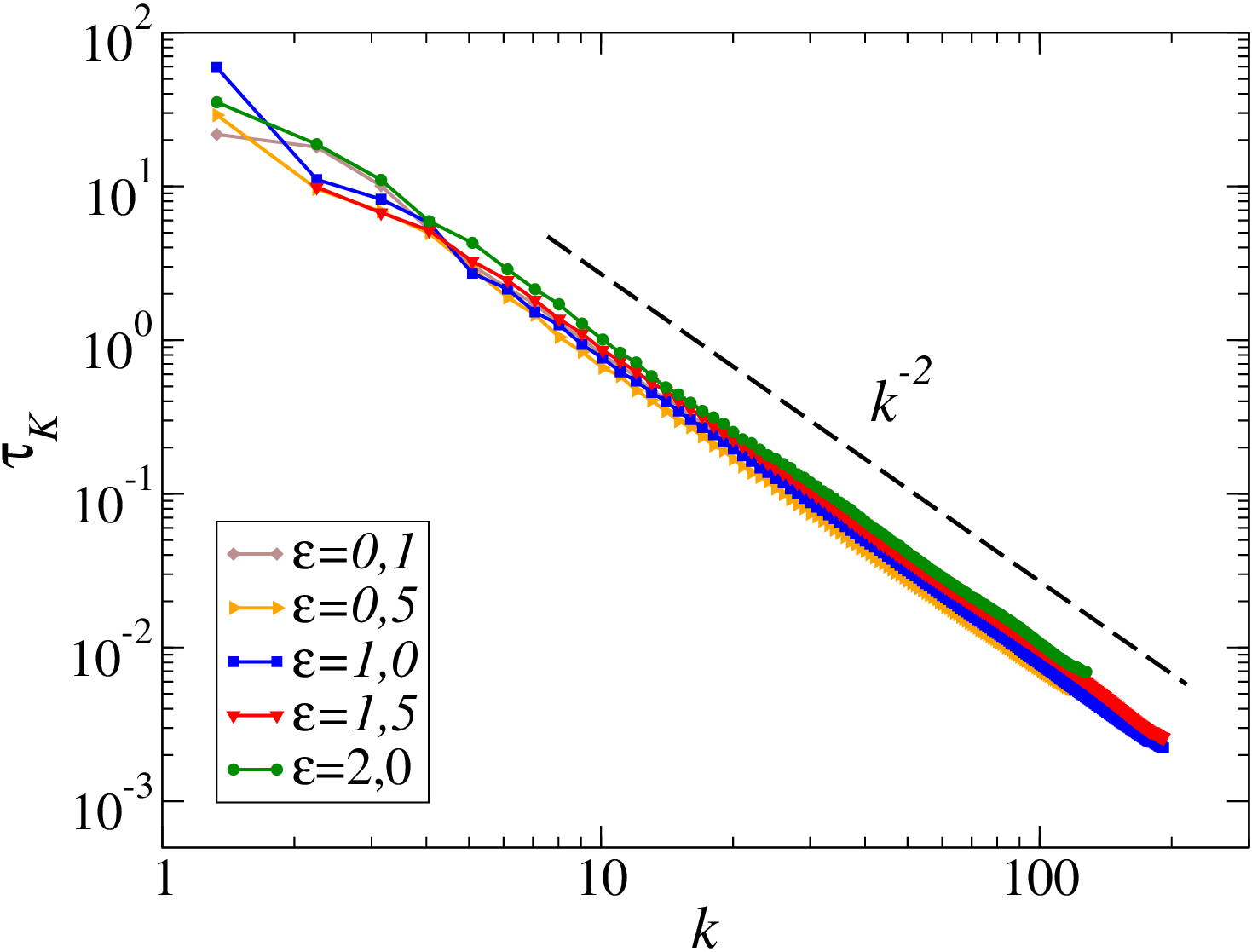}\hspace{0.5cm}
	\includegraphics[width=0.49\linewidth]{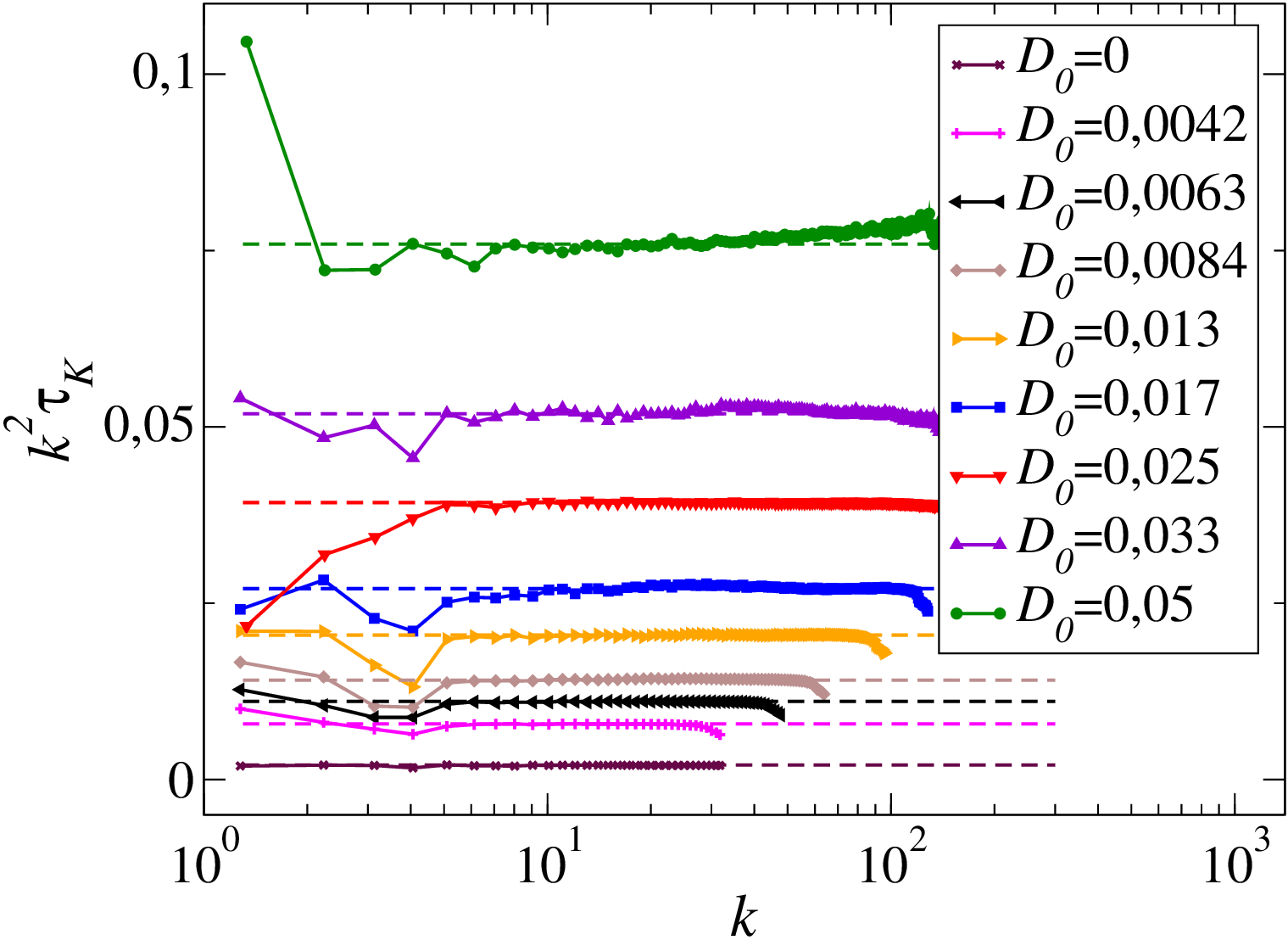}
	\caption{{\it Left panel:} Decorrelation time $\tau_K$ extracted from the exponential fits for different values of the H\"older exponent $\varepsilon$ from 0.1 to 2. {\it Right Panel:} Compensated decorrelation time $k^2 \tau_K$ extracted from the exponential fits, for different values of $D_0$ and $\varepsilon=1$. The renormalised diffusivity $\kappa_{\rm ren}$ for each curve is determined as the fitted value of the plateau represented by dashed lines.}
	\label{fig:tau-kraichnan}
\end{figure}

For each set, the correlation function of the scalar $C_{\theta,{\rm K}}(t,\vk)$ was computed. As illustrated in \fref{fig:C2-Kraichnan}, it always exhibits an exponential decay in time, perfectly modelled by the fitting function $f_{\rm exp}(t) = c\exp(-t/\tau_K)$.
 The decorrelation time $\tau_K$ depends on the wavenumber as $\tau_K\sim k^{-2}$, as expected from \eq{eq:C2-scalK}, and shown in \fref{fig:tau-kraichnan}.

\begin{figure}
	\begin{center}
	\includegraphics[width=0.5\linewidth]{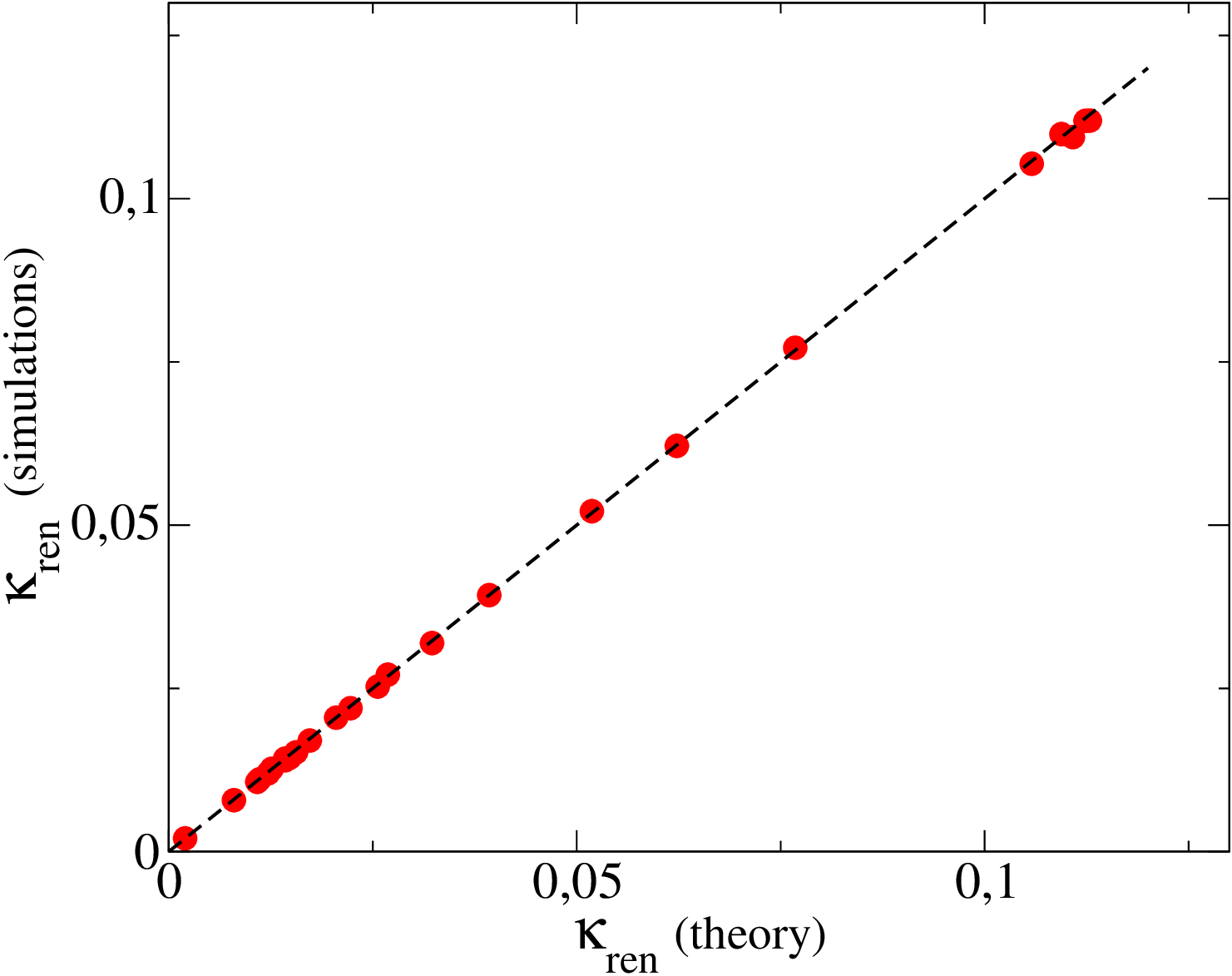}
	  \end{center}
	\caption{Renormalised scalar diffusivity: $\kappa_{\rm ren}$ {\scriptsize (simulations)} is obtained  from the plateau values of $k^2 \tau_K$ extracted from the exponential fits of the scalar correlation function;  $\kappa_{\rm ren}$ {\scriptsize (theory)} is calculated from its theoretical estimate based on  \Eq{eq:def-renormalised-mol-visc}.
    The data are gathered from  the 24 data sets for which the parameters $\varepsilon$, $Sc$, $D_0$, $\kappa_\theta$ are varied independently (see (\cite{Gorbunova2021scalar}) for detailed parameters).}
	\label{fig:kapparen}
\end{figure}

Beyond the global exponential form of the spatio-temporal dependence of the scalar correlation function, the  non-universal factor $\kappa_{\rm ren}$ in the exponential can be computed explicitly for the Kraichnan model within the \FRG framework and is given by \eq{eq:def-renormalised-mol-visc}. Its value depends on the parameters $\varepsilon$, $D_0$ characterising the synthetic velocity and $\kappa_\theta$ characterising the scalar.
All these parameters have been varied independently in the 24 sets of simulations. For each set, $\kappa^{\rm num}_{\rm ren}$ is determined from the numerical data as the plateau value of $\tau_K k^2$, where $\tau_K$ is extracted from the exponential fit, as illustrated in \fref{fig:tau-kraichnan}. Besides, the value of $(\kappa_{\rm ren}-\kappa_\theta)$ depends on the synthetic velocity only and can be computed prior to any simulation as $\kappa^{\rm num}_{\rm ren} -\kappa_\theta= A_\varepsilon D_0$ with $A_\varepsilon = \frac{1}{3} \sum_{\vp} \frac{D_{0}}{\left(p^{2}+m^{2}\right)^{\frac{3}{2}+\frac{\varepsilon}{2}}}$ following the  \FRG expression  \eq{eq:def-renormalised-mol-visc}. The comparison of the two estimations for $\kappa_{\rm ren}$ is displayed in \fref{fig:kapparen}, which shows a remarkable agreement. 
 Let us emphasise that the numerical data presented spans values of $\varepsilon$ up to 1.5, well beyond the perturbative regime $\varepsilon\simeq 0$ (\cite{Adzhemyan98}). 
 This analysis hence provides a thorough confirmation of the \FRG theory,
    including to the precise form of the non-universal prefactor in the exponential.

\subsection{Passive scalars in a time-correlated synthetic flow}
\label{sec:KraichnanTe-DNS}

Let us now turn to the extended Kraichnan model iii), \ie a scalar field advected by a synthetic velocity field with finite time correlations $D_{T_e}(t-t')$.
The simplest implementation of these finite time correlations in the \DNS is to 
 define a correlator $D_{T_e}(t) = \frac{1}{T_e}\theta(t-T_e)$ where $\theta(t)$ is the Heavyside step function. Hence, when $T_e$ is negligible compared to the other dynamical time scales of the mixing ($\tau_A \sim \eta_B /U_{\textrm{rms}}$ for advection, and  $\tau_\kappa \sim \eta_B^2/\kappa_\theta$ for diffusion, with  $\eta_B$ the Batchelor scale, \ie the smallest variation scale of the scalar), the velocity field can be considered as white-in-time (pure Kraichnan model). In contrast, when $T_e$ becomes comparable with the typical decorrelation time of the scalar -- say $\tau_K$ estimated from the pure Kraichnan model at intermediate wavenumbers -- then it realises a correlated synthetic velocity field.

\begin{figure}
	\includegraphics[width=0.48\linewidth]{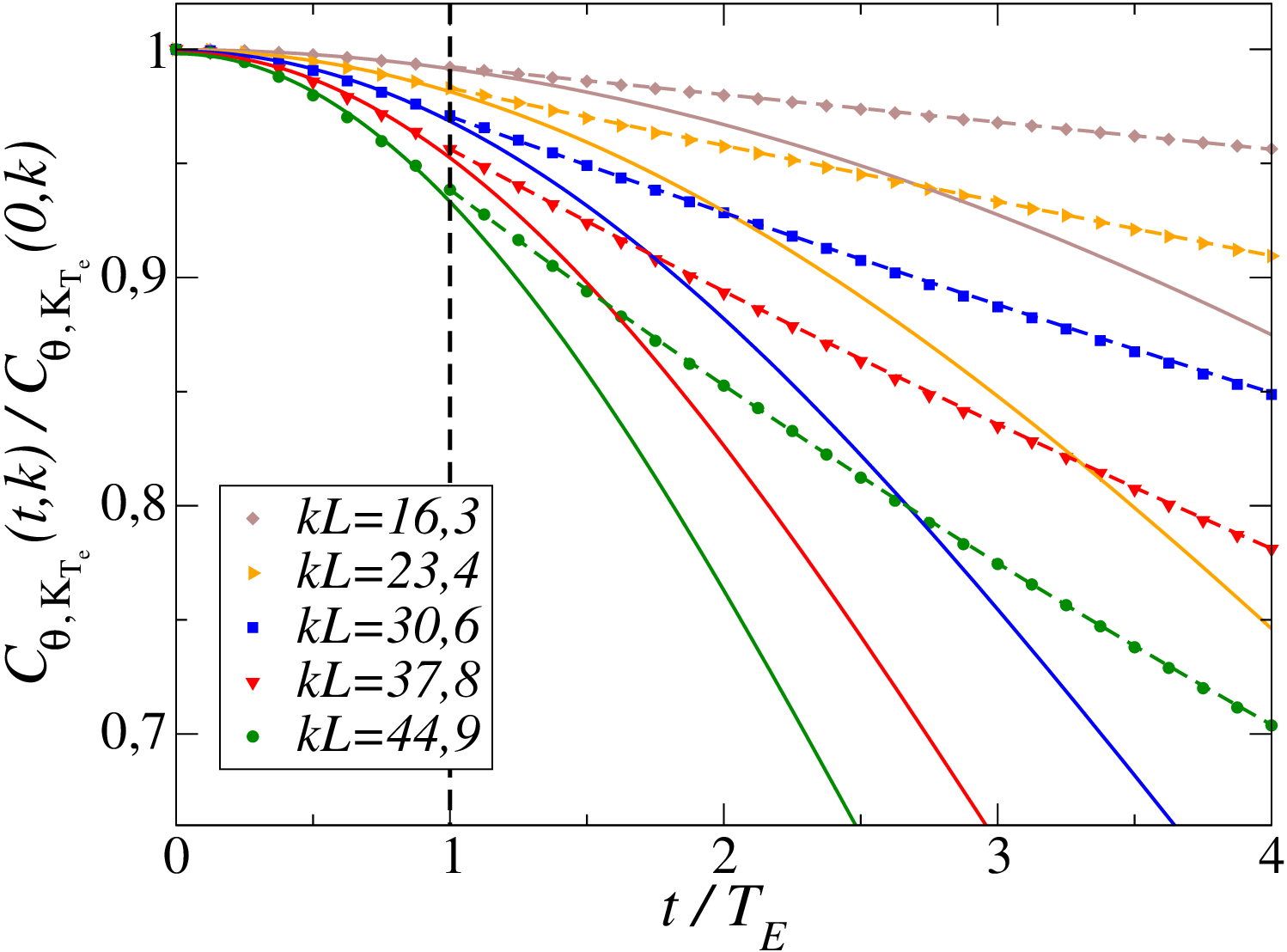}\hspace{0.5cm}
	\includegraphics[width=0.49\linewidth]{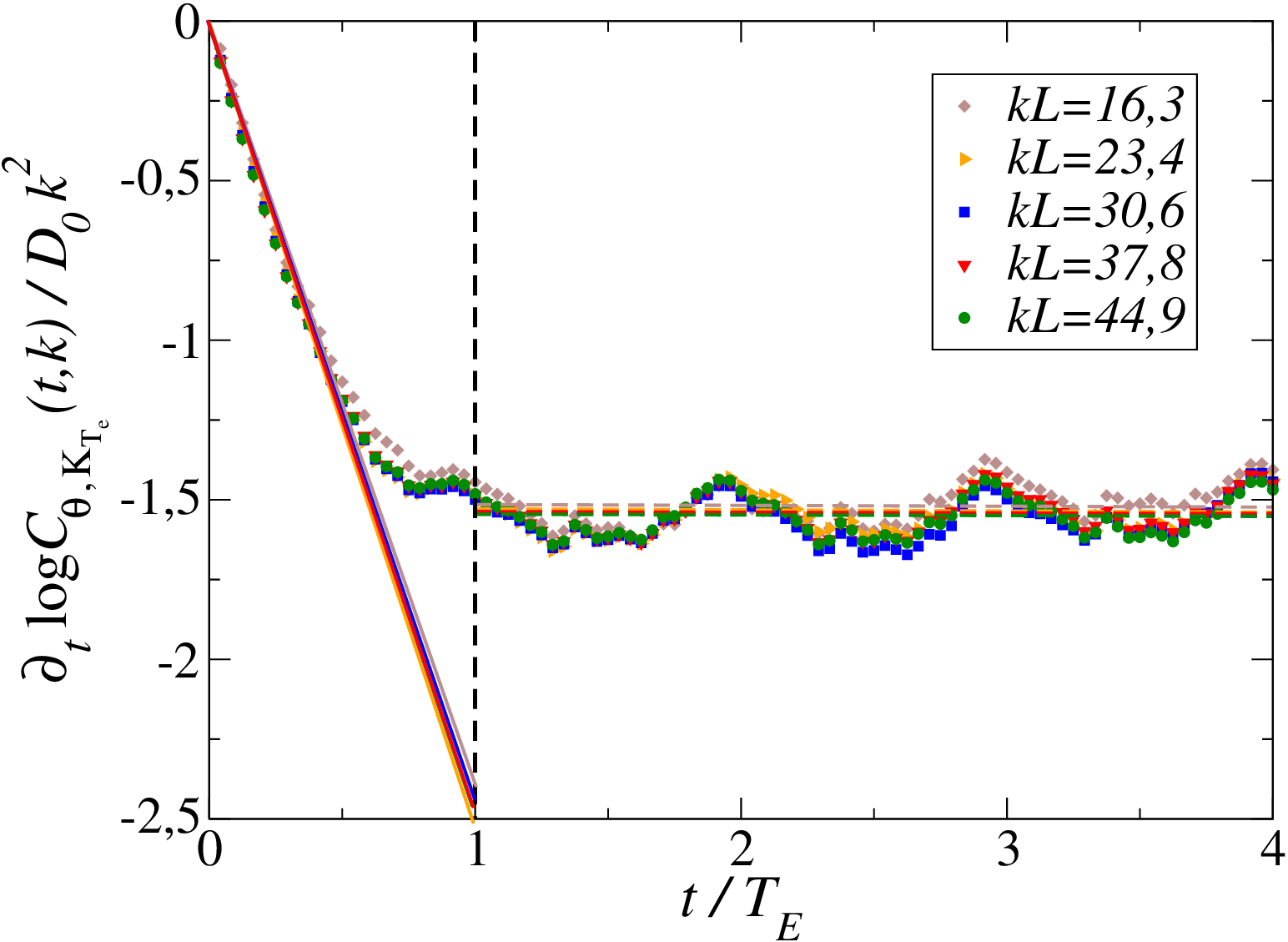}
	\caption{{\it Left panel}: Time dependence of the normalised two-point correlation function \(C_{\theta,{\rm K_{T_e}}} (t,k)\) at different wavenumbers \(k\). Data from the numerical simulation are denoted with dots, their Gaussian fits calculated for $t\leq T_e$ with continuous lines, their exponential fits calculated for $t\geq T_e$ with dashed lines. {\it Right panel}: time derivative of the logarithm of the same data (including the fits), divided by $D_0 k^2$.}
	\label{fig:C2-syn}
\end{figure}

The correlation function of the scalar \(C_{\theta,{\rm K_{T_e}}}(t,\vk)\), obtained from  \DNS of model iii), is shown in \fref{fig:C2-syn}. At small time delays $t \leq T_e$, the different curves for each wavenumber are well fitted by Gaussian curves, whereas at large time delays, they depart from their Gaussian fits and decay much more slowly. The slower decay is precisely modelled by exponential fits. Hence, the crossover from the small-time Gaussian regime to the large-time exponential one, predicted by \FRG, can be observed in this model. It can be further evidenced by computing the time derivative of $\log C_{\theta,{\rm K_{T_e}}}(t,\vk)$. According to the \FRG results, one should obtain
\begin{equation}
\frac{1}{k^2}\frac{\p \log C_{\theta,{\rm K_{T_e}}}(t,k)}{\p t} = \left\{
\begin{array}{l l}
-2 \alpha_0 t & t \ll T_e\\
- \alpha_\infty    &  t \gg T_e
\end{array}
 \right.\, ,
 \label{eq:synth}
\end{equation}
where $\alpha_{0,\infty}$ are the non-universal prefactors depending on the velocity characteristics only.
This derivative is displayed in \fref{fig:C2-syn}. All the curves collapse  when divided by $k^2$, and show a linear decay with negative slope at small time, which crosses over to a negative constant at large time, in agreement with \eq{eq:synth}.
Note that at higher wavenumbers (not shown), the large-time regime becomes indiscernible as the scalar field decorrelates fast (because of the $\sim k^2$ dependence) down to near-zero noisy values before $T_e$ is reached. This prevents us from resolving the crossover for these wavenumbers. This hints that a similar effect hinders the large-time regime in NS flows.

\section{Conclusion and Perspectives}
\label{sec:perspectives}

The purpose of this {\it JFM Perspectives} was to give an overview of what has been achieved using \FRG methods in the challenging field of turbulence, and to provide the necessary technical elements for the reader to grasp the basis and the scope of these results. An important aspect is that the \RG is in essence  conceived to build effective theories from ``first principles'', in the sense that it is a tool to compute statistical properties of a system, by averaging over fluctuations, from its underlying microscopic or fundamental description, that is, in the case of turbulence, from the \NS equations, without phenomenological inputs. An essential ingredient in this method is the symmetries, and extended symmetries, which can be fully exploited within the field-theoretical framework through exact Ward identities, that we expounded. So far, the main aspects of turbulence  addressed with the \FRG methods are the universal statistical properties of  stationary, homogeneous and isotropic turbulence. In this context, the main achievements of this approach are twofold.
 
 The first one is the evidence for the existence of a fixed-point of the \RG flow, for physical  forcing concentrated at large scales, which describes fully developed turbulence. This was not accessible using perturbative \RG approaches, because they are restricted to power-law forcing which changes the nature of the turbulence.  The very existence of a fixed-point demonstrates two essential properties of turbulence: universality (independence of the precise forcing and dissipation mechanisms) and power-law behaviours. So far, this fixed-point has been studied within a simple approximation of the \FRG, called LO (leading-order), which amounts to neglecting all vertices of order $n\geq 3$ but the one present in the \NS field theory. Nonethless, this rather simple approximation allows one to recover the exact result for the third-order structure function, and yields the correct K41 scaling for the energy spectrum and the second-order structure function, with accurate estimates of the Kolmogorov constant. This work calls for further calculations, within improved approximations, in order to determine whether anomalous exponents arise when including higher-order vertices in the \FRG ansatz. This is a first route to be explored, possibly coupled with the introduction of composite operators, to achieve this goal. This program has already been started in shell models, which are simplified models of turbulence. It indeed allows to compute intermittency corrections to the exponents of the structure functions  (\cite{Fontaine2022sabra}).

 The second achievement using \FRG, which is the most striking one, is the obtainment of the general expression, at large wavenumbers, of the spatio-temporal dependence of any $n$-point correlation and response function in the turbulent stationary state. The valuable feature of this expression is that it is asymptotically exact in the limit of large wavenumbers. Given the scarcity of rigorous results in 3D turbulence,  this point is remarkable and worth highlighting. All the $n$-point correlation functions are endowed with a common structure for their temporal behaviour, which can be simply summarised, on the example of the two-point function $C(t,\vp)$ for conciseness, as follows. At small time delays $t$, $C(t,\vp)$ exhibits a fast Gaussian decay in the variable $p t$, while at large time delays, it shows a crossover to a  slower exponential decay in the variable $p^2 |t|$. The prefactors $\alpha_0$ entering the Gaussian, and $\alpha_\infty$ entering the exponential, respectively, are non-universal, but  are equal for any $n$-point correlations. Similar expressions have been established in the case of passive scalars transported by turbulent flows, in the three cases where the turbulent velocity field is either the solution of the \NS equations, or a synthetic random velocity field (Kraichnan model and its time-correlated extension). For the \NS turbulent flow, the behaviour of the scalar follows in the inertial-convective range the one of the velocity field with fully analogous spatio-temporal correlations, including the non-universal prefactors.
In the case of the white-in-time Kraichnan model, the temporal decay of the scalar correlations is purely exponential in $p^2 t$, but a Gaussian regime opens up at small time delays when introducing a finite time-correlation in the velocity covariance. 


The \FRG results can be compared with \DNS, which allows one in particular to quantify the precise range of validity of the ``large wavenumber limit'' which underlies the \FRG calculations. Several \DNS have been conducted both for \NS turbulence and passive scalar turbulence with the different models for the advecting velocity field. They  brought a thorough and accurate confirmation of the \FRG predictions in the small-time regime, including high-precision tests in the case of the Kraichnan model. In contrast, the large-time regime has remained elusive in most \DNS, due to the steepness of the Gaussian regime, which leads to a fast damping of the correlations to near-zero levels and renders very challenging its numerical detection. The crossover has been evidenced in the time-correlated synthetic velocity field, because in this case, the different time-scales  are more easily adjustable in order to reach the large-time exponential decay before the signal gets too low and indiscernible from numerical noise. Although challenging, it would be very interesting to push further the \DNS investigations in order to access and study the large-time regime for the \NS flow. Another direction which remains unexplored
 so far is to test in \DNS the \FRG predictions for 2D turbulence, for which explicit expressions for the spatio-temporal correlations at large wavenumbers are also available (\cite{Tarpin2019}). 

The temporal behaviour of the $n$-point correlation functions are obtained within \FRG from the leading term in the large wavenumber expansion of the flow equations. Although this leading term is exact at large $p$, it vanishes at coinciding times $t=0$, such that it does not carry any information on the structure functions, and in particular on their possible anomalous exponents. Another route in the quest for the intermittency corrections is thus to compute within the large wavenumber expansion, the first non-vanishing term at equal times.  Preliminary results were obtained in this direction for 2D turbulence (\cite{Tarpin2019}), but it remains to be completed and extended to study 3D turbulence.

The calculation of intermittency corrections within the \FRG framework is clearly the next challenge to be addressed and probably the most significant, and recent results (\cite{Fontaine2022sabra}) already pave the way for this quest. However, another promising line of research is the use of \FRG in turbulence modelling.  The \FRG is designed by essence to construct a sequence of scale-dependent effective descriptions of a given system from the microscopic scale to the macroscopic one. In particular, if stopped at any chosen scale, the \FRG flow  provides the effective model at that scale, including the effect of fluctuations on all smaller (distance-)scales (larger wavenumbers). As mentioned in the introduction, computing such effective models from the fundamental equations, \eg \NS equations, via a controlled procedure of integration of fluctuations would be very valuable in many applications. Probably the most relevant one is to improve current models used as a ``sub-grid'' description -- that is model for the non-resolved scales -- in many numerical schemes, such as LES, meteorological or climate models. Of course, a long path lies before arriving at competitive results, since one should first extend the \FRG framework to describe more realistic situations, such as non-homogeneous, or anisotropic, or non-stationary conditions, the presence of shear, fluxes at the boundaries, \etc\dots Nonetheless,  we believe that this path is worth exploring, and should be the subject of future efforts.

\backsection[Acknowledgements]{L.C. would like to sincerely thank all her collaborators involved in the works presented here, whose implication was essential to obtain all these results. She thanks in particular  Bertrand Delamotte and Nicol\'as Wschebor, with whom the \FRG approach to turbulence was initiated, and largely developed. She is also grateful to Guillaume Balarac for all the work with \DNS,  and Gregory Eyink and Vincent Rossetto, who greatly contributed in analysing the data and the results of \DNS and experiments. Her special and warm thanks go to  the PhD students,  Anastasiia Gorbunova and Malo Tarpin, and postdoc Carlo Pagani, whose contributions were absolutely pivotal on \FRG or \DNS works.
L.C. would also like to thank Mickael Bourgoin,  Nicolas Mordant, B\'ereng\`ere Dubrulle and her  team for their great help with experimental data.}

\backsection[Funding]{This work was supported by the Agence Nationale pour la Recherche (grant ANR-18-CE92-0019 NeqFluids) and by the Institut Universitaire de France.}

\backsection[Declaration of interests]{ The author reports no conflict of interest.}

\appendix

\section{Additional extended symmetries and Ward identities}
\label{app:Ward}

\subsubsection{Local shift symmetries of the pressure fields}
\label{sec:Wardpressure}

An evident symmetry of the \NS action \eq{eq:NSaction} is the invariance under
 a time-dependent shift of the pressure field $\spr(t,\vx)\to \spr(t,\vx)+\varepsilon(t)$
  since the latter only appears with a gradient. 
   Let us now consider instead an infinitesimal time- {\it and} space-dependent shift  $\spr(t,\vx)\to \spr(t,\vx)+\varepsilon(t,\vx)$.
  This field transformation yields a variation of ${\cal S}_{\rm NS}$ which is linear in $\bar\vvv$. Since the field transformation is simply a change of variables in the functional integral \eq{eq:ZNS}, it must leave it unchanged. One deduces to first order in $\varepsilon$ the equality
\begin{equation}
 0= \int_{t,\vx} \Big\langle -\bar \sv_\alpha(t,\vx) \partial_\alpha \varepsilon(t,\vx) + K(t,\vx)\varepsilon(t,\vx)\Big\rangle \,,
 \label{eq:wardFP}
\end{equation}
where the first term comes from the variation of the action and the second one from the variation of the source term. Since this equality holds for any arbitrary infinitesimal $\varepsilon(t,\vx)$, one deduces the following Ward identity
\begin{equation}
 \frac{\delta \Gamma}{\delta \langle \spr(t,\vx)\rangle}\equiv K(t,\vx)  =-\frac{1}{\rho} \big\langle \partial_\alpha   \bar{v}_\alpha(t,\vx)\big\rangle =\frac{\delta {\cal S}_{\rm NS}}{\delta \spr(t,\vx)}\Big|_{\langle\Phi_k\rangle}\; ,
\end{equation}
where the first equality simply stems from the Legendre conjugate relations \eq{eq:Legendre-conjugate}, the second from the relation \eq{eq:wardFP}, and the third can be deduced from the NS action \eq{eq:NSaction}.
This entails that the dependence of the effective action $\Gamma$ in the pressure field $\spr$  remains the same as the one of the original action ${\cal S}_{\rm NS}$, \ie it keeps the same form in terms of the average fields $\frac 1 \rho \int_{t,\vx} \langle \bar v\rangle_\alpha \partial_\alpha \langle \spr\rangle$, or otherwise stated, this term is not renormalised -- it is not modified by the fluctuations.

The same analysis can be carried over for the response pressure field, by considering
the infinitesimal  field transformation   $\bar \spr(t,\vx)\to \bar \spr(t,\vx)+\bar \varepsilon(t,\vx)$. This yields the Ward identity
\begin{equation}
 \frac{\delta \Gamma}{\delta\langle \bar \spr(t,\vx)\rangle} \equiv \bar K(t,\vx) = \big\langle\p_\alpha \sv_\alpha(t,\vx) \big\rangle= \frac{\delta {\cal S}_{\rm NS}}{\delta  \bar \spr(t,\vx)}\Big|_{\langle\Phi_k\rangle}\;,
\end{equation}
which means that the corresponding term in the effective action is not renormalised either and keeps its original form $\int_{t,\vx} \langle \bar \spr\rangle \partial_\alpha \langle \sv_\alpha \rangle$. Hence the effective action remains linear in the pressure and response pressure fields, and there are no mixed pressure-velocity vertices beyond quadratic order, \ie no vertices for $n\geq 3$ with a $p$ or a $\bar p$ leg. As a consequence, the pressure fields only enter in the propagator, and the whole pressure sector essentially decouples, as will be manifest in the \FRG framework. 

Since the pressure fields are not renormalised, we have used throughout the paper the same notation $\spr$ for both the pressure and the average pressure, and similarly  $\bar{\spr}$ for both the response pressure and its average. This is of course not the case for the velocity sector which contains non-trivial fluctuations,  and we have used different notations $\vu \equiv\langle \vvv\rangle$ and $\bar\vu \equiv\langle \bar\vvv\rangle$ for the fields and average fields.

\subsubsection{Extended symmetries for the passive scalar fields}

Let us now consider the symmetries and extended symmetries of the passive scalar action \eq{eq:action-NS-scalar}. This action possesses two extended symmetries related to time-dependent shifts of the scalar field : $\theta(t,\vx)\to \theta(t,\vx) +\varepsilon(t)$ and the joint transformation for the response field
 \begin{align}
 &\bar\theta(t,\vx)\to \bar\theta(t,\vx) +\bar \varepsilon(t)\nonumber\\
 &\bar \spr(t,\vx) \to \bar \spr(t,\vx) + \bar \varepsilon(t)\theta(t,\vx)\, . \label{eq:shiftjaugetheta}
 \end{align}
Each of these transformations leads to a variation of the action which is linear in the fields.
 The corresponding functional Ward identities are straightforward to derive and read respectively (\cite{Pagani2021}). 
 \begin{eqnarray}
&& \int_{\vx} \frac{\delta \Gamma}{\delta \theta(t,\vx)} = -\int_{\vx}\p_t\bar\theta(t,\vx) \, \nonumber\\
 &&\int_{\vx} \frac{\delta \Gamma}{\delta \bar \theta(t,\vx)} = \int_{\vx} \Big\{\p_t\theta(t,\vx)+\sv_\beta(t,\vx)\p_\beta \theta(t,\vx) \Big\} \, .
 \end{eqnarray}
 These identities imply that the two terms $\int \bar\theta \p_t\theta$ and $\int \bar \theta \sv_\beta \p_\beta \theta$ are not renormalised, and they entail that any vertex with one vanishing wavevector carried either by a $\theta$ or a $\bar\theta$ field vanishes -- except  for $\Gamma^{\theta\bar\theta}$ which keeps its original  form given by ${\cal S}^{\theta\bar\theta}$ -- \ie
  \begin{align}
  & \Gamma^{(n_v,n_{\bar v}, n_\theta\geq 1, n_{\bar\theta})}(\cdots, \omega_\theta,\vp_\theta=0,\cdots) = 0 \nonumber\\
  &   \Gamma^{(n_v,n_{\bar v}, n_\theta, n_{\bar\theta}\geq 1)}(\cdots, \omega_{\bar\theta},\vp_{\bar \theta}=0,\cdots) = 0 \nonumber\\
 &  \Gamma^{(0,0,1, 1)}(\omega_\theta,\vp_\theta=\vzero,\omega_{\bar\theta},\vp_{\bar\theta}) = i\omega_\theta\;(2\pi)^{d+1}\delta(\omega_\theta+\omega_{\bar\theta})\delta^d(\vp_\theta+\vp_{\bar\theta})\, .
  \end{align}
These identities are very similar to \Eqs{eq:wardShift}{eq:wardShift21} ensuing from the time-gauged shift of the response fields.

Moreover, the time-dependent Galilean symmetry is also an extended symmetry of the total action ${\cal S}_{\rm NS}+{\cal S}_{\theta}$. Indeed,  the passive scalar and its response field behave as Galilean scalar densities under time-dependent Galilean transformations \eq{eq:gaugedGal}, \ie  $\delta\theta(t,\vx) = \varepsilon_\beta(t) \partial_\beta \theta(t,\vx)$ and  $\delta\bar\theta(t,\vx) = \varepsilon_\beta(t) \partial_\beta \bar\theta(t,\vx)$, which yield a variation of the total action linear in the field.	
The associated Ward identity constrains the vertices for which one
of the velocity has a zero wavenumber, 
\begin{eqnarray}
\Gamma_{\alpha_{1}\cdots\alpha_{n_{v}+n_{\bar{v}}+1}}^{\left(n_{v}+1,n_{\bar v},n_{\theta},n_{\bar{\theta}}\right)}\Bigr(\cdots,\underbrace{\omega_\ell,\vp_\ell=0}_{\ell={\rm velocity \;index}},\cdots\Bigr)  &=&
-\sum_{i=1}^{n}\frac{p_{i}^{\alpha_\ell}}{\omega_\ell}\Gamma_{\alpha_{1}\cdots\alpha_{n_{v}+n_{\bar{v}}}}^{\left(n_v,n_{\bar v},n_{\theta},n_{\bar{\theta}}\right)}\Bigr(\cdots,\underbrace{\omega_{i}+\omega_\ell,\vp_{i}}_{i^{{\rm th}}{\rm field}},\cdots\Bigr),\nonumber\\
&=&  {\cal D}_{\alpha_1}(\omega_\ell)\Gamma_{\alpha_{1}\cdots\alpha_{n_{v}+n_{\bar{v}}}}^{\left(n_v,n_{\bar v},n_{\theta},n_{\bar{\theta}}\right)}\Bigr(\cdots, \omega_i, \vp_i, \cdots\Bigr)
\label{eq:wardGalNtheta}
\end{eqnarray}
where $\alpha_{1}\cdots\alpha_{n_{v}+n_{\bar{v}}+1}$ are the spatial
indices of the velocity (and response velocity) fields,  $n=n_{\theta}+n_{\bar{\theta}}+n_{v}+n_{\bar v}$, and ${\cal D}_\alpha$ is the same operator as in \Eq{eq:wardGalN}.

To summarise, the key point is that the Ward identities for the scalar advected by the NS flow share essentially the same structure as the ones for the \NS action. They have a strong implication: when at least one wavevector of any vertex is set to zero, then it vanishes, except if the wavevector is carried by a velocity field, in which case it is controlled by \eq{eq:wardGalN} or \eq{eq:wardGalNtheta}, \ie given in terms of a linear combination of lower-order vertices through the operator ${\cal D}_\alpha$. This constitutes the cornerstone of the large wavenumber closure expounded in \sref{sec:general-flow}.

\section{Flow equations in the LO approximation}
\label{app:floweqLO}

We report in this Appendix additional details on the derivation of the flow equations for \NS, within the LO approximation presented in \sref{sec:fixed-point}, for the functions $f_{\kappa,\perp}^\nu$ and $f_{\kappa,\perp}^D$, and their explicit expressions. We refer to \crefs ~(\cite{Canet2016}) for the full derivation. 

From the general form of the effective average action and of the regulator term $\Delta {\cal S}_\kappa$, one can infer the general structure of the propagator matrix, defined as the inverse of 
the Hessian of $\Gamma_\kappa + \Delta {\cal S}_\kappa$. It is more conveniently expressed in Fourier space where it is diagonal in wavevector and frequency, such that it is just the inverse of a matrix
\begin{equation}
 \bar G_\kappa(\omega,\vp) = \left[\bar \Gamma_\kappa^{(2)} +{\cal R}_\kappa \right]^{-1}(\omega,\vp)\,.
\end{equation}
 As usual,  because of rotational and parity invariance,
   any generic two-(space)index function (in Fourier space) 
$F_{\alpha\beta}(\omega,\vp)$ can be decomposed  into a longitudinal and a 
transverse part as $F_{\alpha\beta}(\omega,\vp) = F_{\parallel}(\omega,\vp)P_{\alpha\beta}^\parallel(\vp) +  F_{\perp}(\omega,\vp)P_{\alpha\beta}^\perp(\vp)$ where the transverse and longitudinal projectors are defined by
\begin{equation}
 P_{\alpha\beta}^\perp(\vp)=\delta_{\alpha\beta}-\frac{p_\alpha p_\beta}{\vp\,^2}, \hspace{.5cm}\mathrm{and}\hspace{.5cm} P_{\alpha\beta}^\parallel(\vp)=\frac{p_\alpha p_\beta}{\vp\,^2}.
\end{equation}
Inverting the matrix $\left[\bar \Gamma_\kappa^{(2)} +{\cal R}_\kappa \right]$, one obtains  the general propagator 
\begin{equation}
\label{eq:G}
 \bar  G_{\kappa,\alpha\beta}(\omega,\vp)=
 \bordermatrix{~  & u_\beta & \bar u_\beta & \spr & \bar \spr\cr
u_\alpha      & \bar G^{u u}_{\kappa,\alpha\beta}(\omega,\vp) &\bar G^{u \bar 
u}_{\kappa,\alpha\beta}(\omega,\vp) & 0 & i p_\alpha 
  \bar G_\kappa^{u\bar \spr}(\omega,\vp) \cr
\bar u_\alpha &  \bar  G^{u \bar u}_{\kappa,\alpha\beta}(-\omega,\vp) & 0 & i p_\alpha  \bar  G_\kappa^{\bar u \spr}(\omega,\vp) & 0 \cr
\spr & 0& -i p_\beta \bar G_\kappa^{\bar u \spr}(-\omega,\vp) &  
\bar G_\kappa^{\spr \spr}(\omega,\vp) & \bar G_\kappa^{\spr \bar \spr}(\omega,\vp)\cr
\bar \spr & -i p_\beta\bar  G_\kappa^{u\bar \spr}(-\omega,\vp)& 0 &  \bar G_\kappa^{\spr\bar \spr}(-\omega,\vp) & 0}
\end{equation}
with in the pressure sector
\begin{equation}
  \bar  G_\kappa^{\spr \spr}(\omega,\vp)=-\frac {\rho} {p^2}\bar \Gamma^{(0,2)}_{\kappa,\parallel}(\omega,\vp)\quad,\qquad
 \bar G_\kappa^{\spr \bar \spr}(\omega,\vp)=\frac{\rho}{p^2}\bar \Gamma_{\kappa,\parallel}^{(1,1)}(-\omega,\vp)\, ,
\end{equation}
and  in the mixed velocity-pressure sector 
\begin{equation}
 \bar  G_\kappa^{u \bar \spr}(\omega,\vp) =-\frac 1 {p^2} \quad,\qquad  \bar G_\kappa^{\bar u \spr}(\omega,\vp)=\frac \rho {p^2}\,.
\end{equation} 
Furthermore, the components of the propagator in the velocity sector are purely transverse 
(that is, all the longitudinal parts vanish), as a consequence of incompressibility, and given by
\begin{align}
 \bar  G^{u \bar u}_{\kappa,\alpha\beta}(\omega, \vq)&= P_{\alpha\beta}^\perp(\vq) 
\frac{1}{\bar \Gamma^{(1,1)}_{\kappa,\perp}(-\omega, \vq) + R_\kappa(\vq)}\nonumber\\
 \bar G^{u u}_{\kappa,\alpha\beta}(\omega, \vq)&= -P_{\alpha\beta}^\perp(\vq) 
\frac{\bar \Gamma^{(0,2)}_{\kappa,\perp}(\omega, \vq) -2 N_\kappa(\vq)}{\left|\bar \Gamma^{(1,1)}_{\kappa,\perp}(\omega, \vq)+R_\kappa(\vq)\right|^2}.
\label{eq:propGk}
\end{align}

One can then insert the expression \eq{eq:G} into  \Eq{eq:dkgam2}, and compute the matrix product and trace to obtain the flow of $\bar\Gamma_\kappa^{(2)}$. Within the LO approximation, only the  vertex $\bar\Gamma_\kappa^{(2,1)}$ remains and is given by \eq{eq:gamma3L0}, such that in \Eq{eq:dkgam2}, one may set $\Bar\Gamma_\kappa^{(4)}=0$, and only a few elements are non-zero in the matrices $\bar\Gamma_{\kappa,\ell}^{(3)}$. 

According to the expressions \eq{eq:gam2LO} for $\bar\Gamma_{\kappa,\alpha\beta}^{(0,2)}$ and  $\bar\Gamma_{\kappa,\alpha\beta}^{(1,1)}$, the
 flow equations of the transverse functions $f_{\kappa,\perp}^D$ and $f_{\kappa,\perp}^\nu$ may 
be defined as
 \begin{align}
\partial_\kappa f_{\kappa,\perp}^{\nu}(\vp)&  = \frac{1}{(d-1)}\, 
P_{\alpha\beta}^\perp(\vp) \p_\kappa \Gamma^{(1,1)}_{\kappa,\alpha\beta}(\nu=0,\vp)  \nonumber\\
\partial_\kappa f_{\kappa,\perp}^{D}(\vp)&  = -\frac{1}{2(d-1)} 
\,P_{\alpha\beta}^\perp(\vp)\p_\kappa  
\Gamma^{(0,2)}_{\kappa,\alpha\beta}(\nu=0,\vp)\, .
\end{align}
After some  calculations (\cite{Canet2016}), one deduces the following flow equations
 \begin{align}
 \partial_s f_{\kappa,\perp}^{\nu}(\vp)&= \frac{1}{(d-1)}\int_{\vq}\Bigg\{\frac{\partial_s R_\kappa(\vq) \,\tf_\perp^D(\vp+\vq)}{ 
\tf_\perp^{\nu}(\vp +\vq) \big(\tf_\perp^{\nu}(\vq) + 
\tf_\perp^{\nu}(\vp+\vq)\big)^2} \nonumber\\
&\times \Big[\Big(-\vp\,^2+\frac{(\vp \cdot 
(\vp+\vq))^2}{(\vp+\vq)^2}\Big)(d-1)-2 \vp\cdot \vq\Big(1-\frac{(\vp\cdot \vq)^2}{\vq\,^2 \vp\,^2}\Big) 
\Big]\nonumber\\
&+ \frac{1}{\tf_\perp^{\nu}(\vq) \big( \tf_\perp^{\nu}(\vq) +  
\tf_\perp^{\nu}(\vp+\vq)\big)} \Big[ \partial_s R_\kappa(\vq) \frac{ 
\tf_\perp^D(\vq)\big(2  \tf_\perp^{\nu}(\vq) + \tf_\perp^{\nu}(\vp+\vq)\big)}{\tf_\perp^{\nu}(\vq) \big(  \tf_\perp^{\nu}(\vq) +  
\tf_\perp^{\nu}(\vp+\vq)\big)}- \partial_s N_\kappa(\vq)\Big]\nonumber\\
&\times \Big[\Big(-\vp\,^2+\frac{(\vp \cdot \vq)^2}{\vq\,^2}\Big)(d-1)+2 \frac{\vp\cdot (\vp+\vq)}{(\vq+\vp)^2}\Big(\vq\,^2-\frac{(\vp\cdot \vq)^2}{\vp\,^2}\Big) \Big]\Bigg\}
\label{eq:dkfnu}\\
 \partial_s f_{\kappa,\perp}^D(\vp)&
 =-\frac{1}{2(d-1)}\int_{\vq}  \Bigg\{
\frac{2 \tf_\perp^D(\vq+\vp)}{\tf_\perp^{\nu}(\vp+\vq) 
\tf_\perp^{\nu}(\vq)\big( \tf_\perp^{\nu}(\vq) +  \tf_\perp^{\nu}(\vp+\vq)\big)}\nonumber\\
 &\times \Big[ \partial_s R_\kappa(\vq) \frac{\tf_\perp^D(\vq)\big(2  \tf_\perp^{\nu}(\vq) + \tf_\perp^{\nu}(\vp+\vq)\big) 
}{\tf_\perp^{\nu}(\vq)\big(  \tf_\perp^{\nu}(\vq) + \tf_\perp^{\nu}(\vp+\vq)\big) }-  \partial_s N_\kappa(\vq)\Big]\nonumber\\
&\times\Big[2 
\frac{1}{\vq\,^2 (\vp+\vq)^2}\Big(\vq\,^2-\frac{(\vp\cdot \vq)^2}{\vp\,^2}\Big) 
 \Big(\vp\,^2 \vp\cdot \vq+2 (\vp\cdot \vq)^2-\vp\,^2 \vq\,^2\Big)\nonumber\\
 & + \Big(2\vp\,^2-\frac{(\vp \cdot (\vp+\vq))^2}{(\vp+\vq)^2}-\frac{(\vp \cdot \vq)^2}{\vq\,^2}\Big)(d-1)\Big]\Bigg\}
 \label{eq:dkfd}
\end{align}
where $\p_s\equiv \kappa \p_\kappa$,  
$\tf_\perp^{\nu}(\vp)\equiv f_{\kappa,\perp}^{\nu}(\vp)+ R_\kappa(\vq)$ and
 $\tf_\perp^{D}(\vp)\equiv f_{\kappa,\perp}^{D}(\vp)+  N_\kappa(\vq)$.
Note that a typo has been corrected in the last line of \Eq{eq:dkfd} compared to \crefs ~(\cite{Canet2016}).
 
\section{Next-to-leading order term in 2D Navier-Stokes equation}
\label{app:nexttoleading}

The large wavenumber expansion underlying the derivation of \Eq{eq:flowWn} is inspired by the \BMW approximation scheme. This scheme can be in principle improved order by order. There are two possible strategies to achieve this. The first one is to increase the order at which the closure using the \BMW expansion is performed. Let us consider for instance the flow equation for the two-point function $\Gamma_\kappa^{(2)}$. At leading order, the three- and four-point vertices entering this equation are expanded around $\vq=0$. At the next-to-leading order, one keeps the full vertices in the flow equation for $\Gamma_\kappa^{(2)}$, and perform the $\vq=0$ expansion for the higher-order vertices entering the  flow equations of $\Gamma_\kappa^{(3)}$ and $\Gamma_\kappa^{(4)}$, and so on. An alternative way to improve the \BMW approximation scheme is to include higher-order terms in the Taylor expansion of the vertices for a given order $n$, \eg
 \begin{align}
 \Gamma^{(n)}_\kappa(\omega,\vq,\varpi_1,\vp_1,\cdots) & \stackrel{\vq\simeq 0}{\simeq} \Gamma^{(n)}_\kappa(\omega,0,\varpi_1,\vp_1,\cdots) +\frac{\partial}{\partial \vq} \Gamma^{(n)}_\kappa(\omega,\vq,\varpi_1,\vp_1,\cdots)\Big|_{\vq=0} \nonumber\\
& +\frac{1}{2} \frac{\partial^2}{\partial \vq^2}\Gamma^{(n)}_\kappa(\omega,\vq,\varpi_1,\vp_1,\cdots)\Big|_{\vq=0} \, .
\label{eq:Taylorexp}
 \end{align}
 This second strategy has been implemented to study 2D turbulence in (\cite{Tarpin2019}). The reason for focusing on 2D rather than 3D turbulence is that the 2D \NS action possesses additional symmetries, which can be used to control exactly the derivatives of vertices at  $\vq=0$.
 
In 2D, the incompressibility constraint allows one to express the velocity field as the curl of a pseudo-scalar field $\psi(t,\vx)$ called the stream function:
$\sv_\alpha = \varepsilon_{\alpha\beta} \p_\beta \psi$. Formulating the action in term of the stream function allows one to simply integrate out the pressure fields, yielding 
\begin{align}
 \mathcal{S}_{\psi}[\psi, \bar \psi] &= \int_{t,\vx} \p_\alpha \bar \psi \Big[ \partial_t \partial_\alpha \psi - \nu \nabla^2 \partial_\alpha \psi +\varepsilon_{\beta\gamma} \,\partial_\gamma \psi \,\partial_\beta \partial_\alpha \psi \Big ] \nonumber\\
 &-\int_{t,\vx,\vx'}   \partial_\alpha \bar\psi(t, \vx) N\left(\frac{|\vx-\vx'|}{L}\right) \partial_\alpha' \bar \psi (t, \vx')  \Big\}\, .
\label{eq:ActionNSstream}
\end{align}
This action possesses all the extended symmetries discussed in \sref{sec:Ward}. In particular, the Galilean transformation reads in this formulation 
\begin{equation}
  \delta\psi = \varepsilon_{\alpha\beta} x_\alpha \dot\eta_\beta(t) + \eta_\alpha(t)\p_\alpha \psi\,, \quad\quad\quad \delta\bar\psi =\eta_\alpha(t)\p_\alpha \bar\psi\, , 
  \label{eq:Galpsi}
\end{equation}
and the time-gauged shift of the response fields now simply amounts to the transformation
\begin{equation}
\delta\psi = 0\,, \quad\quad\quad \delta\bar\psi = x_\alpha \bar\eta_\alpha(t)\, ,
\label{eq:shiftpsi}
\end{equation}
where $\veta$, $\bar\veta$ are the time-dependent parameters of the transformations.
 Note that these transformations are linear in $x$. Thus, in Fourier space, this leads to Ward identity constraining the first derivative of a vertex at one vanishing wavevector. If the zero wavevector is carried by a response stream, one deduces  from \eq{eq:shiftpsi} the following Ward identity
 \begin{equation} 
 \frac{\partial}{\partial q_{m+1}^\alpha} \Gamma_\kappa^{(m,n)}(\cdots,\varpi_{m+1},\vq_{m+1},\cdots)\Big|_{\vq_{m+1}=0} = 0\,. 
  \end{equation}
  which is equivalent to \Eq{eq:Wardshift} in terms of the velocity. Similarly, if the zero wavevector is carried by a stream field, the time-gauged Galilean transformation \eq{eq:Galpsi} yields a Ward identity, equivalent to  \Eq{eq:Wardshift}, which exactly fixes  the $\p_q$ derivative of this vertex in the stream formulation. 
   
One can study the action \eq{eq:ActionNSstream} within the \FRG formalism. It is clear that the same closure can be achieved in the limit of large wavenumber. In the stream formulation, the zeroth order in the large wavenumber expansion is trivial. This just reflects the fact that the stream and response stream functions are defined up to a constant function of time, and the functional integral does not fix this gauge invariance, such that the zeroth order carries no information. One thus has to consider the next term (first $\vq$ derivative) in the Taylor expansion of the vertices, which are precisely the ones fixed by the Ward identities. One thus obtains the exact same result as \eq{eq:flowWn} expressed in term of the stream function. One can derive from it in a similar way the expression of the  time dependence for generic $n$-point correlation functions in 2D turbulence. Note that they bare a different form than in 3D, due to the different scaling exponents (in particular  $z$). Contrary to the 3D case, this result has not been tested yet in direct numerical simulations. 

As in 3D, the leading order term in the limit of large wavenumber vanishes at equal times, such than one is left with power laws in wavenumbers with exponents stemming from standard scale invariance (Kraichnan-Leith scaling). However, additional extended  symmetries were unveiled  for the 2D action, which can be exploited to compute the next-to-leading order term (given by the second line of \Eq{eq:Taylorexp}). At equal times, \ie integrated over all external frequencies, this term writes 
\begin{align}
\p_s \int_{\varpi_\ell}{\cal W}^{(n)}_\kappa&\Big|_{\rm next-to-leading} =  -\frac{1}{2}\int_{\omega_1,\vq_1,\omega_2,\vq_2} H_{\kappa,\gamma\delta}\left(-\omega_1,-\vq_1,-\omega_2,-\vq_2\right)\nonumber\\
&\times \int_{\varpi_\ell}\left[\frac{q_a^\mu q_b^\nu q_c^\rho q_d^\sigma}{4!}\frac{\partial^2}{\p q_a^\mu \p q_b^\nu \p q_c^\rho \p q_d^\sigma}\frac{\delta}{\delta u_{\gamma}\left(\omega_1,\vq_1\right)}\frac{\delta}{\delta u_{\delta}\left(\omega_2,\vq_2\right)}{\cal W}^{\left(n\right)}_{\kappa} \right]\Bigg|_{\vq_1=\vq_2=0}
\label{eq:flowWnnexttoleading}
\end{align}
where $a,b,c,d$ take values in $\{1,2\}$, and we have not used explicitly translational invariance so that the $\vq$ derivatives can be expressed unambiguously. One can show that the only non-vanishing contributions are the ones with two $q_1$ and two $q_2$. The contributions with four $q_1$ vanish when evaluated at $\vq_2=0$ because of the extended symmetry, while the ones with three $q_1$ and one $q_2$ are proportional to a ${\cal D}_\mu$ operator and vanish when integrated over frequencies. To go further requires to control second derivatives of vertices. This can partially be achieved in 2D because  the  field theory \eq{eq:ActionNSstream} exhibits two additional extended symmetries, which correspond to transformations quadratic in $x$, and thus yield  Ward identities for the second $\vq$ derivative of vertices in Fourier space.

 The first new symmetry  is a quadratic in $x$ shift of the response fields,  following 
\begin{equation}
\delta\psi = 0\,, \quad\quad\quad \delta\bar\psi =\frac{x^2}{2}\bar\eta(t)\, .
\label{eq:shiftpsibar}
\end{equation}  
One deduces the following Ward identity for second derivatives of a vertex function evaluated at a vanishing wavenumber carried by a response field
\begin{equation} 
 \frac{\partial^2}{(\partial q_{\bar\ell}^\alpha)^2} \Gamma_\kappa^{(m,n)}(\cdots,\omega_{\bar\ell},\vq_{\bar\ell},\cdots)\Big|_{\vq_{\bar\ell}=0} = 0\,,
 \label{eq:Wardshiftquadratic}
  \end{equation}
  where $\bar\ell$ is a response field index.
Remarkably, the transformation \eq{eq:shiftpsibar}, which reads in the velocity formulation 
\begin{equation}
\delta \bar \sv_\alpha =  \varepsilon_{\alpha\beta\gamma} x_\beta \eta_\gamma(t)\, ,\quad \delta \bar \spr = \sv_\alpha  \varepsilon_{\alpha\beta\gamma} x_\beta \eta_\gamma(t)\,,
\end{equation}
is also an extended symmetry of the 3D \NS equation, but it has not been exploited yet.

In analogy with  the extended Galilean transformation, which  can be interpreted as a time-dependent generalisation of space translations, one can write the time-dependent extension of space rotations, which reads in the stream formulation
\begin{equation}
\delta\psi =-\dot\eta(t)\frac{x^2}{2}+\eta(t)\varepsilon_{\alpha\beta}x_\beta\p_\alpha\psi \,, \quad\quad\quad \delta\bar\psi =\eta(t)\varepsilon_{\alpha\beta}x_\beta\p_\alpha \bar\psi\, .
\end{equation}
This transformation leads to a new extended symmetry of the 2D action, while it is not an extended symmetry of the 3D one. It is interesting to notice that time-dependent rotations are also an extended symmetry of the action corresponding to the Kraichnan model in any spatial dimension. As this transformation is also quadratic in $x$, it leads as well to a Ward identity for the second $\vq$ derivative of a vertex in Fourier space, this time when the $\vq$ wavevector is carried by the field itself
\begin{equation} 
 \frac{\partial^2}{(\partial q^\alpha)^2} \Gamma_\kappa^{(m+1,n)}(\omega,\vq,\omega_1,\vq_1\cdots)\Big|_{\vq=0} = {\cal R}(\omega)\Gamma_\kappa^{(m,n)}(\omega_{1},\vq_{1},\cdots)\, ,
 \label{eq:Wardrotation}
  \end{equation}
where ${\cal R}(\omega)$ is an operator achieving finite shifts by $\omega$ of the external frequencies of the function on which it acts, similarly to ${\cal D}_\alpha(\omega)$ in \eq{eq:wardGalN}, at the difference that it also involves a derivative with respect to external wavevectors (see (\cite{Tarpin2019}) for details). 

Using translation and rotation invariance of $H_{\kappa,\gamma,\delta}$ in \eq{eq:flowWnnexttoleading}, one can show that there are only two remaining contributions of the form $\frac{\p^4}{\p q_1^\mu \p q_1^\mu \p q_2^\nu  \p q_2^\nu}$ and $\frac{\p^4}{\p q_1^\mu \p q_1^\nu \p q_2^\mu  \p q_2^\nu}$. The first type of contributions are controlled by the Ward identities \eq{eq:Wardshiftquadratic} and \eq{eq:Wardrotation}. Hence, either they are zero or they are proportional to a ${\cal R}$ operator, and thus vanish when integrated over frequencies.
However, the second type of contributions, the crossed derivatives, are not controlled by these identities. One can argue that they are nevertheless suppressed by a combinatorial factor. One thus arrives at the conclusion that most of the sub-leading terms in the large wavenumber expansion are exactly controlled by symmetries and vanish, while the remaining terms, which are not controlled by the symmetries, appear small based on combinatorics. This suggests that corrections to standard Kraichnan-Leith scaling of equal-time quantities in the large wavenumber regime, \ie for the direct cascade, should be small, although it can only be argued  at this stage.

\bibliographystyle{jfm}


\end{document}